\documentclass[aps,prd, notitlepage, onecolumn,superscriptaddress,floatfix,letterpaper,nofootinbib, longbibliography]{revtex4-1}

\usepackage{amsmath, amsthm, amssymb,slashed,mathtools,tabu}
\usepackage{pifont}

\usepackage{multirow}

\usepackage[usenames, dvipsnames]{color}
\usepackage[svgnames]{xcolor}
\usepackage[colorlinks,citecolor=RoyalBlue, urlcolor=RoyalBlue, linkcolor=RoyalBlue ]{hyperref} 

\usepackage[normalem]{ulem}

\usepackage{booktabs}

\definecolor{mygray}{gray}{0.6}

\usepackage{upgreek}
\usepackage{bbm}

\newenvironment{myfont}[2][]{\csname#2\endcsname[#1]}{}

\usepackage{slashed}
\usepackage[makeroom]{cancel}
\usepackage[normalem]{ulem}
\usepackage{soul}
\newcommand{\stkout}[1]{\ifmmode\text{\sout{\ensuremath{#1}}}\else\sout{#1}\fi}

\usepackage{sseq}
\usepackage[all,cmtip]{xy}
\usepackage{tikz-cd}
\usepackage{tikz}
\usetikzlibrary{matrix}
\usetikzlibrary{decorations.markings}
\usetikzlibrary{tikzmark,decorations.pathreplacing,positioning}
\usepackage{amsfonts}

\newcommand{\bea}{\begin{eqnarray}}
\newcommand{\eea}{\end{eqnarray}}
\def\be{\begin{equation}}
\def\ee{\end{equation}}

\newcommand{\e}{\hspace{1pt}\mathrm{e}}

\newcommand{\ii}{\hspace{1pt}\mathrm{i}\hspace{1pt}}

\definecolor{red}{rgb}{1,0,0}
\definecolor{blue}{rgb}{0,0,1}
\definecolor{dblue}{rgb}{0,0,0.4}
\definecolor{green}{rgb}{0,1,0}
\definecolor{black}{rgb}{0,0,0}
\definecolor{white}{rgb}{1,1,1}

\definecolor{brn}{rgb}{.8,.4,.0}
\definecolor{redo}{rgb}{1,.5,.0}
\definecolor{ddgrn}{rgb}{0,0.4,0}
\definecolor{dgrn}{rgb}{0,0.55,0}
\definecolor{dbl}{rgb}{0,0,0.5}

\usepackage[bbgreekl]{mathbbol}
\usepackage{amscd}

\newcommand{\Z}{\mathbb{Z}}
\newcommand{\C}{\mathbb{C}}
\newcommand{\R}{\mathbb{R}}

\newcommand{\dd}{\hspace{1pt}\mathrm{d}}
\newcommand{\<}{\langle} 
\renewcommand{\>}{\rangle} 

\newcommand{\Refe}[1]{Ref.~[\onlinecite{#1}]}

\newcommand{\eq}[1]{(\ref{#1})} 
\newcommand{\eqq}[1]{eq.~(\ref{#1})}

\newcommand{\Tr}{{\rm Tr}}

\newcommand{\prt}{\partial}

\newcommand{\bpm}{\begin{pmatrix}}
\newcommand{\epm}{\end{pmatrix}}
\newcommand{\bmm}{\begin{matrix}}
\newcommand{\emm}{\end{matrix}}

\newcommand{\cD}{ {\cal D} }

\newcommand{\cL}{ {\cal L} } 
\newcommand{\cP}{ {\cal P} }

\def\CM{{\cal M}}
\def\CN{{\cal N}}
\def\CO{{\cal O}}

\def\CW{{\cal W}}

\def\Z{{\mathbb{Z}}}

\def\R{{\mathbb{R}}}
\def\C{{\mathbb{C}}}

\def\Tr{{\mathrm{Tr}}}

\def \Z{\mathbb{Z}}

\newcommand {\emptycomment}[1]{}

\def\TP{\mathrm{TP}}

\newcommand{\Spin}{{\rm Spin}}
\newcommand{\U}{{\rm U}}
\newcommand{\SU}{{\rm SU}}

\def\bZ{{\mathbf{Z}}}
\usepackage{centernot}
\newcommand{\nn}{{\nonumber}}
\newcommand{\Sec}[1]{Sec.~\ref{#1}}

\newcommand{\diag}{{\rm diag}}
\usepackage{enumitem} 
\usepackage{mathtools,amssymb,varwidth}

\newcommand{\Fig}[1]{Fig.~\ref{#1}} 
\newcommand{\Table}[1]{Table \ref{#1}}

\usepackage{datetime}

\newcommand{\rT}{{\rm T}}
\newcommand{\rR}{{\rm R}}

\newcommand{\QCD}{\mathrm{QCD}}

\newcommand{\rE}{{\rm E}}

\newcommand{\rCS}{{\rm CS}}
\newcommand{\PD}{{\rm PD}}

\newcommand{\su}{\mathfrak{su}}
\renewcommand{\u}{\mathfrak{u}}
\renewcommand{\so}{\mathfrak{so}}

\def\ra{\mathrm{a}}

\newcommand{\SM}{{\rm SM}}

\newcommand{\SMG}{{\rm SMG}}

\newcommand{\re}{{\rm e}} 

\def\cM{{\cal M}}

\begin{document}


\title{Strong CP Problem and Symmetric Mass Solution
}

\author{Juven Wang}
\email[]{jw@cmsa.fas.harvard.edu}

\affiliation{Center of Mathematical Sciences and Applications, Harvard University, MA 02138, USA}

\begin{abstract}

We propose a novel 
solution to the Strong CP problem 
--- to explain why SU(3) strong force  has a nearly zero theta angle $\bar\theta_3 \simeq 0$ 
 for the 4d Standard Model (SM). 
The new ingredient is Symmetric Mass Generation (SMG): symmetry-preserving mass or energy gap 
can be generated without breaking any symmetry $G$ and without any quadratic \emph{mean-field} mass deformation
as long as $G$ is all perturbative local and nonperturbative global anomaly-free. 
In our first model, we propose a disordered \emph{non-mean-field} SMG gap (instead of the ordered Anderson-Higgs-induced mass gap)
for the $u$ quark (or generally a set of quarks and leptons totally anomaly-free in $G$)
generated by multi-fermion interactions or by dynamical disordered mass fields, absorbing $\bar\theta_3$.
Another variant of this first model is the SMG gapping a hypothetical hidden full fourth family of SM fermions.
In our second model, we have a chiral SM and mirror SM together 
to respect the Nielsen-Ninomiya fermion-doubling
and a parity-reflection $\mathbb{Z}_2^{\rm PR}$ symmetry at high energy, so the $\bar\theta_3 = 0$. 
Then the SMG lifts only the mirror SM with a large energy gap 
but leaves the chiral SM at lower energy, 
which not only ``spontaneously'' breaks the parity-reflection symmetry maximally
but also relates our solution to solve another nonperturbative chiral fermion regularization problem by removing the fermion doubling. 
The predictive signature 
of both SMG-based models is that some SM fermions or mirror fermions are highly interacting
(beyond the conventional SM Higgs or SM gauge interactions) mediated through hypothetical direct multi-fermion or disordered mass-field interactions.


\end{abstract}


\maketitle

\tableofcontents

\newpage

\section{Introduction, the Problem, and a New Solution}

\subsection{Strong CP Problem}

The Standard Model (SM) is a chiral gauge theory
of local Lie algebra $\su(3) \times  \su(2) \times \u(1)_Y$
coupling to three families of 15 Weyl fermions of the representation
\bea \label{eq:SMrep}
&&\bar{d}_R \oplus {l}_L  \oplus q_L  \oplus \bar{u}_R \oplus   \bar{e}_R  \cr
&&\sim (\overline{\bf 3},{\bf 1})_{2} \oplus ({\bf 1},{\bf 2})_{-3}  
\oplus
({\bf 3},{\bf 2})_{1} \oplus (\overline{\bf 3},{\bf 1})_{-4} \oplus ({\bf 1},{\bf 1})_{6} 
\eea 
in a 4-dimensional spacetime (4d), with Yukawa-Higgs term to the electroweak Higgs field.
However, their corresponding theta terms inserted into the SM path integral with a weight factor
\bea
\exp(\ii \theta_j n^{(j)}), \quad n^{(j)} \equiv \frac{1}{8 \pi^2} g^2 \int \Tr[F^{(j)} \wedge F^{(j)}]
\eea
(with the instanton number or topological charge $n^{(j)}$ \cite{BelavinBPST1975, tHooft1976ripPRL, tHooft1976instanton} 
appropriately quantized and summed in the path integral for the 2-form field strength $F^{(j)}= \dd A^{(j)} - \ii g A^{(j)} \wedge A^{(j)}$ of 1-form gauge connection $A^{(j)}$
of the Lie algebra sector $j=1,2,3$ for
$\u(1)_Y \times  \su(2) \times \su(3)$\footnote{Carefully we use the mathfrak font 
$\u(1)_Y,  \su(2), \su(3)$ for the local structure Lie algebra. We use the $\U(1)$, $\SU(2)$, $\SU(3)$ for the global structure Lie group.})
have different physical outcomes.
The surprising experimental fact that 
Cabibbo-Kobayashi-Maskawa (CKM) matrix CP violating angle $\delta_{\rm CP}$ is order 1,
but the other CP violating theta angle 
is nearly zero ${|\bar{\theta}_{3}}| < 10^{-10}$ measured by the neutron electric dipole moment (EDM) \cite{Bakerhepex0602020, PDG2018} for
\bea 
\label{eq:theta3}
{\bar{\theta}_{3}} 
\equiv
{\theta_{3}}
+\arg (\det 
(M_{u}
M_{d}))
\eea
(where $M_{u}$ and $M_{d}$ are
two rank-3 
matrices specifying the Yukawa-Higgs coupling, 
for $u,c,t$-type quarks and $d,s,b$-type quarks respectively,
to be introduced later),
is known as the Strong CP problem \cite{Smith1957ht, Bakerhepex0602020, Abel2020PRL2001.11966, Dine2000TASI0011376, Hook2018TASI1812.02669}.

Since there is no particular reason (not even an anthropic reason) 
for the SM to have ${\bar{\theta}_{3}} \simeq 0$,
the typical Strong CP solutions
proposed in the past literature tend to
modify or enlarge the SM 
to include some beyond the SM assumptions. For example,\\
(1) some of the quarks (e.g., up quark) are massless \cite{tHooft1976ripPRL},\\
(2) extra continuous U(1) symmetry, then based on dynamical arguments on the ``spontaneous'' symmetry breaking which relaxes to ${\bar{\theta}_{3}} \simeq 0$, 
e.g., Peccei-Quinn symmetry with axions \cite{PecceiQuinn1977hhPRL,PecceiQuinn1977urPRD, Weinberg1977ma1978PRL, Wilczek1977pjPRL}
(although rigorously speaking, Peccei-Quinn symmetry is not a global symmetry once the Strong force $\su(3)$ is dynamically gauged).\\
(3) extra discrete P or CP symmetry imposed at high energy like Nelson-Barr
\cite{Nelson1983zb1984PLB,Barr1984qx1984PRL} or Parity Solution \cite{BabuMohapatra1989, BabuMohapatra1989rbPRD, BarrChangSenjanovic1991qxPRL, CraigGarciaKoszegiMcCune2012.13416}.

The purpose of this present work is to propose a new type of the Strong CP solution 
by involving the Symmetric Mass Generation (SMG, see a recent overview \cite{WangYou2204.14271})
--- fermions can become massive (also known as gapped) by a symmetric deformation from a massless (aka gapless) theory, 
without involving any symmetry breaking within an anomaly-free symmetry group.
In the context of the global symmetry $G$ being anomaly-free, 
this condition is well-known as the 't Hooft anomaly free in $G$ \cite{tHooft1979ratanomaly}.
In this work, we provide a new Strong CP solution to the 3+1d SM
based on 
the SMG mechanism 
in the SM chiral fermion sector alone,
or in its mirror fermion sector.\footnote{In a companion work, 
we had provided an SMG solution to the analogous CT or P problem for a 1+1d toy model \cite{WangCTorPProblem2207.14813}.}
Before we re-examine the mass origin to ask ``what is mass,''
we shall overview the roles of each ${\theta}_{j}$ for $j=1,2,3$ to set up the problem. 

\subsubsection{$\su(2)$'s $\theta_2$ angle}

Let us recall why the $\su(2)$'s $\theta_2$ can be rotated away to 0. The $\su(2)$ only couples to the left-handed doublet $\psi_L$,
including three colors $r,g,b$ of quarks $q_L$ and leptons $l_L$.
The pertinent Adler-Bell-Jackiw (ABJ)
perturbative local 
anomaly \cite{Adler1969gkABJ, Bell1969tsABJ}
is captured by the triangle Feynman graph with vertices
$\u(1)_L$-$\su(2)^2$. Thus, under the $\u(1)_L$ chiral symmetry transformation 
$$\psi_L \mapsto \e^{\ii \alpha_{{L}}} \psi_L,$$
the ${\theta_{2}}$ is rotated to
\bea
{\theta_{2}} \mapsto
{\theta_{2}} -
\sum_{f = 1,2,3}
(\alpha_{{q_L}_{r f}} + \alpha_{{q_L}_{g f}}
+ \alpha_{{q_L}_{b f}} +
\alpha_{l_{L f}}).
\eea
{Here
$\alpha_{{q_L}_{r f}}$, $\alpha_{{q_L}_{g f}}$, and $\alpha_{{q_L}_{b f}}$ are the left-handed quark $q$ chiral rotation phase 
from $\psi_L \mapsto \e^{\ii \alpha_{{L}}} \psi_L$, for the three kinds of colors (red $r$, green $g$, blue $b$ from the $\su(3)$ fundamental),
with $f=1,2,3$ the index for the three families of SM fermions.
}
We can 
do the baryon $\u(1)_{{\bf B}}$ rotation alone
(with $\alpha_{{q_L}_{r f}} = \alpha_{{q_L}_{g f}} = \alpha_{{q_L}_{b f}} =\frac{1}{3} \alpha_{\bf B}$, and the same angle for each right-handed quark $\alpha_{q_R}$,
thus also $\alpha_{{q_R}_{r f}} = \alpha_{{q_R}_{g f}} = \alpha_{{q_R}_{b f}} =\frac{1}{3} \alpha_{\bf B}$),
such that the $\u(1)_{{\bf B}}$ does not generate complex phase to the mass matrix 
$$
m_q \mapsto m_q \e^{\ii (\alpha_{q_L} - \alpha_{q_R})},
$$
because it is invariant under the vector symmetry rotation due to $\alpha_{q_L}= \alpha_{q_R}$.
The $\u(1)_{{\bf B}}$ also does not generate any extra complex phase to 
the CKM matrix, because 
$\u(1)_{{\bf B}}$ rotates all the left-handed quarks with the same angle
(thus $\alpha_{q_L}$ of $u$ and $d$ types cancel out in the CKM matrix). 
But the $\u(1)_{{\bf B}}$ can rotate ${\theta_{2}}$ to $0$. 
Notice above we leave out the lepton $l$'s lepton number $\u(1)_{{\bf L}}$ rotation (set $\alpha_{l_{L f}}=0$, also $\alpha_{l_{R f}}=0$),
so the argument
${\theta_{2}}=0$ is independent of whether we add the right-handed neutrino or not to the SM.
Namely, ${\theta_{2}}=0$ holds regardless of whether we have 15 or 16 Weyl fermions in any family of the SM. 

\subsubsection{$\u(1)_Y$'s $\theta_1$ angle}

The $\u(1)_Y$'s $\theta_1$ or $\u(1)_{\rm EM}$'s $\theta_{\rm EM}$: 
No finite energy and action configurations
 carry the instanton number $n^{(1)}$ {in the flat space $\R^4$ or $\R^{3,1}$}; thus typically people do not
worry $\theta_1$ within the SM vacuum 
(unless we encounter some domain wall or boundary effect that $\theta_1$ jumps, then there could be Witten effect on the other side of vacuum).

More precisely, under the $\u(1)_L$ or $\u(1)_R$ symmetry transformation
for all quarks/leptons of three families,
due to the ABJ perturbative local 
anomaly captured by the triangle Feynman graph with vertices
$\u(1)_L$-$\u(1)_Y^2$ and $\u(1)_R$-$\u(1)_Y^2$,
the ${\theta_{1}}$ is rotated to
\bea
{\theta_{1}} \mapsto
{\theta_{1}} -
\sum_{f = 1,2,3}
(3 \cdot(2 \alpha_{q_{L f}}  -
16 \alpha_{u_{R f}} - 4 \alpha_{d_{R f}}) + 18 \alpha_{l_{L f}}  -36 \alpha_{e_{R f}} ).
\eea
{The $f=1,2,3$ is the index for the three families of SM fermions.
Following the notation \eq{eq:SMrep}, here the chiral rotation phase
$\alpha_{q_{L f}}$ is for the left-handed doublet quark $q_L$, 
while the $\alpha_{u_{R f}}$ is for the right-handed singlet $u$-type quark $u_{R}$,
the $\alpha_{d_{R f}}$ is for the right-handed singlet $d$-type quark $d_{R}$;
the $\alpha_{l_{L f}}$ is for the left-handed doublet lepton $l_L$, 
and the $\alpha_{e_{R f}}$ is for the right-handed singlet electron $e_R$.
}
The continuous baryon minus lepton (${{\bf B} - {\bf L}}$) vector symmetry
(namely,
when $3 \alpha_{q} = \alpha_{\bf B} = - \alpha_{\bf L}$)
is anomaly-free under the chiral $\u(1)_Y$ and the chiral $\su(2)$:
$\u(1)_{{\bf B} - {\bf L}}$-$\u(1)_Y^2$
and 
$\u(1)_{{\bf B} - {\bf L}}$-$\su(2)^2$, 
also under the vector $\su(3)$:
$\u(1)_{{\bf B} - {\bf L}}$-$\su(3)^2$ (namely, the vector $\su(3)$ has no mix anomaly with any vector symmetry). 
This also means that the anomaly-free $\u(1)_{{\bf B} - {\bf L}}$ cannot rotate any $\theta_j$ for $j=1,2,3$.
But either $\u(1)_{\bf B}$ or $\u(1)_{\bf L}$ symmetry alone is anomalous 
thus can rotate the $\theta_1$ and $\theta_2$, 
except a certain linear combination $\theta_1 + 18 \theta_2$ cannot be rotated at all \cite{Tong2017oea1705.01853}.
This $\theta_1 + 18 \theta_2$ is indeed proportional to 
the $\theta_{\rm EM}$ for the electromagnetic $\U(1)_{\rm EM}$ ---
this is related to the fact that the {\bf B} and {\bf L} currents are both conserved under dynamical $\U(1)_{\rm EM}$.
(This is also related to the fact that $\Z_{36N_g,{\bf B}+{\bf L}}$ or $\Z_{2N_g,{\bf B}+{\bf L}}$ 
[more precisely $\Z_{36N_cN_g, {{\bf Q} + {N_c} {\bf L}} }$ or $\Z_{2N_cN_g, {{\bf Q} + {N_c} {\bf L}} }$ in a properly integer quantized 
quark number ${\bf Q}$ with the color number $N_c=3$]
is still respectively preserved under the dynamical $\U(1)$ or $\SU(2)$ gauge fields for the $N_g$ families (or generations) of fermions
\cite{KorenProton2204.01741, WangWanYouProton2204.08393}.)
But people do not concern $\U(1)_{\rm EM}$, since it has no finite energy instanton configuration {in the flat space $\R^4$ or $\R^{3,1}$} stated earlier. 

\subsubsection{$\su(3)$'s $\theta_3$ angle}

The $\su(3)$'s $\theta_3$:
Under the $\u(1)_L$ or $\u(1)_R$ symmetry transformation
for the $u$ or $d$ type of quarks of three families,
due to the ABJ perturbative local 
anomaly captured by the triangle Feynman graph with vertices
$\u(1)_L$-$\su(3)^2$ and $\u(1)_R$-$\su(3)^2$,
the ${\theta_{3}}$ is rotated to
\bea \label{eq:theta3-change}
{\theta_{3}} \mapsto
{\theta_{3}} -
\sum_{f = 1,2,3}
(\alpha_{u_{L f}} + \alpha_{d_{L f}} -
\alpha_{u_{R f}} - \alpha_{d_{R f}}).
\eea
{Again $f=1,2,3$ is the index for the three families of SM fermions.
Here the chiral rotation phase
$\alpha_{u_{L f}}$ is for the left-handed $u$-type quark $u_L$, 
while the $\alpha_{u_{R f}}$ is for the right-handed $u$-type quark $u_{R}$;
the $\alpha_{d_{L f}}$ is for the left-handed $d$-type quark $d_L$, 
while the $\alpha_{d_{R f}}$ is for the right-handed $d$-type quark $d_{R}$.}
Recall that in the quark sector, 
the rank-3 CKM matrix 
$
V_{\rm IJ}^{\rm CKM}=( U_u^{L\dagger}  U_d^L )_{\rm IJ} 
$
occurs in the 
term $W_{\mu}^+ J^{\mu +}_W + W_{\mu}^- J^{\mu -}_W$, 
such that the $W^{+}=W^{- *}$ gauge bosons generate the weak-flavor-changing EM-charged current
$
J^{\mu +}_W \equiv$
$
{\psi}^{{\rm I} \dagger}_{u_L}  \bar{\sigma}^\mu ( U_u^{L \dagger}  U_d^L )_{\rm IJ}  \psi_{d_R}^{\rm J}
$
$
= J^{\mu - *}_W
$
(here the family index ${\rm I}$ runs through three $u$ type quarks $u,c,t$ while ${\rm J}$ runs through three $d$ type quarks $d,s,b$).
The CKM matrix appearance is
due to the diagonalization to the mass eigenstates of quarks in the Yukawa-Higgs term for the Standard Model Higgs field $\phi_H$, 
with a hermitian conjugate term (H.c.),
\bea \label{eq:Yukawa-Higgs}
&&- (\lambda_{{d},{\rm IJ}} {\psi}^{{\rm I}\dagger}_{q_L} \phi_H \psi_{d_R}^{\rm J} 
+ \lambda_{{u},{\rm IJ}} \epsilon^{ab} {\psi}^{{\rm I}\dagger}_{{q_L} a} \phi_{Hb}^* \psi_{u_R}^{\rm J})
  +{\rm H.c.}\cr
 &&\mapsto
  - (\lambda^{\rm D}_{{d},{\rm II}} {\psi}^{{\rm I}\dagger}_{q_L} \phi_H \psi_{d_R}^{\rm I} 
+ \lambda^{\rm D}_{{u},{\rm II}} \epsilon^{ab} {\psi}^{{\rm I}\dagger}_{{q_L} a} \phi_{Hb}^* \psi_{u_R}^{\rm I})
  +{\rm H.c.}
   \;\;\;
\eea
So the generic non-symmetric non-hermitian matrices $\lambda_{d} = U_d^{L} \lambda^{\rm D}_d U_d^{R\dagger}$
and $\lambda_{u} = U_u^L \lambda^{\rm D}_u U_u^{R\dagger}$
can be diagonalized to $\lambda^{\rm D}_d$ and $\lambda^{\rm D}_u$ with positive eigenvalues.
The rank-3 $V^{\rm CKM}$ contains 3 rotational angles of SO(3), and another 6 complex phases.
The 6 left-handed quark chiral rotation
($\alpha_{u_{L f}}$ and $\alpha_{d_{L f}}$ for ${f = 1,2,3}$)
can remove 5 complex phases, but leave 1 complex phase $\delta_{\rm CP}$ remaining.

In this basis where the Yukawa-Higgs mass term is diagonalized,
only the physical
${\bar{\theta}_{3}}$
defined in \eq{eq:theta3} is invariant under any quark chiral rotation.\\
The previous $M_u$ and $M_d$ matrices in
\eq{eq:theta3} are indeed
$M_u=\lambda^{\rm D}_{{u}} \epsilon\phi_H^*$ and 
$M_d=\lambda^{\rm D}_{{d}} \phi_H$, or precisely in a mean-field manner when $\<\phi_H\>\neq 0$ gets a vacuum expectation value (vev),\footnote{To be absolutely clear,
a vacuum expectation value (vev) of any observable $\CO$ is evaluated as the expectation of $\CO$ in the ground state sector
$\<\CO \> \equiv \< \Psi_{\rm g.s.} |\CO |\Psi_{\rm g.s.}\>$, where $|\Psi_{\rm g.s.}\>$ is a ground state of quantum mechanical system, e.g., quantum field theory (QFT) and quantum many-body system.
It could also be evaluated in the path integral with insertion $\<\CO \> \equiv \frac{1}{Z} \int [\cD \cdots]  \CO\exp(\ii S[\cdots] )$, where $Z$ is the partition function (a path integral without insertion), $S$ is a QFT action, and $\cdots$ are some fields that are integrated over.
}
$$
\< M_{{u},{\rm IJ}}\> =\lambda^{\rm D}_{{u},{\rm IJ}} \epsilon \<\phi_H^*\>,
\quad
\<  M_{{d},{\rm IJ}} \> =\lambda^{\rm D}_{{d},{\rm IJ}}  \<  \phi_H\>.
$$

\noindent
$\bullet$ 
The charge-conjugation-parity CP transformation (similarly also under the time-reversal T) 
sends the instanton number $n^{(j)} \mapsto -n^{(j)}$ for all $j=1,2,3$. 
Thus given the periodicity 
${\theta}_{j} = {\theta}_{j} + 2 \pi {\rm N}_j$
(where ${\rm N}_j=1$ typically, but ${\rm N}_j$ can be other integers, for SM with different $p=1,2,3,6$, or in the presence of fractional instantons \cite{AharonyASY2013hdaSeiberg1305.0318})
implies that only the SM vacuum at  
${\theta}_{j}=0$ or ${\theta}_{j}=  \pi {\rm N}_j$ is kinematically CP and T invariant.

The aforementioned fact says that the ${\theta}_{1}$ and ${\theta}_{2}$ play no important roles,
so there are only two CP violating sources in the SM quark sector: the $\delta_{\rm CP}$ due to the weak force
and the ${\bar{\theta}_{3}}$ due to the strong force.
The experimental fact that 
CKM matrix CP violating angle $\delta_{\rm CP}$ is order 1,
but ${\bar{\theta}_{3}}$ is nearly zero \cite{Bakerhepex0602020, PDG2018},
is the Strong CP problem. Another way to phrase 
the Strong CP problem is that why does quantum chromodynamics (QCD) alone preserve CP symmetry? 

\subsection{A New Solution based on Symmetry Mass Generation}
\label{sec:I-SMG}

In the next sections, 
we will provide a Strong CP solution based on a mass-generating mechanism known as
Symmetric Mass Generation (SMG, see a recent overview \cite{WangYou2204.14271}). 
It turns out that our SMG solution to the Strong CP problem applies to
all four versions of SM gauge groups 
%
\bea
G_{\SM_p} \equiv({\SU(3)} \times {\SU(2)} \times {\U(1)_Y})/{\Z_p},
\eea 
namely, it works for any $p=1,2,3,6$  \cite{AharonyASY2013hdaSeiberg1305.0318, Tong2017oea1705.01853, Wan2019sooWWZHAHSII1912.13504}.
Here in this subsection \Sec{sec:I-SMG},
we will quickly and briefly sketch the key essence of the SMG solution: 
Why does the SMG helps to solve the Strong CP problem?\footnote{In 3+1d, 
the theta term of non-abelian Yang-Mills field strength $\frac{\theta}{8 \pi^2} \int \Tr[F \wedge F]$
 is C even, P odd, and T odd. 
In 1+1d, the theta term of abelian field strength $\frac{\theta}{2 \pi} \int F$ is
C odd, P odd, and T even. 
We could also say that the small 3+1d theta term implies the P, T, CP, or CT problem in 3+1d; typically it is called the Strong CP problem for the SU(3) strong force.
We could say that the small 1+1d theta term implies the C, P, CT, or PT problem \cite{WangCTorPProblem2207.14813}.
So in general even-dimensional spacetime (both $4n$ or $4n+2$ dimensions), 
the small theta term implies universally the P or CT problem.
\Refe{WangCTorPProblem2207.14813} provides the SMG solution to the 1+1d P or CT problem.
In this present work, we tackle the 3+1d P or CT problem known as the Strong CP problem.
 }

{The conceptual idea of our solution focus on the re-examination and re-interpretation of the role of the mass matrix $M$ in \eq{eq:theta3}.
We point out that the previous studies and solutions to the Strong CP problem
in the past literature 
only or mainly rely on the mean-field mass such that mass matrix $M$ is obtained via the 
mean-field expectation value $\< M\>$. 
So what \eq{eq:theta3} really means schematically is
\bea 
\label{eq:theta}
{\bar{\theta}_{}} 
\equiv
{\theta_{}}
+\arg (\det 
\< M\>).
\eea
Meanwhile, based on our understanding of the neutron electric dipole moment (EDM) 
\cite{Bakerhepex0602020, PDG2018} (the available physical observable measuring ${\bar{\theta}_{}}$), 
we also find that the neutron EDM only measures the mean-field contribution (see a mean-field explanation in \cite{Hook2018TASI1812.02669}), 
agreeing with our speculation 
that
${\bar{\theta}_{}} 
\equiv
{\theta_{}}
+\arg (\det 
\< M\>)$. 
Henceforth,
we will take \eq{eq:theta} as our assumption and corrected definition of the ${\bar{\theta}_{}}$.

Here in comparison to the Strong CP problem in \eq{eq:theta3}, we can set the ${\bar{\theta}_{}}={\bar{\theta}_{3}}$ and ${\theta_{}}={\theta_{3}}$, 
while the fermion mass matrix $M$ consists of the contribution of quark mass matrix $M_u$ and $M_d$.
Namely, the experimental measurement of the ${\bar{\theta}_{}}$ (such as the neutron EDM in the SM)
really involves the mean-field mass matrix $\< M\>$.
There the mass matrix $M$ is schematically obtained from the fermion bilinear or quadratic term in the Lagrangian 
\bea
\label{eq:mass-term}
\xi_{\rm I} M_{\rm IJ} \psi_{\rm J} + {\rm H.c.},
\eea 
where $\xi$ and $\psi$ are fermion fields in appropriate representations. The $M_{\rm IJ}$ may receive a contribution from the dynamical fields like Higgs $\phi_H$, etc., that gets a
mean-field vacuum expectation value (vev) as $\<\phi_H \> \neq 0$.
The whole $\xi_{\rm I} M_{\rm IJ} \psi_{\rm J}$ is a singlet scalar in a trivial representation of the Lorentz group.}

{Now, our new key idea is that SMG mechanism \cite{WangYou2204.14271} can go beyond the mean-field mass,
such that the SMG deformation
\bea \label{eq:SMG-mass}
\xi_{\rm I} \CO_{\rm SMG, IJ} \psi_{\rm J} + {\rm H.c.}, \text{ such that } \< \CO_{\rm SMG, IJ}\>=0,
\eea 
receives no mean-field value in $\< \CO_{\rm SMG, IJ}\>=0$ and also $\< \xi_{\rm I} \psi_{\rm J}\>=0$ but this \eq{eq:SMG-mass} can still give 
symmetry-preserving \emph{non-mean-field mass} energy gaps to the full set of fermions by preserving an anomaly-free symmetry $G$. 
The \eq{eq:SMG-mass} may involve the \emph{multi-fermion interaction}, e.g., 
\bea
\xi_{\rm I} \CO_{\rm SMG, IJ} \psi_{\rm J} +  {\rm H.c.}
=
g_{\rm SMG} \big( \xi_{\rm I}  ( \psi_{q' } \dots \psi_{q''} + \dots  )  \psi_{\rm J}  +  {\rm H.c.}\big)
 \eea 
(where interactions between $\xi_{\rm I}$, $\psi_{q'}$, $\psi_{q''}$, and $\psi_{\rm J}$, etc., with the SMG interaction strength $g_{\rm SMG}$)
and \emph{disordered mass field interaction} 
(where $\CO_{\rm SMG, IJ}$ 
may be regarded as  the field operator that becomes disordered in the order parameter target space), 
beyond the quadratic fermion interaction \cite{WangYou2204.14271}.
The reason that $\< \CO_{\rm SMG, IJ}\>=0$ is due to that the condensation of these $\CO_{\rm SMG, IJ}$ operators 
would often break the $G$ symmetry (so nonzero $\< \CO_{\rm SMG, IJ}\> \neq 0$ often implies no SMG) in the low-energy ground state sector. 
In our approach, we shall generalize the mass matrix 
in \eq{eq:theta} and \eq{eq:mass-term} to $\mathbf{M}$ so
to include both the \emph{mean-field} mass $M=\< M \>$ (e.g., from Higgs) and the SMG's \emph{non-mean-field} mass $\< \CO_{\rm SMG, IJ}\>=0$: 
\bea \label{eq:def-mass}
\mathbf{M}_{\rm IJ} &=&M_{\rm IJ}+\CO_{\rm SMG, IJ},\cr
\<\mathbf{M}_{\rm IJ} \>&=& \< M_{\rm IJ} \> , \quad \text{ while } \<\CO_{\rm SMG, IJ}\>=0,\cr
\xi_{\rm I} \mathbf{M}_{\rm IJ} \psi_{\rm J} + {\rm H.c.}  &=& \xi_{\rm I} (M_{\rm IJ} + \CO_{\rm SMG, IJ}) \psi_{\rm J} + {\rm H.c.}, \cr 
{\bar{\theta}_{}} 
&\equiv&
{\theta_{}}
+\arg (\det 
\< \mathbf{M}\>)
=
{\theta_{}}
+\arg (\det 
\< M\>).
\eea 
Here we should also provide a definition to extract the mass matrix $\mathbf{M}_{\rm IJ}$ from the QFT partition function $Z$ and 
its action $S= \int \dd^4 x \cL$ with lagrangian $\cL$. 
In 4d, we can write every fermion in terms of the left-handed Weyl fermions (here denoted as $\xi_{\rm I}$ and $\psi_{\rm J}$),
then\footnote{Here $\mathbf{M}_{\rm IJ} = - \frac{\delta^2 \cL}{(\delta \xi_{\rm I}) (\delta \psi_{\rm J})}$ extracts the generalized mass matrix but
avoids extracting the Dirac operator. Because the Dirac operator is obtained from the diagonal and the same species of 
Weyl fermions in particle and anti-particle paired in 
$ \frac{\delta^2 \cL}{(\delta \psi^\dagger_{\rm I}) (\delta \psi_{\rm I})}$,
or in the left-handed or right-handed Weyl fermion basis, $\frac{\delta^2 \cL}{(\delta \psi^\dagger_{L}) (\delta \psi_{L})}$ or 
$\frac{\delta^2 \cL}{(\delta \psi^\dagger_{R}) (\delta \psi_{R})}$. Here we write $- \frac{\delta^2 \cL}{(\delta \xi_{\rm I}) (\delta \psi_{\rm J})}$ all in the left-handed particle basis,
thus extract only the $\mathbf{M}_{\rm IJ}$ but not the Dirac operator.}
\bea
\mathbf{M}_{\rm IJ} = - \frac{\delta^2 \cL}{(\delta \xi_{\rm I}) (\delta \psi_{\rm J})}, \text{ where $\cL$ contains a generic term $-(\xi_{\rm I} \mathbf{M}_{\rm IJ} \psi_{\rm J} + {\rm H.c.})$.}
\eea
Then our solution to the Strong CP problem requires at least \emph{any one} of the fermions (call this fermion $\zeta$) in the full theory  
to receive no mean-field mass at all (so there is at least one zero eigenvalue of the mean-field mass matrix $\< M_{\rm IJ} \>$) 
but this fermion $\zeta$ can still be massive or gapped due to the SMG contribution
(namely $\CO_{\rm SMG, IJ}\neq 0$ but $\<\CO_{\rm SMG, IJ}\>=0$).
Thus, at least one of the eigenvalues of the mean-field mass matrix $\< M\>$ being zero
implies that the $\det \<\mathbf{M} \>= \det \< M\> =0$.
In that case,
${\bar{\theta}_{}} 
={\theta_{}}.$ 
Next, we can do the $\u(1)_{\rm chiral}$ chiral transformation only on this specific chiral fermion 
\bea 
\zeta \to \e^{\ii \upalpha } \zeta
\eea that has no mean-field mass, but $\zeta$ can receive the non-mean-field SMG mass.
{As long as there is an ABJ perturbative local 
anomaly between the $\u(1)_{\rm chiral}$ and the gauge Lie algebra of the theta term
(in the case of QCD's ${\theta_{3}}$, it is captured by the triangle Feynman graph with vertices
$\u(1)_{\rm chiral}$-$\su(3)^2$)}, this chiral transformation will send 
\bea
{\theta_{}} \mapsto {\theta_{}} - \upalpha,
\eea
while this 
also sends
\bea
M_{\rm IJ} + \CO_{\rm SMG, IJ} \mapsto M_{\rm IJ}(\upalpha) + \CO_{\rm SMG, IJ}(\upalpha)
\eea
presumably with $\upalpha$ dependence.
However, the $$\< M_{\rm IJ}\>=\<M_{\rm IJ}(\upalpha)\>$$ is not changed because the $\zeta$ has no mean-field mass.
The mean-field $$\<\CO_{\rm SMG, IJ}(\upalpha)\>=0$$ anyway so it does not contribute to the ${\bar{\theta}}$.
So we end up redefining ${\bar{\theta}_{}}$ by a chiral transformation, 
with $\det \<\mathbf{M} (\upalpha) \>=0$ still, 
so 
\bea \label{eq:bartheta0}
{\bar{\theta}_{}}= {\theta_{}} - \upalpha =0
\eea can be appropriately chosen to be zero.
This provides the solution of the Strong CP problem: The ${\bar{\theta}_{}}$ is zero for the entire theory.
The ${\bar{\theta}_{}}=0$ in principle solves the Strong CP problem 
at some energy scale, then we provide arguments how ${\bar{\theta}_{}}$ remains zero or small at the IR low energy theory.
}

{\bf Order to Disorder the ${\theta_{}}$ and mass field}:
{There is another interpretation to look at our Strong CP solution setting ${\bar{\theta}_{}}=0$. 
We are looking at the \emph{disordered} phase of the dynamical ${\bar{\theta}_{}}$ 
(which includes the \emph{disordered} dynamical ${\theta_{}}$ and dynamical complex phase of mass field 
$\arg (\det \< \mathbf{M}\>)$).
Instead, Peccei-Quinn solution with axions \cite{PecceiQuinn1977hhPRL,PecceiQuinn1977urPRD, Weinberg1977ma1978PRL, Wilczek1977pjPRL}
looked at the \emph{ordered} phase of the dynamical ${\bar{\theta}_{}}$ (i.e., the small fluctuation around the vacuum expectation value $\< {\bar{\theta}_{}}\>$ gives rise to axion mode). The \emph{ordered} phase to the \emph{disordered} phase of the ${\bar{\theta}_{}}$ is analogous to the superfluid-to-insulator type of phase transition in the condensed matter  \cite{Fisher1989zza}.\footnote{Even more precisely, the continuous Peccei-Quinn symmetry
would \emph{not} be a global symmetry once the internal symmetry of the gauge group $G$ is dynamically gauged. 
Due to the mixed anomaly between the $G$ and Peccei-Quinn symmetry,
the classical Peccei-Quinn symmetry is broken down by the Adler-Bell-Jackiw (ABJ) anomaly 
\cite{Adler1969gkABJ, Bell1969tsABJ} to its discrete subgroup.
So rigorously speaking, neither \emph{superfluid} nor \emph{algebraic superfluid} exists 
as Peccei-Quinn symmetry-breaking phase in the $G$ gauge theory.
Nonetheless, at least in the weakly gauge or the global symmetry limit of $G$,
we expect that the physical intuitive picture of the superfluid to insulator transition analogy still applies.
\label{ft:Peccei-Quinn-transition}
}\\
$\bullet$ When the ${\bar{\theta}_{}}$ is in the ordered phase, it makes sense to ask the value of $ \< \exp(\ii{\bar{\theta}_{}}) \>$, which determines the
orientation of the ${\bar{\theta}_{}}$-clock and how it affects the CT or P breaking (also T or CP breaking in 3+1d).\\
$\bullet$ However, when the ${\bar{\theta}_{}}$ is in the disordered phase, 
it makes no sense to extract the mean-field value of $\< \exp(\ii{\bar{\theta}_{}}) \>=0$ --- because the
${\bar{\theta}_{}}$-clock becomes fully disoriented, with no CT or P breaking (also T or CP breaking in 3+1d).
The appropriately designed \emph{disordered} phase of the dynamical ${\bar{\theta}_{}}$ can also give the non-mean-field mass gap to a set of fermions via the SMG. 
}

To solve the Strong CP problem,
{we will implement this particular non-mean-field massive fermion $\zeta$ in two approaches:}

In  \Sec{sec:FirstSolution}, we only need the original \emph{chiral fermion} theory in the SM, without adding any \emph{mirror fermion} sector.
In this case, some of the original \emph{chiral} fermion say $\zeta$ in the SM receives its full mass from the SMG, not from the Higgs condensation.
Another possibility is that if there is an extra family of SM fermions (e.g. the 4th family), 
such that the extra family of SM fermions receive only the SMG mass (but no other mean-field mass), 
then it can also make ${\bar{\theta}_{3}}=0$.

In \Sec{sec:SecondSolution}, instead, we will choose the $\zeta$ as a new set of \emph{mirror fermions} (namely, not the SM fermions)
being fully gapped by SMG. The \emph{mirror fermion} sector is the \emph{fermion doubling} of the original \emph{chiral fermion} theory.

{\bf Quark masses}: The crucial issue is --- what do we really mean by the quark masses?
Since in the real-world experiment, we do \emph{not} directly measure the quark mass as the quarks are confined in the hadrons such as mesons or baryons
due to the color confinement,
we have to clarify how do we measure the quark mass? 
At which energy scale and by which method do we extract the quark mass?

In particular, 't Hooft massless $u$ quark solution \cite{tHooft1976ripPRL} that requires $m_u =0$
has been ruled out by the lattice QCD data.
So whatever new Strong CP solution that we will provide,
we must interpret the lattice QCD data correctly without conflicting with our new Strong CP solution.
According to Particle Data Group (PDG)
\cite{PDG-quark}, the current quark mass $m_u=2.16$ MeV
is an input to the QCD lagrangian at the energy scale $E_{\overline{\text{MS}}} =  2$ GeV
based on the modified minimal subtraction ($\overline{\text{MS}}$) renormalization scheme. 
Other current quark masses $m_q$ are also nonzero,
in order to produce the correct hadron mass 
(e.g. meson such as the pion mass 135 MeV or baryon such as the proton mass 938 MeV)
observed at the low energy (say the $\Lambda_{\QCD}$ scale 200 MeV)
in Nature confirmed by experiments  \cite{PDG-quark}.
We thus know this lattice data indeed rules out the 't Hooft massless $u$ quark solution \cite{tHooft1976ripPRL}
because this solution requires to have zero current $u$ quark masses $m_u=0$
or zero Yukawa-Higgs-$u$-quark coupling
for the QCD bare lagrangian under the electroweak Higgs mechanism
at any energy scale above the chiral symmetry breaking scale 
(namely, at any energy scale
below the electroweak scale $E_{{\text{EW}}} \sim 246$ GeV
and above the QCD confinement scale $\Lambda_{\QCD}$ scale $\sim$ 200 MeV,
before the chiral condensate $\< \bar{\psi}_{q} \psi_q \> \neq 0$ kicks in).

We will advocate candidate resolutions to the Strong CP problem compatible with the up-to-date lattice QCD data \cite{PDG-quark}.
One possibility is that the $u$ quarks obtain its current quark mass $m_u$ 
\emph{not from} the electroweak Higgs mechanism \emph{but from} the SMG in the real-world Nature.
If so, we propose to study a toy model such that a QCD bare lagrangian including the SMG interactions between a set of quarks that preserve an anomaly-free symmetry.
We imagine in this new scenario, 
such that even if both the $\su(3)$ color Yang-Mills gauge force and the electroweak Yukawa-Higgs couplings are turned off to zero,
there are still intrinsic direct multi-quark or multi-fermion (between quarks and leptons) interactions via
the SMG interactions.
Namely, in this new scenario, 
the familiar QCD lagrangian is modified to
\bea \label{eq:modified-QCD-SMG}
-\frac{1}{4} F_{\mu\nu}^\ra F_{}^{\ra \mu\nu}
 {-\, \frac{\bar{\theta}  }{{64} \pi^2} {g}^2 \epsilon^{\mu\nu \mu'\nu'} F_{\mu\nu}^\ra F_{\mu'\nu'}^\ra}
 +\bar \psi(\ii \gamma^\mu D_\mu -
  {
 \begin{pmatrix}
\lambda^{\rm D}_{{u}} \< \phi_H \> & \\
& \lambda^{\rm D}_{{q}, {\rm I I}} \delta_{\rm I J}  \< \phi_H \>
  \end{pmatrix}
} )\psi 
+ 
\big(\psi' \CO_{\rm SMG} \psi +  {\rm H.c.}\big)
\eea
where the Lie algebra generator index $\ra$ is summed over 
and the $\mu,\nu$ are spacetime vector indices are contracted summed.
This modified QCD lagrangian consists of three kinds of interactions between quarks:\\
\noindent
{\bf {(1)}} The $\su(3)$ color Yang-Mills gauge force via
$\bar \psi(\ii \gamma^\mu D_\mu) \psi$ with the nonabelian covariant derivative $D_\mu \equiv \prt_\mu - \ii g A_\mu^\ra T^\ra$.\\
\noindent
{\bf {(2)}} The electroweak Yukawa-Higgs couplings
$\bar \psi \lambda^{\rm D} \< \phi_H \> \psi$ that pairs left-handed and right-handed quarks.
With an appropriate choice of basis within the SM matter content, now the diagonalized Yukawa-Higgs coupling 
$\lambda^{\rm D}_{{u}}$, $\lambda^{\rm D}_{{q}, {\rm I I}}$, $\< \phi_H \> \in \R$ 
all are chosen to be real. The $m_u = \lambda^{\rm D}_{{u}} \langle \epsilon\phi_H^* \rangle$
with the SU(2) doublet $\epsilon\phi_H^*$ notation is simplified to 
$m_u = \lambda^{\rm D}_{{u}} \langle \phi_H \rangle$ once we choose $\< \phi_H \> \in \R$ via a basis choice based on the appropriate $\su(2) \times \u(1)_Y$
gauge transformation. Similarly, this basis choice  works for other $u$ type quarks.
\\
\noindent
{\bf {(3)}} 
The schematic SMG interaction 
\bea
\psi' \CO_{\rm SMG} \psi +  {\rm H.c.}
\equiv
g_{\rm SMG} \big( 
 \psi'  ( \psi_{q' } \dots \psi_{q''} + \dots  )  \psi +  {\rm H.c.}
 \big)
 \eea can consist 
of the
direct multi-quark or multi-fermion interactions (here interactions between $\psi'$, $\psi_{q'}$, $\psi_{q''}$, and $\psi$, etc.), 
being non-mean-field with $\<\CO_{\rm SMG}\>=0$ 
and $\< \psi' \psi \>=0$. The SMG interaction is model-dependent and must be carefully tailor-made and designed. 
In \Sec{sec:toy-SM}. we will provide an explicit form of the SMG interaction for the full SM matter contents.

A possible resolution to the lattice QCD data \cite{PDG-quark}
that seems to require $m_u \neq 0$
is that we can follow 't Hooft's $\lambda^{\rm D}_{{u}}=0$ and the $u$ quark has
a zero current mass $m_u=\lambda^{\rm D}_{{u}} \< \phi_H \> =0$,
the SMG interaction $\psi' \CO_{\rm SMG} \psi +  {\rm H.c.}$
can still give the $u$ quark a nonzero SMG mass.
Other fermions involved in the SMG also get a portion of SMG mass, other than the electroweak Higgs-induced mass
$\lambda^{\rm D}_{{q}, {\rm I I}} \< \phi_H \> \neq 0$. 
%
Some comments on this \eq{eq:modified-QCD-SMG} solution:\\
$\bullet$ Theoretically, this is a candidate solution --- as long as we can verify that by re-running the lattice QCD simulation, 
the new QCD lagrangian \eq{eq:modified-QCD-SMG}
with a zero current mass $m_u=\lambda^{\rm D}_{{u}} \< \phi_H \> =0$ but with the appropriate SMG interaction
at $E_{\overline{\text{MS}}} =  2$ GeV                                                                                                                                                                                                                                                                                                                                                                                                                                                                                                                                                                                                                                                                                                                                                                                                                                                                                                                                                                                                                                                                                                                                                                                                                                                                                                                                                                                                                                                                                                                                                                                                                                                                                     
can still generate the correct hadron masses at low energy $\Lambda_{\QCD}$ scale.
If so, this will be sufficient to satisfy the lattice simulation constraint and the low energy hadron mass constraint.
In \Sec{sec:toy-SM}. we will provide an SMG interaction, and then also explain why this solves the Strong CP problem (at least theoretically).
\\
$\bullet$  Phenomenologically, this candidate solution may suffer constraints by experiments.                                                                                              
For example, if the SM's $u$ quark and other quarks are indeed involved in the SMG interaction,
then at a sufficiently high enough energy, 
the non-renormalizable multi-fermion interaction becomes nonperturbatively dominant.
On one hand, this direct multi-fermion interaction
indicates some of the quarks may not be genuinely \emph{asymptotic-free} at arbitrary high energy, which indicates some falsifiable experimental predictions.
On the other hand, this also means that there could be a different ultra-violet  (UV) completion of SMG interaction by a parent UV theory 
lurking behind at very high energy.

The main purpose of this article is to provide a \emph{theoretical SMG solution} to the Strong CP problem.
Although we will try to elaborate on how this theoretical solution can fit into phenomenologically constraints to be a \emph{phenomenological SMG solution},
we will not pursue detailed phenomenological parameter fittings in this initial work, 
as those fittings can be very data dependent and model-dependent.

Our strategy is that if this specific SMG solution \eq{eq:modified-QCD-SMG} (applied to the chiral fermion)
elaborated in \Sec{sec:FirstSolution} is not favored due to (known or unknown) phenomenological constraints,
we will look for a different SMG solution (applied to the mirror fermion) in \Sec{sec:SecondSolution}.
In general, there must be a certain version of SMG solution to the Strong CP problem 
being not only theoretically but also phenomenologically viable. The readers can decide which version is more promising
to be realized in Nature  
based on their own taste.

The remaining part of the article is organized as follows:

In \Sec{sec:Mass}, we overview and contrast between
the familiar {\bf symmetry-breaking} mass
and the {\bf Symmetric Mass Generation} (SMG via a {\bf symmetry-preserving} mechanism) 
or the {\bf Topological Mass Generation} (TMG, also symmetry-preserving by via a {\bf symmetry-extension} mechanism \cite{Wang2018Symmetric1705.06728}).

In \Sec{sec:FirstSolution} and \Sec{sec:SecondSolution},
these two aforementioned approaches (without or with mirror fermions) provide two different solutions to the Strong CP problem.

In \Sec{sec:Conclusion}, we conclude and make comparisons with other Strong CP solutions.

\newpage
\section{Mass Mechanisms: Old and New}
 \label{sec:Mass}
 
\subsection{Properties of Mass}
Below we recall some different concepts of masses, in particular the fermion masses, 
following \cite{WangYou2204.14271}.
\begin{enumerate}[leftmargin=-0mm, label=\textcolor{blue}{\arabic*}., ref={\arabic*}]

\item \label{rest-mass}
{\bf 
Rest mass}: We can define the mass as an energy gap between the excitation energy $E_{\rm{excited}} $ 
above the ground state $E_0$,
\bea
m_{\rm{rest}}= \Delta E = E_{\rm{excited}} - E_0.
\eea
This definition works in general for many-body interacting quantum systems that can include arbitrary interactions 
without assuming any mean-field or quadratic Hamiltonian realization.
This definition also works for both relativistic or nonrelativistic systems, also for both quantum field theories or not.

When the excitation is a fermion in a translational symmetry invariant theory where the momentum ${\vec{k}}$ is a good quantum number,
we could define the rest mass as the energy gap between the minimal value of energy band $E_{\vec{k}}$ above the chemical potential $\mu$
as $m_{\rm{rest}}={\rm{min}}(E_{\vec{k}})- \mu$.

\item \label{inertial-mass}
{\bf 
Inertial mass}: If there is an energy band $E_{\vec{k}}$ description,
we can define a different type of mass --- 
the inverse curvature of the fermion energy band dispersion:
\bea
m_{\rm{inertial}}= \lim_{{\vec{k}} \to {\vec{k}}_{\rm{min}}} (\nabla^2_{\vec{k}}(E_{\vec{k}}))^{-1}.
\eea
Note that this mass $m_{\rm{inertial}}$ is an inertial mass,
where we can read its effective mass from the energy band dispersion
$
E_{\vec{k}}=\sqrt{{\vec{k}}^2+|m|^2}=|m|+ \frac{{\vec{k}}^2}{2m} +\dots
$

\item \label{correlator}
{\bf 
Mass read from the correlator}: There is also the mass $m$ read from the inverse of the correlation length $\xi = \frac{1}{m}$.
The correlation length $\xi$ can be read from the exponentially decayed two-point 
fermion-fermion correlation function:
\bea
\<\bar\psi (x_1) \psi (x_2)\> {+ {\rm H.c.}} \propto \exp(- |m| |x_1- x_2|) = \exp(- \frac{|x_1- x_2|}{|\xi|}).
\eea
Here $\bar\psi$ and $\psi$ are the shorthands of appropriate fermion creation and annihilation operators;
for example the $\bar\psi=\psi^\dagger\gamma^0$ for a Dirac fermion.
In 4d spacetime, for Dirac fermion,
we have $\bar\psi (x_1) \psi (x_2)=\psi_L^\dagger (x_1) \psi_R (x_2) + \psi_R^\dagger (x_1) \psi_L (x_2)$.
For Majorana fermion,
we have to pair the same Weyl fermion by $\psi_L^\dagger (x_1) (\ii \sigma^2) \psi^{*}_L (x_2)+ {\rm H.c.}$ with the standard Pauli matrices 
$(\ii \sigma^2)_{ab}=\epsilon_{ab}$.

\item \label{bilinear} 
{\bf 
Bilinear mean-field mass condensation}:
We can also read the mean-field mass $\<m\>$ from the two-point function when the two points $x_1$ and $x_2$ coincide:
\bea
\<\psi'(x) \psi(x)\>= \<m\> f(|m|),
\eea
where the function $f$ is model dependent.
This relation is due to the linear response theory on $m \psi'(x) \psi(x)$
at the mean-field level contains $\<m\>\<\psi'(x) \psi(x)\>$ such that $\<m\> \propto \<\psi'(x) \psi(x)\>$.
Here $\psi'$ and $\psi$ may be paired by Dirac or Majorana type mass pairing.
\end{enumerate}

The above different kinds of definitions of masses may not always be equivalent.
Here are some comments on these definitions of masses:\\
\noindent
$\bullet$ In a relativistic quantum field theory with translational symmetry (so the momentum ${\vec{k}}$ is a good quantum number),
the rest mass (\ref{rest-mass}), the inertial mass (\ref{inertial-mass}), 
and the mass read from the correlator (\ref{correlator}) all become the same mass.
However, in a nonrelativistic quantum system, 
the rest mass (\ref{rest-mass}) and the inertial mass (\ref{inertial-mass}) are generally different ---
the rest mass (\ref{rest-mass}) is about the excitation energy gap $m_{\rm{rest}}= \Delta E$,
the inertial mass (\ref{inertial-mass}) is the inverse curvature of energy band dispersion.
\\
\noindent
$\bullet$ The rest mass (\ref{rest-mass}) and the mass read from the correlator (\ref{correlator}) are the more general form of mass in a quantum theory, 
without or with relativity, beyond the mean-field theory, beyond a quadratic Hamiltonian, 
and beyond the energy band theory. When we refer to the mass hereafter, we mean these two (\ref{rest-mass}) and (\ref{correlator}).\\
\noindent
$\bullet$ The bilinear mass condensation (\ref{bilinear}) detects the \emph{mean-field mass} (like Higgs-induced mass), but cannot detect the SMG-induced \emph{non-mean-field mass}.\\
\noindent 
$\bullet$ {\bf Fermionic Green's Function on \emph{mean-field mass} vs \emph{non-mean-field mass}}:\\
Consider a fermionic Green's Function $\mathcal{G}(k)=\mathcal{G}(E,\vec{k} )$ in the energy $E$-momentum $\vec{k}$ space
with $k^\mu k_\mu= E^2 -\vec{k}^2$, in the time and spatial translational invariant system.
Given a generic mass matrix $\mathbf{M}$ read from \eq{eq:def-mass},\\
(i) Green's Function with the mean-field mass $\<\mathbf{M}\>\equiv \<M\>$ shows
\bea \label{eq:mean-Green's}
\mathcal{G}(k)=\frac{\gamma^\mu k_\mu + \<\mathbf{M} \>}{k^\mu k_\mu -\<\mathbf{M}\>^2}
=\frac{\gamma^\mu k_\mu + \<M\> }{k^\mu k_\mu -\<M\>^2}
\eea
There are well-known {\bf Green's function poles} 
$\mathcal{G}(E= \pm \sqrt{\vec{k}^2 + \<M\>^2},\vec{k} ) \to \infty$
around the dispersion relation ${k^\mu k_\mu -\<M\>^2}$ $=$ $E^2 - \vec{k}^2 -\<M\>^2$ $=$ $0$.
Crossing the locations of these poles (along the dispersion $E= \pm \sqrt{\vec{k}^2 + \<M\>^2}$,
the $\det(\mathcal{G})$ flips sign, between $\det(\mathcal{G})>0$ and $\det(\mathcal{G})<0$, where the determinant is evaluated in the spinor and flavor component space.
\\
(ii) Green's Function with the SMG-induced non-mean-field mass ($\<\CO_{\rm SMG}\>=0$)
shows
\bea \label{eq:SMG-Green's}
\mathcal{G}(k)=\frac{\gamma^\mu k_\mu + \<\mathbf{M} \>}{k^\mu k_\mu -\<\mathbf{M} \>^2}
=\frac{\gamma^\mu k_\mu + \<\CO_{\rm SMG}\> }{k^\mu k_\mu -|m|^2}
=\frac{\gamma^\mu k_\mu  }{k^\mu k_\mu -|m|^2}.
\eea
Here $|m|$ is the absolute magnitude of the SMG-induced  non-mean-field mass
that can be read from 
the rest mass (\ref{rest-mass}) and the correlator (\ref{correlator}).
There is a {\bf universal Green's function zero} when approaching $k_\mu=0$
as  
$\mathcal{G}(k= 0)=\mathcal{G}(E= 0,\vec{k}= 0 )=0$
or $\det(\mathcal{G}(k= 0))=0$ \cite{You1403.4938,Xu2021Greens}.
Crossing the locations of these zeros (along the dispersion $E= \pm |\vec{k}|$),
the $\det(\mathcal{G})$ flips sign, between $\det(\mathcal{G})>0$ and $\det(\mathcal{G})<0$.

So this Green's function zero in \eq{eq:SMG-Green's} 
is a distinct feature, 
not happening to the mean-field mass (in contrast to \eq{eq:mean-Green's}), 
but only for SMG-induced non-mean-field mass.

\subsection{Mass Generating Mechanism}

Here we enumerate and re-examine ``How do we gain the mass? What are possible mass-generating mechanisms?'' 
Conventionally, there are mean-field quadratic fermion bilinear mass terms
that we can write down:

\begin{enumerate}[leftmargin=-0mm, label=\textcolor{blue}{\arabic*}., ref={\arabic*}]

\item
{\bf 
Bilinear pair two Weyl fermions} ($\psi_L$ and $\psi'_L$): 
$$
m(\psi_L^\dagger (\ii \sigma^2) \psi'^{*}_L
+{\psi'^{\rT}_L} (-\ii \sigma^2) \psi_L ) 
\equiv - m((\psi'_L \psi_L)^\dagger +\psi'_L \psi_L)
$$
is most general, say in 4d,
 including both Dirac mass (when $\psi_R= (\ii \sigma^2) \psi'^{*}_L$) 
or Majorana mass (when $\psi_L=\psi_L'$).

\item
{\bf 
Anderson-Higgs mechanisms}: Developed by
Nambu-Goldstone-Anderson-Higgs \cite{Nambu1960PRQuasiparticles, Nambu1961PRJonaLasinio, Goldstone1961NuovoCim, GoldstoneSalamWeinberg1962,
Anderson1963pcPRPlasmons, EnglertBrout1964PRL, Higgs1964PRL},
the vacuum expectation value (vev) of a scalar Higgs $\< \phi_H \>\neq 0$
and the mean-field behavior of Yukawa-Higgs term $\phi_H ((\psi'_L \psi_L)^\dagger +\psi'_L \psi_L)$
induces $m \propto \< \phi_H \> \neq 0$.

\item
{\bf 
{Chiral symmetry breaking}}:
Even without any explicit scalar Higgs field,
the dynamical quark chiral condensate $\< \bar{\psi}_{q} \psi_q \> \neq 0$ in QCD
generates a mass scale  \cite{Nambu1960PRQuasiparticles, Nambu1961PRJonaLasinio, tHooft1979ratanomaly}:  
$$
 L_{\text{ChSB}}= \< \bar{\psi}_{q} \psi_q \> ((\psi'_L \psi_L)^\dagger +\psi'_L \psi_L),
$$
which breaks a chiral symmetry but maintains a vector symmetry. 

All the above mass generating mechanisms
are based on a mean-field bilinear pair of fermion fields, with a lagrangian form
\bea
\CO \psi' \psi + {\rm H.c.}
\eea
such that the mean-field expectation value of $\< \CO \> \neq 0$ is nonzero to generate a mean-field mass gap.
Similarly, the $\< \psi' \psi \> \neq 0$.

Recently it becomes clear that, beyond the conventional mean-field quadratic (less-interacting) mass listed above,
there are other mass-generating mechanisms --- non-mean-field, preserving more symmetry, 
furthermore interacting and nonperturbative, with a close analogy in the many-body interacting quantum matter contemporary developments 
\cite{Senthil1405.4015, Witten1508.04715, Wen2016ddy1610.03911}:

\item
{\bf 
{Symmetric Mass Generation}} (SMG) \cite{WangYou2204.14271}:
Given a $G$-{anomaly-free} theory, there always exists a $G$-symmetric deformation 
such that we can deform a $G$-symmetric gapless phase to a $G$-symmetric trivial gapped phase (a short-range entangled tensor product state).
For example, $G$-symmetry can be a chiral symmetry, 
such that we can deform to gapped confinement without chiral symmetry breaking (known as the smooth s-confinement \cite{Seiberg19949402044, Seiberg19949411149}).
Other examples include gapping 8 or multiple Majorana fermions in one dimensional spacetime \cite{FidkowskifSPT1,FidkowskifSPT2}
or gapping multiple chiral Weyl fermions in even spacetime dimensions 
(history back to Eichten-Preskill \cite{Eichten1985ftPreskill1986},
but SMG only appears in the recent advance  
\cite{Wen2013ppa1305.1045, Wang2013ytaJW1307.7480, You2014oaaYouBenTovXu1402.4151, YX14124784,
BenTov2015graZee1505.04312, Wang2018ugfJW1807.05998, WangWen2018cai1809.11171, RazamatTong2009.05037, Tong2104.03997}), 
see more references in the overview \cite{Tong20198Majorana1906.07199, WangYou2204.14271}.

In the case of fermions, the non-mean-field SMG deformation requires that
\bea 
\label{eq:SMG-fermions}
 \psi' \CO_{\rm SMG} \psi +  {\rm H.c.}
\equiv
 \psi'  ( \psi_{q' } \dots \psi_{q''} + \dots  )  \psi +  {\rm H.c.},  
  \text{ with } \<\CO_{\rm SMG}\>=0  
 \text{ and }
\< \psi' \psi \>=0.
\eea

\item
\noindent 
{\bf 
{Symmetric Gapped Topological Order}} (SGTO):
For certain $G$-{anomalous} theory for spacetime dimension $d \geq 3$
(i.e., precisely a certain type of 't Hooft anomalous $G$-symmetry that must have a nonperturbative global anomaly
\cite{Witten2016cio1605.02391, 
Wang2017locWWW1705.06728,Wan2018djlW2.1812.11955,Cordova2019bsd1910.04962,Cordova1912.13069}, instead of any perturbative local anomaly), 
there still exists a $G$-symmetric deformation 
such that we can deform a $G$-symmetric gapless phase to a $G$-symmetric gapped topologically ordered phase 
(now long-range entangled, with a low-energy  topological quantum field theory [TQFT] below the finite energy gap).
Originated from condensed matter examples (see a review \cite{Senthil1405.4015}),
one can formulate a  systematic construction based on {\bf Symmetry Extension} \cite{Witten2016cio1605.02391,Wang2017locWWW1705.06728} instead of Symmetry Breaking;
where $G$-anomaly is trivialized to none in the pullback extended symmetry group $\tilde{G}$. 
Typically the TQFT's gauge group is a finite group, 
constructed from gauging a normal subgroup $K$ of $\tilde{G}$ with the quotient $\tilde{G}/K=G$
via a typical group extension (here a short exact sequence) 
$1 \to K \to \tilde{G} \to \tilde{G}/K=G \to 1$.
\end{enumerate}
In the following, we apply the SMG to provide novel solutions to the Strong CP problem.

\section{First Solution: Symmetric Mass 
Gap within the Chiral Fermion}
\label{sec:FirstSolution}

\subsection{Toy Model for QCD}
\label{sec:toy-QCD}

Now we aim to apply the SMG for a new Strong CP solution.
To motivate our solution, we first look at a QCD toy model --- 
which turns out to fail our purpose eventually, 
but it serves to offer the key intuitions behind it.
For simplicity, take $\SU(N_c)$ QCD lagrangian with $N_c$ color and $N_f$ flavor Dirac fermions $\psi$ with equal $m$:
\bea
\label{eq:Lagrangian-QCD}
 L=-\frac{1}{4} F_{\mu\nu}^\ra F_{}^{\ra \mu\nu}
 {-\, \frac{{\theta}}{{64} \pi^2} {g}^2 \epsilon^{\mu\nu \mu'\nu'} F_{\mu\nu}^\ra F_{\mu'\nu'}^\ra} 
 +\bar \psi(\ii \gamma^\mu D_\mu -m{\mathbb{I}_{N_f}} \e^{\ii \theta' \gamma_5} )\psi,
\eea
where $\ra$ is the Lie algebra generator index.
The ${\mathbb{I}_{N_f}}$ is a rank-${N_f}$ identity matrix and $\psi$ is an $N_f$-multiplet of Dirac fermions.
The ${\theta_{3}}$ in \eq{eq:theta3} is now called the $\theta$.\\
 \noindent
$\bullet$ $\U(1)_A$ axial symmetry $\psi \mapsto \e^{\ii \alpha \gamma^5}\psi$
is anomalous, whose transformation sends
$\theta \mapsto {\theta}- {2 \upalpha N_f}$ and $\theta'  \mapsto { \theta'+2 \upalpha}$, but keeps 
$
\theta+  N_f\theta' \mapsto
({\theta}- {2 \upalpha N_f}) + N_f ({ \theta'+2 \upalpha})
=\theta+  N_f\theta'$ 
invariant. \\
$\bullet$ $\U(1)_A$ has a mixed anomaly under the ABJ anomaly with the vertices $\U(1)_A$-$\SU(N_c)^2$ captured by a triangle Feynman graph. 
Since $[\SU(N_c)]$ is dynamically gauged, 
$\U(1)_A$ is broken down to a discrete $\Z_{2N_f,A}$ due to the $\SU(N_c)$ instanton, 
the overall internal symmetry including the gauged $[\SU(N_c)]$ is
\bea \label{eq:sym-group-after-gauged}
G_{\rm QCD}\equiv\frac{[\SU(N_c)] \times \SU(N_f)_{\rm L}  \times \SU(N_f)_{\rm R}  \times \U(1)_V  }{
\Z_{N_c} \times \Z_{N_f} },
\quad
\eea
where $\Z_{2N_f,A}$ is secretly already part of its 
subgroup.

The invariant
${\bar{\theta}} 
\equiv
{\theta}
+\arg (\det 
({\mathbb{I}_{N_f}} \e^{\ii \theta'}))=\theta+  N_f\theta'$
cannot be rotated away via the axial symmetry and is generally nonzero --- this
is exactly the Strong CP problem of this QCD model.
Below we show that n\"aively we can solve this problem, if a set of the quarks (in particular the $u$ quark)
obtained its mass not from Anderson-Higgs mechanism, 
but from SMG by preserving the $\SU(N_c)$ and other vector symmetries of
$G_{\rm QCD}$ in \eq{eq:sym-group-after-gauged}. 
{Whether $u$ or other quarks obtain some portion of its mass via the chiral condensate $\< \bar{\psi}_{q} \psi_q \>$ does not alter our argument.
Because the chiral condensate occurs at a lower energy below the confinement scale $\Lambda_{\QCD}$,
but here we concern a QCD lagrangian at a higher energy much above $\Lambda_{\QCD}$.
So the chiral condensate does not affect the fact whether we can rotate the ${\bar{\theta}}$ to zero or not in the QCD lagrangian.}
The mass $m$ is now treated as a dynamical field $\mathfrak{m}$.
The conventional 
Higgs-induced mass has $m = \langle \mathfrak{m} \rangle \propto \langle \phi_H \rangle \neq 0$.
In contrast, the SMG via {a \emph{disordered}} field means that
\bea
\<  \mathfrak{m} \>=0, \quad 
\text{ but only $\langle |{ \mathfrak{m}}|^{\rm n} \rangle \neq 0$,}
\eea 
of a higher power n of 
condensate 
driven by the disordered
Yukawa-Higgs or Yukawa-${\mathfrak{m}}$ multi-fermion interaction term.
SMG says that even though $\<  \mathfrak{m} \>=0$, 
 appropriate disordered ${\mathfrak{m}}$ configurations can indeed give an energy gap, via:
(i) smooth fluctuation (the correlation length $\xi_{\mathfrak{m}}$ satisfies $l_{\rm UV} \ll \xi_{\mathfrak{m}} \ll l_{\rm system}$
where $l_{\rm UV}$ is the UV cutoff of quantum field theory [QFT], or Planck scale in quantum gravity [QG], or some effective ``lattice'' constant),
 (ii) intermediate strength coupling (not too weak nor too strong, thus in a nonperturbative regime).
 
We can use the chiral rotation of all quarks $\psi \mapsto \e^{\ii \alpha \gamma^5}\psi$ to rotate $\theta'$ to $0$ while redefine $\theta$ by $\bar\theta \equiv\theta+  N_f\theta'$.
Next, the $u$ quark's chiral rotation alone
$\psi_u \mapsto \e^{\ii \upalpha_{u} \gamma^5 } \psi_u$,
sends the path integral $\int [\cD \psi][\cD \bar\psi][\cD \mathfrak{m}]
 \e^{\ii \int\dd^4x L}$ to
\begin{widetext}
\bea\hspace{-8mm}
\label{eq:chiral-transformation-on-quark}
\int [\cD \psi][\cD \bar\psi][\cD \mathfrak{m}] 
\e^{\ii \int\dd^4x (-\frac{1}{4} F_{\mu\nu}^\ra F_{}^{\ra \mu\nu}
+
\upalpha_{u}\prt_{\mu} J^{\mu 5}_u
 {-\, \frac{(\bar{\theta} - 2 \upalpha_{u} )}{{64} \pi^2} {g}^2 \epsilon^{\mu\nu \mu'\nu'} F_{\mu\nu}^\ra F_{\mu'\nu'}^\ra}
 +\bar \psi(\ii \gamma^\mu D_\mu -
  {
 \begin{pmatrix}
 \mathfrak{m}_u \e^{\ii { 2 \upalpha_{u} } \gamma_5}  & \\
& (\mathfrak{m}_q + \phi_H )   {\mathbb{I}_{N_f-1}}  
  \end{pmatrix}
} )\psi 
+ \dots)}.
\eea
\end{widetext}
Assuming the $u$ quark only contains the interacting non-mean-field mass from the SMG,
so
\bea
\text{$\<  {m}_u \>= \<  \mathfrak{m}_u \>=0$,  only $\langle |{ \mathfrak{m}}_u|^{\rm n} \rangle \neq 0$}
\eea
for some higher ${\rm n} > 1$.
The crucial input is that
$$
\bar \psi_u \langle \mathfrak{m}_u \rangle \e^{\ii 2 \upalpha_{u} \gamma_5}\psi_u =0,
$$  
so we can choose $\upalpha_{u}$ to rotate
$\bar{\theta} - 2 \upalpha_{u}$ to 0.

In contrast, other quark mass $m_q$ contains the mean-field mass induced by Higgs condensation
$\langle \phi_H \rangle \neq 0$,
$$
\<  {m}_q \>=  \langle \mathfrak{m}_q \rangle + \langle \phi_H \rangle =  \langle \phi_H \rangle \neq 0,
$$
in addition to other possible non-mean-field SMG contribution,
such that $\<  \mathfrak{m}_q \>=0$,  only $\langle |{ \mathfrak{m}}_q|^{\rm n} \rangle \neq 0$
for some higher ${\rm n} > 1$.

Thus this SMG solves the Strong CP problem {within QCD via setting $\bar{\theta} - 2 \upalpha_{u}=0$}.
Some comments on this solution:

\noindent
{\bf {(1)}} 
Compare to the \emph{massless quark solution} ($m_u=0$ \cite{tHooft1976ripPRL}) of the Strong CP problem,
our $u$ quark now instead has a finite energy gap (although $\<  \mathfrak{m}_u \>=0$) induced by
{a non-mean-field disordered energy gap from a dynamical mass field $\mathfrak{m}_u$}.

\noindent
{\bf {(2)}} 
Compare to the \emph{axion solution} \cite{PecceiQuinn1977hhPRL,PecceiQuinn1977urPRD, Weinberg1977ma1978PRL, Wilczek1977pjPRL}
which makes the $\theta$ field dynamical, 
our solution instead makes the mass field $\mathfrak{m}$ dynamical.
Instead of making the dynamical $\theta$ 
localized and \emph{ordered} by a potential \cite{PecceiQuinn1977hhPRL,PecceiQuinn1977urPRD, Weinberg1977ma1978PRL, Wilczek1977pjPRL}
to induce pseudo-Goldstone mode,
our solution makes the dynamical $\mathfrak{m}$ \emph{disordered} to generate an energy gap, namely here an interacting mass gap.

\noindent
{\bf {(3)}} 
This $u$ quark energy spectrum above the disordered gap
can still produce the conventional dispersion $E_u \propto \sqrt{m_u^2 + p_u^2}$ with momentum $p_u$  \cite{WangYou2204.14271}.
Gapping the $u$ and other quarks by SMG, and gapping other quarks also by the Higgs $ \langle \phi_H \rangle$ 
may also change the quark multiplet from $N_f$ flavor to $N_f-1$ flavor. Thus this solution is pending to fit the SM phenomenology.

\noindent
{\bf {(4)}} In SM,
the $u$ quark is in a representation of the chiral $\su(2) \times \u(1)_Y$ that carries a nonzero anomaly index to cancel the SM's perturbative local anomaly.
Because the SMG cannot be applied 
to a non-vanishing perturbative local anomaly, in order to apply the SMG to this SM's $u$ quark,
either we need to include more quarks or leptons (e.g., 15 or 16 Weyl fermions \cite{WangWen2018cai1809.11171, RazamatTong2009.05037}) 
to gap them altogether, or we need to break down some continuous symmetry to discrete symmetry. 
The discrete symmetry may change the anomaly becoming a nonperturbative global anomaly, 
then we may obtain the SGTO with a low energy TQFT to match the nonperturbative global anomaly.

Because the issue {\bf {(3)}} and {\bf {(4)}} listed above may lead to disagreements with the SM phenomenology,
this solution may \emph{not} be favored by nature. 
Yet another issue is that the dynamics of the disorder field 
driving to a gapped phase are not analytically well-controlled, 
but only numerically verified (e.g., \cite{DeMarcoWen1706.04648}).
In principle, we have to modify the disorder mass field interaction term 
\bea  \label{eq:SMG-dynamical-m}
\bar \psi_q  \mathfrak{m}_q  \psi_q
\text{ with }
\<  \mathfrak{m}_q\>= 0
\eea
to
the
SMG interaction for the set of quarks 
schematically as
\bea
 \label{eq:SMG-quark}
\bar \psi_q \CO_{\rm SMG} \psi_q
\equiv
g_{\rm SMG} \bar \psi_q  ( \psi_{q' } \dots \psi_{q''} + \dots +  {\rm H.c.})  \psi_q, 
 \quad\quad 
 \text{ with } \<\CO_{\rm SMG}\>=0.
\eea
The set of quarks (labeled by $q,q',q''$, etc.) would interact
through the multi-fermion interactions even in the absence of a dynamical $[\SU(N_c)]$ gauge field.
The multi-fermion interaction can involve more than four-fermion interactions \cite{Wang2013ytaJW1307.7480}.
The anomaly-free condition within the vector symmetry part of $G_{\rm QCD}$
guarantees us to find the SMG interaction such that
(i) the interaction preserves at least the vector symmetry part of $G_{\rm QCD}$,
(ii) $\<\CO_{\rm SMG}\>=0$, and 
(iii) the full \eq{eq:SMG-quark} is a singlet ${\bf 1}$ in the trivial representation
of the vector symmetry of $G_{\rm QCD}$.
We can also interpret that by 
integrating out the dynamical mass field in \eq{eq:SMG-dynamical-m},
we can induce the interaction \eq{eq:SMG-quark} as a consequence.

Related designs of the SMG interaction terms similar to \eq{eq:SMG-quark} 
are explored examples by examples in recent works, e.g., \cite{WangYou2204.14271, Tong2104.03997}.
We will not go into the further detail of the interaction design of \eq{eq:SMG-quark}, 
since this QCD toy model 
unfortunately fails in the context of solving the Strong CP problem for the full SM.
Instead, we use this n\"aive QCD toy model in \Sec{sec:toy-QCD} to motivate 
our SMG solution for the SM presented next in \Sec{sec:toy-SM}.

\subsection{Toy Model for the $\su(3) \times  \su(2) \times \u(1)$ Standard Model}
\label{sec:toy-SM}

The quarks alone cannot fully cancel the anomaly
for the chiral $\su(2) \times \u(1)_Y$ symmetry
in the full SM's gauge sector $\su(3) \times  \su(2) \times \u(1)_Y$.
So the QCD toy model in \Sec{sec:toy-QCD} fails 
to generate the SMG gap in the context of the full SM. 
%

In particular, we have to at least include the quarks and the leptons, 
involving at least the 15 Weyl fermions $(\bar{d}_R \oplus {l}_L  \oplus q_L  \oplus \bar{u}_R \oplus   \bar{e}_R)$ in
\eq{eq:SMrep} in one family,
to cancel the anomaly within the full SM's gauge sector $\su(3) \times  \su(2) \times \u(1)_Y$.
In this subsection, we modify the QCD toy model in \Sec{sec:toy-QCD}
to the new toy model for the full SM, which will become the first candidate Strong CP solution based on the SMG.

Before we present the detail of the model, a few more comments on this model follow:

\noindent
$\bullet$ Although we have known by far the $N_g=3$ {families} of quarks and leptons,
our model does not necessarily require all 3 families of quarks and leptons
to get involved in the SMG process.
It could be that the SMG happens in one family of SM, such as the \emph{first family} (or any one family out of the \emph{three families}) of quarks and leptons.
It could also be that the SMG happens in one new family of SM, such as the \emph{fourth family} of hypothetical quarks and leptons.
As long as \emph{at least one family} of quarks and leptons
get a portion of its mass by SMG, then we can readily apply a chiral rotation of a quark (say the $u$ quark) 
to solve the Strong CP problem, at least theoretically. However, we must say that there are phenomenological constraints
that may favor or disfavor this theoretical Strong CP solution.
So without loss of generality, we will implement the SMG only for one family of quarks and leptons.

\noindent
$\bullet$ We could consider the model with either the 15 Weyl fermions or 
the 16 Weyl fermions
$(\bar{d}_R \oplus {l}_L  \oplus q_L  \oplus \bar{u}_R \oplus   \bar{e}_R \oplus \bar{\nu}_R)$
including the right-hand neutrino ${\nu}_R$.
Both the 15 or 16 Weyl fermions can have the full anomaly cancellation with the full SM's $\su(3) \times  \su(2) \times \u(1)_Y$,
so we could implement the SMG for both cases.

Now we present the step-by-step construction of the first candidate Strong CP solution based on SMG:

\begin{enumerate}[leftmargin=-0mm, label=\textcolor{blue}{\arabic*}., ref={\arabic*}]

\item We follow the Razamat-Tong model \cite{RazamatTong2009.05037}
to embed the 15 or 16 Weyl-fermion SM into the 27 Weyl-fermion
with the following fermion matter content:
\bea\label{eq:SM-nonSUSY}
\hspace{-6mm}
\setlength{\extrarowheight}{1pt}
\begin{array}{| l | c | ccc | c}
\cline{1-5}
(\overline{\bf 3},{\bf 1})_{2} & ({\bf 1},{\bf 2})_{-3}  &\multicolumn{1}{c|}{({\bf 3},{\bf 2})_{1}}   & \multicolumn{1}{c|}{ (\overline{\bf 3},{\bf 1})_{-4}} & ({\bf 1},{\bf 1})_{6} & \\
\cline{3-6}
(\overline{\bf 3},{\bf 1})_{2} & ({\bf 1},{\bf 2})_{-3}  & &  & & \multicolumn{1}{c|}{ {({\bf 1},{\bf 1})_{0}}}\\
\cline{1-2}
({\bf 3},{\bf 1})_{-2} & ({\bf 1},{\bf 2})_{+3}  & &  & & \multicolumn{1}{c|}{{({\bf 1},{\bf 1})_{0}}}\\
\cline{1-2} \cline{6-6}
\end{array}.
\eea
In \eq{eq:SM-nonSUSY}, the first row's
$$
(\overline{\bf 3},{\bf 1})_{2} \oplus ({\bf 1},{\bf 2})_{-3}  \oplus
({\bf 3},{\bf 2})_{1} \oplus (\overline{\bf 3},{\bf 1})_{-4} \oplus ({\bf 1},{\bf 1})_{6}
\sim
(\bar{d}_R \oplus {l}_L  \oplus q_L  \oplus \bar{u}_R \oplus   \bar{e}_R)
$$ 
has the original 15 Weyl fermions
of the SM.\footnote{{For the left and right notations,
we use the italic font $L$ and $R$ to denote that of spacetime symmetry,
while use the text font L and R for that of internal symmetry.}
\label{ft:LR}
}
However, the $\bar{d}_R$ and ${l}_L$ are doubled (in the first and the second rows)
to become a doublet under a new $\su(2)_{\rm R}$. 
The anti-particle of right-handed neutrino $\bar{\nu}_R$ written as the left-handed particle, 
is also introduced, but also further doubled as the $\su(2)_{\rm R}$ doublet of ${({\bf 1},{\bf 1})_{0}}$ (in the second and the third rows).
There are also extra new fermions $({\bf 3},{\bf 1})_{-2}$ and $({\bf 1},{\bf 2})_{+3}$
denoted as ${\bar{d}'}_R$ and $l'_L$.

So the SM's original $G_{\SM_p}$ and $\su(3) \times  \su(2) \times \u(1)_Y
\equiv \su(3) \times  \su(2)_{\rm L} \times \u(1)_{\rm L}$
is enlarged to 
\bea \label{eq:GSMLR}
&&G_{\text{SM-LR}_{p,p'}} \equiv
\frac{G_{\SM_p}\times\SU(2)_\mathrm{R}\times\U(1)_\mathrm{R}}{\Z_{p'}}
\equiv
\frac{\SU(3)\times\SU(2)_\mathrm{L}\times\SU(2)_\mathrm{R}\times\U(1)_\mathrm{L}\times\U(1)_\mathrm{R}}{\Z_p\times\Z_{p'}}
\eea
with $p, p' \in \{1,2,3,6\}$. 
All of $G_{\text{SM-LR}_{p,p'}}$ are compatible with the Lie algebra 
$$
\su(3) \times  \su(2) \times \u(1)_Y \times  \su(2)_{\rm R} \times \u(1)_{\rm R}
\equiv{
\su(3) \times  \su(2)_{\rm L} \times  \su(2)_{\rm R} 
\times \u(1)_{\rm L}  \times \u(1)_{\rm R}
}$$
such that the 27-fermion
representation in \eq{eq:SM-nonSUSY}
becomes
\bea \label{eq:SMrepLR-chiral}
&&\hspace{-2mm}
\big(\bar{d}_R \oplus {l}_L  \oplus q_L  \oplus \bar{u}_R \oplus   \bar{e}_R \oplus \bar{\nu}_R \big) 
\oplus d'_L   \oplus {\bar{l}'}_R \\
&&\sim \big((\overline{\bf 3},{\bf 1},{\bf 2})_{2,-1} \oplus ({\bf 1},{\bf 2},{\bf 2})_{-3,3}  
\oplus
({\bf 3},{\bf 2},{\bf 1})_{1,-2} \oplus (\overline{\bf 3},{\bf 1},{\bf 1})_{-4,2}  
\oplus ({\bf 1},{\bf 1},{\bf 1})_{6,-6} 
\oplus
{({\bf 1},{\bf 1},{\bf 2})_{0,-3}}\big)
\oplus
({\bf 3},{\bf 1},{\bf 1})_{-2,4}  
\oplus ({\bf 1},{\bf 2},{\bf 1})_{3,0}
.\nn
\eea 

\item The 27 to 16 Weyl fermions' deformation is done by a new scalar Higgs field
$h_\mathrm{R}=(\mathbf{1},\mathbf{1},\mathbf{2})_{0,-3}$ that couples to the Weyl fermions by the following Higgs term
\bea \label{eq:27-16Higgs}
{\big(\epsilon_{\rm R} h_\mathrm{R}(\bar{d}_R d'_L+
\bar{l}'_R l_L)+(h_\mathrm{R}^\dagger \bar{\nu}_R)(h_\mathrm{R}^\dagger \bar{\nu}_R)\big)+\text{H.c.}.}
\eea
We shall write down explicitly the anti-symmetric tensor $\epsilon_{\rm R}$ in the $\SU(2)_\mathrm{R}$ doublet subspace.
But the $\epsilon_{\rm L}$ and $\epsilon$ tensor for the $\SU(2)_\mathrm{L}$ and Lorentz $\su(2)$ subspace are omitted.
In the 
Higgs condensed $\langle h_\mathrm{R}\rangle \neq 0$ phase,
the \emph{lower} half of both $\SU(2)_\mathrm{R}$ doublets of $\bar{d}_R$ and $l_L$ get mass,
while the dimension-5 term $(h_\mathrm{R}^\dagger\bar{\nu}_R)(h_\mathrm{R}^\dagger\bar{\nu}_R)$ 
gives Majorana mass to only the \emph{upper} half of the $\SU(2)_\mathrm{R}$ doublet $\bar{\nu}_R$.
Thus $\langle h_\mathrm{R}\rangle \neq 0$ lifts 11 Weyl fermions with a mass gap, 
leaving 16 Weyl fermions at low energy, and breaks the symmetry from $G_{\text{SM-LR}_{p,p'}}$ down to $G_{\SM_p}\times\U(1)_{\bf B - L}$.

\item The 27 to 15 Weyl fermions' deformation is done by adding yet another new scalar Higgs 
field $h'=(\mathbf{1},\mathbf{1},\mathbf{1})_{0,6}$,
\bea \label{eq:27-15Higgs}
\big(\epsilon_{\mathrm{R}} h_\mathrm{R}(\bar{d}_R d'_L+\bar{l}'_R l_L)+
h'({\epsilon_{\mathrm{R}}} \bar{\nu}_R {\bar{\nu}_R)}\big)+\text{H.c.}
\eea
The Higgs condensation $\langle h_\mathrm{R}\rangle \neq 0$ and $\langle h'\rangle\neq 0$ 
leave 15 Weyl fermions at low energy ---
the $h'(\epsilon_R \bar{\nu}_R {\nu}_R)$ gives a
Dirac mass to both upper and lower components of the $\SU(2)_\mathrm{R}$ doublet $\bar{\nu}_R$. 
This breaks the symmetry from $G_{\text{SM-LR}_{p,p'}}$ down to $G_{\SM_p}$
with no $\U(1)_{\bf B-L}$, 
and drives the transition from the 27 to the 15 Weyl-fermion phases.

\item
The 27 Weyl fermion to the SMG deformation: 
The 27 Weyl-fermion model
can be fully gapped by preserving \emph{not only} 
the SM internal symmetry group $G_{\SM_p}$ for $p=1,2,3,6$,
\emph{but also} an additional continuous baryon minus lepton symmetry $\U(1)_{\bf B - L}$.

$\bullet$ \emph{Mean-field} SMG that \emph{cannot} solve the Strong CP problem:
Tong \cite{Tong2104.03997} suggests fully gapping the 27 Weyl fermions 
to achieve the SMG by adding a scalar field
${\phi}$ in ${({\bf 1},{\bf 2},{\bf 2})_{-3,3}}$
when the generic condensation $\langle \phi \rangle \neq 0$ occurs in this deformation
\bea 
\label{eq:Yukawa-full-gap-phi-DT}
&& 
 \epsilon_{\rm R} \epsilon_{\rm L} \big(\phi  (
  {{\bar{d}_R}}  {{q_L}} 
+ {{\bar{\nu}_R}} {{\bar{l}'_R}}
+ {{\bar{e}_R}} {{l_L}})
+\phi^{\dagger2}  {{{\bar{u}_R}}{{d'_L}} }
\big) 
+ {\rm H.c.}
\equiv \big(\phi (
   {{\bar{d}_R}}  {{q_L}} 
+ {{\bar{\nu}_R}} {{\bar{l}'_R}}
+ {{\bar{e}_R}} {{l_L}})
+\phi^{\dagger2}  {{{\bar{u}_R}}{{d'_L}} }
\big) 
+ {\rm H.c.}
\eea
Hereafter
the $\epsilon_{\rm L}$, $\epsilon_{\rm R}$ and $\epsilon$ tensors 
for the $\SU(2)_\mathrm{L}$, $\SU(2)_\mathrm{R}$ and Lorentz $\su(2)$ subspaces may be omitted.
The deformation term \eq{eq:Yukawa-full-gap-phi-DT} in a lagrangian
is a singlet under all the representations.

The $\langle \phi \rangle \neq 0$ phase still preserves the deformed SM internal symmetry
$\frac{\SU(3) \times \SU(2)_{\text{diagonal}}  \times \U(1)_{\text{diagonal}}}{\Z_p} \times \U(1)_{\bf B-L}$.
However, the $\langle \phi \rangle \neq 0$ violates the no mean-field condensation condition that requires $\langle \phi \rangle = 0$ in 
\eq{eq:SMG-dynamical-m} and \eq{eq:SMG-quark}.
It can also be checked that this deformation cannot solve the Strong CP problem, due to the $\bar{\theta}$ 
(including the contribution from $\theta$ and the generic complex phase of $\langle \phi \rangle$)
cannot be rotated away to $\bar{\theta}=0$.

$\bullet$  \emph{Non-mean-field} SMG and a multi-fermion interaction deformation: We propose to modify Tong's \cite{Tong2104.03997} deformation to the 
case of the dynamical $\phi$ 
with a random disorder configuration that we integrate over $\int [\cD\phi]$ such that 
\bea \label{eq:Yukawa-full-gap-phi-non-mean-field}
\text{$\langle \phi \rangle = 0$, \quad $\langle \phi^2 \rangle = 0$,} 
\eea so there is no mean-field mass contribution to \eq{eq:Yukawa-full-gap-phi-DT}.  
This condition satisfies the no-mean-field version of the SMG criterion \eq{eq:SMG-quark}.
Integrating over $\int [\cD\phi]$ in the path integral
also induces the multi-fermion interaction version of the deformation:
\bea 
\label{eq:Yukawa-full-gap-multi-fermion}
({{\bar{d}_R}}  {{q_L}} 
+ {{\bar{\nu}_R}} {{\bar{l}'_R}}
+ {{\bar{e}_R}} {{l_L}})^2({{{\bar{u}_R}}{{d'_L}} })+ {\rm H.c.}
\eea
There are also additional density-density interactions induced by integrating over $\int [\cD\phi]$, such as
\be
\label{eq:Yukawa-full-gap-multi-fermion-2}
( {{\bar{d}_R}}  {{q_L}} 
+ {{\bar{\nu}_R}} {{\bar{l}'_R}}
+ {{\bar{e}_R}} {{l_L}}) (
     {{\bar{q}_L}}  {{{d}_R}}
+  {{{l}'_R}} {{{\nu}_R}}
+  {{\bar{l}_L}} {{{e}_R}}) 
+  ({{{\bar{u}_R}}{{d'_L}} })({ {{\bar{d}'_L}} {{{u}_R}} }).
\ee

Any higher power (larger than 1) of 
any Grassmann-number-valued fermion in a lagrangian density
needs to be point-split, such as 
${\psi}_{R}^2 \equiv {\psi}_{R} \sigma^\mu \prt_\mu  {\psi}_{R}$
or ${\psi}_{R} \sigma^\mu D_\mu  {\psi}_{R}$ in the covariant derivative form, etc. 
If we like to have a lattice regularization to do the numerical simulation, 
we also have to point split ${\psi}_{R}^n(x) \equiv {\psi}_{R}(x) {\psi}_{R}(x+ \varepsilon) \dots {\psi}_{R}(x+ (n-1)\varepsilon)$ 
on neighbor sites where $\varepsilon$ is the lattice constant.

The interaction terms, both the six-fermion \eq{eq:Yukawa-full-gap-multi-fermion} and the four-fermion \eq{eq:Yukawa-full-gap-multi-fermion-2} interactions,
are irrelevant in the sense of perturbative renormalization group (RG). 
In order to drive from a nearly gapless phase to a gapped SMG phase,
we require to turn on their coupling strength to a nonperturbative scale.

To justify that both the disorder scalar phase of the deformation (\eq{eq:Yukawa-full-gap-phi-DT} and \eq{eq:Yukawa-full-gap-phi-non-mean-field})
and the multi-fermion interaction deformation with an appropriate nonperturbative coupling strength 
(\eq{eq:Yukawa-full-gap-multi-fermion} and \eq{eq:Yukawa-full-gap-multi-fermion-2})
can fully gap the 27 Weyl fermions and can achieve the SMG, we apply the argument given in \cite{Wen2013ppa1305.1045}:

``The $G$-symmetry preserving SMG can be obtained from the disorder scalar phase, if the following sufficient conditions are hold.
First, breaking $G$ to $G_{\rm sub}$ via the scalar condensation $\langle \phi \rangle \neq 0$
can give the fully gapped symmetry-breaking theory.
Second, there are no topological defects that can trap zero modes robust under any nonperturbative $G$-symmetry-preserving deformation.
Namely, any zero modes can be gapped by some $G$-symmetry-preserving deformation.''

The first condition holds, because Tong \cite{Tong2104.03997} already shows that $\langle \phi \rangle \neq 0$ gives the fully gapped phase.
The second condition holds under the condition given in \cite{Wen2013ppa1305.1045} 
if the homotopy group of the quotient space
$\pi_d(G/G_{\rm sub})=0$ for all the dimensions $0 \leq d \leq D+1$
for the total spacetime dimension $D=4$ here.

In the case of the $\so(10)$ Grand Unified Theory (GUT) with
a Spin(10) internal symmetry group, 
 \Refe{Wen2013ppa1305.1045} verifies that
 $\pi_d(\frac{\Spin(10)}{\Spin(9)})=\pi_d(S^9)=0$ for $0 \leq d \leq 5$ is true.
However, this is only a sufficient condition not a necessary condition.
The 1+1d $\U(1)$-symmetry chiral fermion model violates the sufficient condition
because $\pi_d(\U(1))=\pi_d(S^1)$ gives $\Z$ for $d=1$. But a 1+1d $\U(1)$-symmetry anomaly-free chiral fermion model
can indeed have the SMG achieved by the multi-fermion interaction 
\cite{Wang2013ytaJW1307.7480,  Wang2018ugfJW1807.05998, Zeng2202.12355}
or by the disorder scalar interaction \cite{WangYou2204.14271, Wang2207.14813}.
So what we really need is that any zero modes trapped by the defects (such as those from $\pi_d(G/G_{\rm sub})$)
can be gapped by some $G$-symmetry-preserving deformation.

In the case of $G_{\text{SM-LR}_{p,p'}}$ in \eq{eq:GSMLR},
the defect of the
$\langle \phi \rangle \neq 0$ phase can be classified by the homotopy group
\begin{multline}
\label{eq:homotopy}
\hspace{-6mm}
\pi_d(\frac{G_{\text{SM-LR}_{p,p'}}}{G_{\rm sub}})
=\pi_d(\frac{\SU(2)_\mathrm{L}\times\SU(2)_\mathrm{R}\times\U(1)_\mathrm{L}\times\U(1)_\mathrm{R}}{
\U(1) \times \U(1) \times 1 \times 1})\\
=\pi_d (\frac{S^3}{S^1} \times \frac{S^3}{S^1} \times S^1 \times S^1) 
=\pi_d (S^2 \times S^2 \times S^1 \times S^1)
=
\left\{
\begin{array}{ll}
0, & d=0\\
\Z^2, & d=1\\
\Z^2, & d=2\\
\Z^2, & d=3\\
\Z_2^2, & d=4\\
\Z_2^2, & d=5
\end{array}
\right..
\end{multline}
All we require is that any zero modes trapped in these defects
 can be gapped by some $G$-symmetry-preserving deformation \cite{YX14124784}.
This is always doable, if and only if the starting theory is anomaly free within the $G$-symmetry
in the given spacetime dimension $D$, so the $G$-symmetry preserving SMG phase exists.
Here we only need to verify that the cobordism group $\TP_{D+1}(G)$ proposed in Freed-Hopkins \cite{2016arXiv160406527F}
that classifies the $D$-dimensional anomaly 
and the $D+1$-dimensional invertible topological field theory.
We check that at $D=4$ to find that \cite{GarciaEtxebarriaMontero2018ajm1808.00009, DavighiGripaiosLohitsiri2019rcd1910.11277, WW2019fxh1910.14668, JW2006.16996}
\bea \label{eq:cobordism-SM}
\TP_{5}(\Spin \times G_{\SM_p}) &=& 
\left\{
\begin{array}{ll}
\Z^5 \times \Z_2, & p=1,3\\
\Z^5 , & p=2,6
\end{array}
\right..
\\
\TP_{5}(\Spin \times {G_{\text{SM-LR}_{p,p'}}}) && \text{contains $\Z$ and $\Z_2$ classes.} \nn
\eea
The integer $\Z$ class corresponds to the perturbative local anomaly classification captured by perturbative Feynman graphs and infinitesimal gauge-diffeomorphism transformations.
The finite group class (here $\Z_2$) captures the nonperturbative global anomaly classification captured only by large gauge-diffeomorphism transformations.
Indeed, we can check that 
the matter content \eq{eq:SMrep}, with or without $\bar{\nu}_R$, is still anomaly-free with the anomaly index 0,
within the cobordism group $\TP_{5}(\Spin \times G_{\SM_p})$.
Similarly, the 27 Weyl-fermion matter content
\eq{eq:SMrepLR-chiral} is also anomaly-free with the anomaly index 0,
within the cobordism group $\TP_{5}(\Spin \times {G_{\text{SM-LR}_{p,p'}}})$.
From this anomaly-free perspective, we now understand the distinctions between Tong's model and our proposal:\\

$\bullet$ Tong's model \cite{Tong2104.03997} on the $\langle \phi \rangle \neq 0$ phase
although counts as the SMG by preserving a modified 
$\frac{\SU(3) \times \SU(2)_{\text{diagonal}}  \times \U(1)_{\text{diagonal}}}{\Z_p} \times \U(1)_{\bf B-L}$,
 it is indeed a symmetry-breaking phase partially breaking the entire ${G_{\text{SM-LR}_{p,p'}}}$ down to 
 $\frac{\SU(3) \times \SU(2)_{\text{diagonal}}  \times \U(1)_{\text{diagonal}}}{\Z_p} \times \U(1)_{\bf B-L}$.
The mass gap is generated by the symmetry-breaking mean-field fermion bilinear mass term, so 
Tong's model \cite{Tong2104.03997} would not directly solve the Strong CP problem.

$\bullet$ Our proposal, however, means to access the entire ${G_{\text{SM-LR}_{p,p'}}}\equiv$
$\frac{G_{\SM_q}\times\SU(2)_\mathrm{R}\times\U(1)_\mathrm{R}}{\Z_{p'}}$
$\equiv$
$\frac{\SU(3)\times\SU(2)_\mathrm{L}\times\SU(2)_\mathrm{R}\times\U(1)_\mathrm{L}\times\U(1)_\mathrm{R}}{\Z_p\times\Z_{p'}}$-preserving SMG.
Our model is based on 
the disorder scalar phase of the deformation (\eq{eq:Yukawa-full-gap-phi-DT} and \eq{eq:Yukawa-full-gap-phi-non-mean-field})
or the multi-fermion interaction deformation with an appropriate nonperturbative coupling strength 
(\eq{eq:Yukawa-full-gap-multi-fermion} and \eq{eq:Yukawa-full-gap-multi-fermion-2}).
With the criterion that any zero modes trapped in the defects \eq{eq:homotopy}
can be gapped by some $G$-symmetry-preserving deformation,
we can assure that our deformation leads to the SMG phase with no mean-field condensate (e.g., $\langle \phi \rangle = 0$)
but with disorder configuration under $\int [\cD\phi]$ in the path integral.

{In particular, for the defects classified by \eq{eq:homotopy}, 
a configuration $\phi \neq 0$ has its core occur at $\phi=0$.
This means that the codimension-$(d+1)$ defects (namely the $D-(d+1)$-dimensional defects)
in the spacetime is classified by $\pi_{d}(\frac{G_{\text{SM-LR}_{p,p'}}}{G_{\rm sub}})$.
There are two $\Z$ classes 0d point defects from $\pi_{3}$,
two $\Z$ classes 1d line defects from $\pi_{2}$,
two $\Z$ classes 2d surface defects from $\pi_{1}$, etc.
There are also possible $(\Z_2)^2$ 
classes of Wess-Zumino-Witten (WZW) terms constructed out of $\pi_{5}$.
But our construction demands that the zero modes trapped must be symmetry-preserving gappable.
Our construction also needs to forbid the induction of any WZW terms.
We can check that these conditions are fulfilled based on the method in \cite{YX14124784}.
}

Overall, all these conditions can be satisfied by the 27 Weyl fermion model \eq{eq:SMrepLR-chiral}, 
because it is anomaly-free in ${G_{\text{SM-LR}_{p,p'}}}$
and it is in the trivial cobordism class (the anomaly index 0) in $\TP_{5}(\Spin \times {G_{\text{SM-LR}_{p,p'}}})$.

\item In order to use the above model with the SMG mechanism to solve the Strong CP problem, 
our model must fit all the known constraints from the experiments in Nature, and the QCD lattice numerical simulation.

One major challenge to overcome is that according to Particle Data Group (PDG)
\cite{PDG-quark}, the current quark mass $m_u$ of the $u$ quark has $m_u=2.16$ MeV
as an input to the  QCD lagrangian at the energy scale $E_{\overline{\text{MS}}} =  2$ GeV
based on the modified minimal subtraction ($\overline{\text{MS}}$) renormalization scheme. 
Other current quark masses $m_q$ are also nonzero \cite{PDG-quark}. (See \Table{tab:current-mass} later in \Sec{sec:hadron} for a summary of mass data.)
Some comments about this current quark masses $m_q$:

$\bullet$
Note that this current quark mass $m_q$ is already a renormalized mass with no divergence
by the $\overline{\text{MS}}$ scheme, away
from the bare quark mass $m_0$ that potentially suffered from divergence.

$\bullet$ The current quark mass $m_q$ is an input to the lattice QCD lagrangian at the energy scale $E_{\overline{\text{MS}}} = 2$  GeV.
So the current quark mass $m_q$ can receive contributions from any higher energy UV theory -- 
such as: 

\begin{enumerate}[
label=\textcolor{blue}{(\arabic*)}., ref={(\arabic*)}]

\item  
The Higgs condensation $\langle \phi_H \rangle \neq 0$
by the electroweak symmetry breaking $\su(2) \times \u(1)_Y \to \u(1)_{\rm EM}$
happened at a higher energy at the Fermi electroweak scale $E_{{\text{EW}}} \sim 246$ GeV.   

\item What we hypothesize is that the SMG also contributes to the current quark mass $m_q$ and the lepton mass, 
at least for one of three families, or for the new family (e.g., the fourth family) of SM matter content.
The SMG happens at much higher energy above the electroweak scale $E_{{\text{EW}}}$, or happens at the earlier universe 
before the Higgs condensation $\langle \phi_H \rangle \neq 0$. 

\item
However, the current quark mass $m_q$ at $E_{\overline{\text{MS}}}$ 
receives no contribution from the QCD chiral condensate $\< \bar{\psi}_{q} \psi_q \>$.
Because at the $E_{\overline{\text{MS}}}=  2$ GeV scale, 
the QCD chiral symmetry is not yet broken since $\< \bar{\psi}_{q} \psi_q \> =0$,
while the $\< \bar{\psi}_{q} \psi_q \> \neq 0$
only happens at much lower energy like $\Lambda_{\QCD}$ around 200 MeV.

\end{enumerate}

This lattice QCD evidence on the
nonzero current quark mass $m_u \neq 0$ \cite{PDG-quark}
rules out the famous 't Hooft massless $u$ quark solution \cite{tHooft1976ripPRL} that requires $m_u =0$.
Because under  $\psi \mapsto \e^{\ii \alpha \gamma^5}\psi$, from \eq{eq:chiral-transformation-on-quark},
\be \label{eq:J5-mass}
\hspace{-2mm}
\upalpha_{u}\prt_{\mu} J^{\mu 5}_u
 {-\, \frac{(\bar{\theta} - 2 \upalpha_{u} )}{{64} \pi^2} {g}^2 \epsilon^{\mu\nu \mu'\nu'} F_{\mu\nu}^\ra F_{\mu'\nu'}^\ra}
 -\bar \psi_u 
   ({m}_u \e^{\ii { 2 \upalpha_{u} } \gamma_5}  )\psi_u, 
\ee
once the mean-field mass $m_u = \lambda^{\rm D}_{{u}} \langle \epsilon\phi_H^* \rangle
\neq 0$ or in an appropriate basis $m_u = \lambda^{\rm D}_{{u}} \langle \phi_H \rangle
\neq 0$  is nonzero, due to 
the nonzero diagonalized Yukawa-Higgs coupling $\lambda^{\rm D}_{{u}} \neq 0$ in \eq{eq:Yukawa-Higgs}
and due to the Higgs condensate $\langle \phi_H \rangle \neq 0$,
then it is impossible to rotate the $\bar{\theta} - 2 \upalpha_{u}=0$ without introducing the complex phase
$\e^{\ii { 2 \upalpha_{u}}}$ to the mean-field ${m}_u$.

Another way to explain the failure of the massless $u$ quark solution \cite{tHooft1976ripPRL}
is that under the $\upalpha_{u}$ variation on the above action $S$ in \eq{eq:J5-mass}, 
then taking its mean-field expectation
$\< \frac{\delta S}{\delta \upalpha_{u}}\> \vert_{\upalpha_{u}=0}$
(i.e., not taking it as the operators or the equations),
we have
\be
\label{eq:J5-mass-mean-field}
 \<\prt_{\mu} J^{\mu 5}_u\> 
=
 -{\frac{ 2 }{{64} \pi^2} {g}^2 \epsilon^{\mu\nu \mu'\nu'} \< F_{\mu\nu}^\ra F_{\mu'\nu'}^\ra \>}
 + {2 \ii
\< {m}_u 
\bar \psi_u  \gamma_5  \psi_u \>}
 =
 -{\frac{ 2 }{{64} \pi^2} {g}^2 \epsilon^{\mu\nu \mu'\nu'} \< F_{\mu\nu}^\ra F_{\mu'\nu'}^\ra \>}
 + {2 \ii
\< {m}_u \>
\<  \bar \psi_u  \gamma_5  \psi_u \>}.
\ee
The evaluation of the vev of $\< {m}_u 
\bar \psi_u  \gamma_5  \psi_u \>= 
{m}_u  \<\bar \psi_u  \gamma_5  \psi_u \>
+ \< {m}_u  \> \bar \psi_u  \gamma_5  \psi_u -
 \< {m}_u  \>  \< \bar \psi_u  \gamma_5  \psi_u  \>$ based on the variational principle 
on varying $\bar \psi_u  \gamma_5  \psi_u$  shows that
 $\< {m}_u 
\bar \psi_u  \gamma_5  \psi_u \>=  \< {m}_u  \>  \< \bar \psi_u  \gamma_5  \psi_u  \>$,
since the mean-field 
$\< {m}_u  \> \propto \< \bar \psi_u  \gamma_5  \psi_u  \> \neq 0$ is correlated based on the linear response theory.
The extra mean-field current $u$ quark mass $\< {m}_u \>={m}_u$ 
 fails to obey the relation $ \<\prt_{\mu} J^{\mu 5}_u\> \propto \< F_{\mu\nu}^\ra F_{\mu'\nu'}^\ra \>$,
 thus also fails to redefine $\bar{\theta}$ to $0$ by the $u$ quark chiral transformation.

However, if we hypothesize that the SMG interactions
are involved to give a $u$ quark some non-mean-field mass, then the pertinent terms from SMG 
(say \eq{eq:Yukawa-full-gap-multi-fermion} and \eq{eq:Yukawa-full-gap-multi-fermion-2})
for the $u$ quark include:

\bea \label{eq:u-quark-SMG}
\big( ({{\bar{d}_R}}  {{q_L}} 
+ {{\bar{\nu}_R}} {{\bar{l}'_R}}
+ {{\bar{e}_R}} {{l_L}})^2({{{\bar{u}_R}}{{d'_L}} })+ {\rm H.c.}
\big)
+
(  {{\bar{q}_L}}  {{{d}_R}})( {{\bar{d}_R}}  {{q_L}} )
+
\big(
({{\bar{q}_L}}  {{{d}_R}})
( {{\bar{\nu}_R}} {{\bar{l}'_R}}
+ {{\bar{e}_R}} {{l_L}})
+ {\rm H.c.}
\big)
+  ({{{\bar{u}_R}}{{d'_L}} })({ {{\bar{d}'_L}} {{{u}_R}} }).
\eea

Among \eq{eq:u-quark-SMG}, the term that is transformed up to a phase under the left-handed $L$ or the right-handed $R$ $u$ quark, $u_L$ or $u_R$, 
chiral transformation,
includes
\bea \label{eq:u-quark-SMG-2}
&&\big( ({{\bar{d}_R}}  {{q_L}} 
+ {{\bar{\nu}_R}} {{\bar{l}'_R}}
+ {{\bar{e}_R}} {{l_L}})^2({{{\bar{u}_R}}{{d'_L}} })+ {\rm H.c.}
\big)
+
\big(
({{\bar{q}_L}}  {{{d}_R}})
( {{\bar{\nu}_R}} {{\bar{l}'_R}}
+ {{\bar{e}_R}} {{l_L}})
+ {\rm H.c.}
\big)
.\cr
&&
\equiv {\bar{u}_R} (\CO_{\psi\psi}) u_L +  {\bar{u}_R} (\CO_{\psi\psi\psi\psi}) u_L
+  {\bar{u}_R} (\CO_{\psi\psi\psi\psi\psi}) 
+ {\rm H.c.}
\eea
See the footnote,\footnote{\label{ft:multi-fermion}
Specifically
$\big(\phi (
   {{\bar{d}_R}}  {{q_L}} 
+ {{\bar{\nu}_R}} {{\bar{l}'_R}}
+ {{\bar{e}_R}} {{l_L}})
+\phi^{\dagger2}  {{{\bar{u}_R}}{{d'_L}} }
\big)$ contains
the ${\phi}$ in ${({\bf 1},{\bf 2},{\bf 2})_{-3,3}}$  written in $\SU(2)_{\rm L} \times \SU(2)_{\rm R}$ doublet $\phi_{a,a'}$,
the ${\bar{d}_R}$ in $(\overline{\bf 3},{\bf 1},{\bf 2})_{2,-1}$ written in $\SU(2)_{\rm R}$ 
doublet $\big(\begin{smallmatrix} {\bar{d}_{R,1}} \\ {\bar{d}_{R,2}}\end{smallmatrix} \big)$,
the ${q_L}$ in $({\bf 3},{\bf 2},{\bf 1})_{1,-2}$
written in $\SU(2)_{\rm L}$ doublet $\big(\begin{smallmatrix} {{u}_{L}} \\ {{d}_{L}}\end{smallmatrix} \big)$,
the ${{\bar{u}_R}}$ and ${{d'_L}}$ are in 
$(\overline{\bf 3},{\bf 1},{\bf 1})_{-4,2}$ and $({\bf 3},{\bf 1},{\bf 1})_{-2,4}$
as the $\SU(2)_{\rm L} \times \SU(2)_{\rm R}$ singlet.
A pertinent term with $u$ quark is
$\phi {{\bar{d}_R}}  {{q_L}}=
\epsilon_{\rm L}^{a,b}   \epsilon_{\rm R}^{a',b'}  
\phi_{a,a'}
 {\bar{d}_{R,b'}}
 {{q_{L,b}}}
 =\phi_{1,1}
 {\bar{d}_{R,2}}{{q_{L,2}}}
 +\phi_{1,2}
 {\bar{d}_{R,1}}{{q_{L,2}}}
 +\phi_{2,1}
 {\bar{d}_{R,2}}{{q_{L,1}}}
 +\phi_{2,2}
 {\bar{d}_{R,1}}{{q_{L,1}}}
  =\phi_{1,1}
 {\bar{d}_{R,2}}{{d_{L}}}
 +\phi_{1,2}
 {\bar{d}_{R,1}}{{d_{L}}}
 +\phi_{2,1}
 {\bar{d}_{R,2}}{{u_{L}}}
 +\phi_{2,2}
 {\bar{d}_{R,1}}{{u_{L}}}
$.
Another pertinent term with $u$ quark is $\phi^{\dagger2}  {{{\bar{u}_R}}{{d'_L}} }
= (\phi_{1,1} \phi_{2,2} + \phi_{1,2} \phi_{2,1} - \dots)^\dagger {{{\bar{u}_R}}{{d'_L}} }$
where we collect the $\SU(2)_{\rm L} \times \SU(2)_{\rm R}$ singlet representation out of $\phi^{\dagger2}$.\\
We can deduce that \eq{eq:u-quark-SMG} contains the $u$ quark field explicitly in:\\
%
$\bullet$ $ ({{\bar{d}_R}}  {{q_L}})^2({{{\bar{u}_R}}{{d'_L}} })$ contains $\big( ...( {\bar{d}_{R,2}}{{d_{L}}})( {\bar{d}_{R,1}}{{u_{L}}})+ 
... ({\bar{d}_{R,1}}{{d_{L}}})
 ({\bar{d}_{R,2}}{{u_{L}}})  \big)
({{{\bar{u}_R}}{{d'_L}} })$. \\
%
$\bullet$  $ ({{\bar{d}_R}}  {{q_L}}) 
({{\bar{\nu}_R}} {{\bar{l}'_R}}
+ {{\bar{e}_R}} {{l_L}})({{{\bar{u}_R}}{{d'_L}} })$
  contains $\big( ...( {\bar{d}_{R,2}}{{d_{L}}})({{\bar{\nu}_{R,1}}} {{\bar{l}'_{R,1}}}
+ {{\bar{e}_{R,1}}} {{l_{L,1}}})+ 
... ({\bar{d}_{R,1}}{{d_{L}}})
 ({{\bar{\nu}_{R,2}}} {{\bar{l}'_{R,1}}}
+ {{\bar{e}_{R,2}}} {{l_{L,1}}})
-
...( {\bar{d}_{R,1}}{{u_{L}}})({{\bar{\nu}_{R,2}}} {{\bar{l}'_{R,2}}}
+ {{\bar{e}_{R,2}}} {{l_{L,2}}})
-
... ({\bar{d}_{R,2}}{{u_{L}}})
 ({{\bar{\nu}_{R,1}}} {{\bar{l}'_{R,2}}}
+ {{\bar{e}_{R,1}}} {{l_{L,2}}})
  \big)
({{{\bar{u}_R}}{{d'_L}} })$.\\
%
$\bullet$  $ (   {{\bar{\nu}_R}} {{\bar{l}'_R}}
+ {{\bar{e}_R}} {{l_L}})^2({{{\bar{u}_R}}{{d'_L}} })$ contains
$\big( ...({{\bar{\nu}_{R,2}}} {{\bar{l}'_{R,2}}}
+ {{\bar{e}_{R,2}}} {{l_{L,2}}})({{\bar{\nu}_{R,1}}} {{\bar{l}'_{R,1}}}
+ {{\bar{e}_{R,1}}} {{l_{L,1}}})+ 
... ({{\bar{\nu}_{R,1}}} {{\bar{l}'_{R,2}}}
+ {{\bar{e}_{R,1}}} {{l_{L,2}}})
 ({{\bar{\nu}_{R,2}}} {{\bar{l}'_{R,1}}}
+ {{\bar{e}_{R,2}}} {{l_{L,1}}})
-
...({{\bar{\nu}_{R,1}}} {{\bar{l}'_{R,1}}}
+ {{\bar{e}_{R,1}}} {{l_{L,1}}})
({{\bar{\nu}_{R,2}}} {{\bar{l}'_{R,2}}}
+ {{\bar{e}_{R,2}}} {{l_{L,2}}})
-
... ({{\bar{\nu}_{R,2}}} {{\bar{l}'_{R,1}}}
+ {{\bar{e}_{R,2}}} {{l_{L,1}}})
 ({{\bar{\nu}_{R,1}}} {{\bar{l}'_{R,2}}}
+ {{\bar{e}_{R,1}}} {{l_{L,2}}})
  \big)
({{{\bar{u}_R}}{{d'_L}} })$.
\\
We collectively gather these terms into \eq{eq:u-quark-SMG-2}
as four-fermion
${\bar{u}_R} (\CO_{\psi\psi}) u_L$ and 
six-fermion ${\bar{u}_R} (\CO_{\psi\psi\psi\psi}) u_L
+  {\bar{u}_R} (\CO_{\psi\psi\psi\psi\psi})$ interactions.
}
where we provide the explicit expression of the 
four-fermion
${\bar{u}_R} (\CO_{\psi\psi}) u_L$ and 
six-fermion ${\bar{u}_R} (\CO_{\psi\psi\psi\psi}) u_L
+  {\bar{u}_R} (\CO_{\psi\psi\psi\psi\psi})$ interactions.
The $\CO_{\psi\psi}$ is some two-fermion term,
the $\CO_{\psi\psi\psi\psi}$ is some four-fermion term,
and the $\CO_{\psi\psi\psi\psi\psi}$  is some five-fermion term, written in the footnote \ref{ft:multi-fermion}.

In summary, the analogous 
lagrangian term of \eq{eq:chiral-transformation-on-quark} under the $u$ quark's left $L$ and right $R$ chiral transformations 
($u_L \to  \e^{\ii \alpha_L} u_L$ and $u_R \to  \e^{\ii \alpha_R} u_R$)
becomes:
\begin{multline}
\label{eq:Ju-mass}
\hspace{-8mm}
(\upalpha_{u_L}  \prt_{\mu} J^{\mu}_{u_L}  + \upalpha_{u_R}  \prt_{\mu} J^{\mu}_{u_R}  ) 
 {-\, \frac{(\bar{\theta} - ( - \upalpha_{u_L} + \upalpha_{u_R}) )}{{64} \pi^2} {g}^2 \epsilon^{\mu\nu \mu'\nu'} F_{\mu\nu}^\ra F_{\mu'\nu'}^\ra}
 -  {m}_u (  \e^{\ii (  \upalpha_{u_L} - \upalpha_{u_R})}   u_R^\dagger u_L  +  {\rm H.c.})  
 \\
    +  g_{\rm SMG} \Big( \big( 
  \e^{\ii (  \upalpha_{u_L} - \upalpha_{u_R})}  {\bar{u}_R} (\CO_{\psi\psi}) u_L +  
\e^{\ii (  \upalpha_{u_L} - \upalpha_{u_R})}   {\bar{u}_R} (\CO_{\psi\psi\psi\psi}) u_L
+ \e^{\ii (  - \upalpha_{u_R})}  {\bar{u}_R} (\CO_{\psi\psi\psi\psi\psi}) 
+ {\rm H.c.} \Big)+ \dots.\quad \\
\end{multline}
The omitted $\dots$ terms are extra terms in \eq{eq:Yukawa-full-gap-multi-fermion} and \eq{eq:Yukawa-full-gap-multi-fermion-2}
that are invariant under $\upalpha_{u_L}$ and $\upalpha_{u_R}$.
Under the $\upalpha_{u}$ variation on the above action $S$ in \eq{eq:Ju-mass}, 
then taking its mean-field expectation $\< \frac{\delta S}{\delta \upalpha_{u}}\> \vert_{\upalpha_{u}=0}$, we obtain:
\bea
\label{eq:Ju-mass-mean-field-total}
&&\<\prt_{\mu} J^{\mu }_{u_L}\> =
 -{\frac{ 1 }{{64} \pi^2} {g}^2 \epsilon^{\mu\nu \mu'\nu'} \< F_{\mu\nu}^\ra F_{\mu'\nu'}^\ra \>}
 - \big( \< {m}_u 
 \ii u_R^\dagger u_L  \> + {\rm H.c.}  \big)
 + g_{\rm SMG} 
 \big(\ii \< 
 {\bar{u}_R} (\CO_{\psi\psi}  + \CO_{\psi\psi\psi\psi} ) u_L\> 
 + {\rm H.c.}  \big) .\cr
&&
\<\prt_{\mu} J^{\mu }_{u_R}\> =
 +{\frac{ 1 }{{64} \pi^2} {g}^2 \epsilon^{\mu\nu \mu'\nu'} \< F_{\mu\nu}^\ra F_{\mu'\nu'}^\ra \>}
 +\big( \< {m}_u 
 \ii u_R^\dagger u_L  \> + {\rm H.c.} \big)
 - g_{\rm SMG} 
 \big(\ii \< 
 {\bar{u}_R} (\CO_{\psi\psi}  + \CO_{\psi\psi\psi\psi} ) u_L\> 
 +\ii \< 
 {\bar{u}_R} ( \CO_{\psi\psi\psi\psi\psi} )\>
 + {\rm H.c.}  \big) 
 .\cr
&&
\eea
Again  
$\< {m}_u  \ii u_R^\dagger u_L  \>=\< {m}_u \>
\<  \ii u_R^\dagger u_L  \> $
when ${m}_u=\< {m}_u \>$ is just a mean-field mass.
In this specific Strong CP solution in \Sec{sec:toy-SM}, we take ${m}_u=\< {m}_u \>=0$.
Next we look at the third term on the right-hand side of \eq{eq:Ju-mass-mean-field-total},
this term is part of the SMG interaction that gives no mean-field mass,
thus $ \<   u_R^\dagger u_L  \> =0$, 
same for its linear response
$\CO_{\psi\psi}  = \CO_{\psi\psi\psi\psi} =0$. 
Similarly, other fermion bilinear pairings also have zero mean-field values. 
Namely, in the schematic SMG deformation \eq{eq:SMG-mass},
we have $\<\xi_{\rm I}  \psi_{\rm J} \>=\< \CO_{\rm SMG, IJ} \>=0$ 
but we are left to evaluate the vev of full SMG term
$\<\xi_{\rm I} \CO_{\rm SMG, IJ} \psi_{\rm J} \>+ {\rm H.c.}$.\\

Once $g_{\rm SMG}$ is turned on, the small $\<\xi_{\rm I} \CO_{\rm SMG, IJ} \psi_{\rm J} \> \neq 0$ starts to gradually develop.
In fact, for a wide range of $g_{\rm SMG} < g_{c,{\rm SMG}}$ below the critical SMG strength $g_{c,{\rm SMG}}$,
the $\<\xi_{\rm I} \CO_{\rm SMG, IJ} \psi_{\rm J} \> \neq 0$ is very small but nonzero.
The $\<\xi_{\rm I} \CO_{\rm SMG, IJ} \psi_{\rm J} \> \neq 0$ in general when $g_{\rm SMG} \neq 0$.\\

Overall, we have to keep $\<\xi_{\rm I} \CO_{\rm SMG, IJ} \psi_{\rm J} \> \neq 0$ contribution in \eq{eq:Ju-mass-mean-field-total}. 
Eq.\eq{eq:Ju-mass-mean-field-total} 
becomes
\bea
\label{eq:Ju-mass-mean-field}
&&\<\prt_{\mu} J^{\mu }_{u_L}\> =
 -{\frac{ 1 }{{64} \pi^2} {g}^2 \epsilon^{\mu\nu \mu'\nu'} \< F_{\mu\nu}^\ra F_{\mu'\nu'}^\ra \>}
 + g_{\rm SMG} 
 \big(\ii \< 
 {\bar{u}_R} (\CO_{\psi\psi}  + \CO_{\psi\psi\psi\psi} ) u_L\> 
 + {\rm H.c.}  \big) .\cr
&&
\<\prt_{\mu} J^{\mu }_{u_R}\> =
 +{\frac{ 1 }{{64} \pi^2} {g}^2 \epsilon^{\mu\nu \mu'\nu'} \< F_{\mu\nu}^\ra F_{\mu'\nu'}^\ra \>}
 - g_{\rm SMG} 
 \big(\ii \< 
 {\bar{u}_R} (\CO_{\psi\psi}  + \CO_{\psi\psi\psi\psi} ) u_L\> 
 +\ii \< 
 {\bar{u}_R} ( \CO_{\psi\psi\psi\psi\psi} )\>
 + {\rm H.c.}  \big) 
 .\cr
&&
\eea
Some small SMG multi-fermion interaction condensates $\<\xi_{\rm I} \CO_{\rm SMG, IJ} \psi_{\rm J} \> \neq 0$ 
that can mildly violate the anomalous current conservation (here $\<\prt_{\mu} J^{\mu }_{u_L}\>$ and $\<\prt_{\mu} J^{\mu }_{u_R}\>$)
when $g_{\rm SMG} < g_{c,{\rm SMG}}$, but violate significantly and nonperturbatively when $g_{\rm SMG} \geq g_{c,{\rm SMG}}$.
It will be interesting to compare these SMG multi-fermion interactions with those 't Hooft vertices with multi-fermion insertions 
\cite{ tHooft1976ripPRL, tHooft1976instanton} in the future.
Those $\<\xi_{\rm I} \CO_{\rm SMG, IJ} \psi_{\rm J} \> \neq 0$ contributions have some other physical effects,\footnote{We 
expect to find physical observables related to $\<\xi_{\rm I} \CO_{\rm SMG, IJ} \psi_{\rm J} \> \neq 0$ 
in the dual variable of ${\bar{\theta}_{}}$ in the SMG disordered side of
the story along the discussion in \cite{WangCTorPProblem2207.14813}.} 
but \emph{not} on the Strong CP problem's 
${\bar{\theta}_{}}$ 
defined in \eq{eq:def-mass}
\bea
{\bar{\theta}_{}} 
\equiv
{\theta_{}}
+\arg (\det 
\< \mathbf{M}\>)
=
{\theta_{}}
+\arg (\det 
\<- \frac{\delta^2 \cL}{(\delta \xi_{\rm I}) (\delta \psi_{\rm J})}\>)
=
{\theta_{}}
+\arg (\det 
\< M\>)
\eea 
--- because the ${\bar{\theta}_{}}$ receives a zero contribution from SMG's fermion bilinear 
$\<\xi_{\rm I}  \psi_{\rm J} \> =  0$ and $\<\CO_{\rm SMG, IJ}\>=0$.
Thus, for now, we can switch gears to show how ${\bar{\theta}_{}}$ can be set to zero.
As long as some of the quarks have their entire mass from no mean-field mass, then $\det \< M \>=0$.
Here in \Sec{sec:toy-SM}, we assume that the $u$ quark gains its mass only from SMG, but not Higgs mechanism;
hence, the mean-field mass eigenvalue for $u$ quark is zero so $\det \< M \>=0$.

We end up redefining ${\bar{\theta}_{}}$ by a chiral transformation,
with $\det \<\mathbf{M} (\upalpha) \>=0$ still, because the $u$ quark mean-field mass eigenvalue is zero, 
so 
\bea
{\bar{\theta}_{}}= {\theta_{}} - \upalpha =0
\eea 
is appropriately chosen to be zero.
This provides the Strong CP solution: The ${\bar{\theta}_{}}$ is zero for the entire theory.

The SMG multi-fermion interaction condensates $\<\xi_{\rm I} \CO_{\rm SMG, IJ} \psi_{\rm J} (\upalpha) \> \neq 0$ 
has $\upalpha$ dependence, and can be nonperturbatively nonzero above $g_{\rm SMG} \geq g_{c,{\rm SMG}}$.
This has a consequence on the vortex dual variable 
of ${\bar{\theta}_{}}$ in the ${\bar{\theta}_{}}$-disordered phase (see \cite{WangCTorPProblem2207.14813}). 
But it does not affect the ${\bar{\theta}_{}}=0$ as far as the Strong CP solution  is concerned. 

 \end{enumerate}
 
%
\subsection{Constraint and Prediction on the Hadron Data}
 \label{sec:hadron}

 't Hooft massless $u$ quark solution \cite{tHooft1976ripPRL} that requires $m_u =0$
has been ruled out by the lattice QCD data.
So whatever new Strong CP solution that we provide,
we must reconcile the lattice QCD data correctly without conflicting with our new Strong CP solution.
According to Particle Data Group (PDG)
\cite{PDG-quark}, the current quark mass $m_u=2.16$ MeV
is an input to the QCD lagrangian at the energy scale $E_{\overline{\text{MS}}} =  2$ GeV
based on the modified minimal subtraction ($\overline{\text{MS}}$) renormalization scheme. 
Other current quark masses $m_q$ are also nonzero,
in order to produce the correct hadron mass 
(e.g. meson such as the pion mass 135 MeV or baryon such as the proton mass 938 MeV)
observed at the low energy (say the $\Lambda_{\QCD}$ scale 200 MeV)
in Nature confirmed by experiments  \cite{PDG-quark}.

In summary of \Sec{sec:toy-SM}'s solution, 
In the conventional scenario (\Table{tab:current-mass} (a)),
all the quarks and other fermions obtain their masses from the Higgs mechanism.
In contrast, in \Sec{sec:toy-SM}'s solution scenario (\Table{tab:current-mass} (b)),
we hypothesized that the $u$ current quark mass is fully due to SMG,
other fermions also obtain a small portion of SMG mass, in addition to the major current mass contribution is from the Higgs mechanism.
Namely, in \Table{tab:current-mass} (b), we replace the $u$ current quark mass $m_u=2.16$ MeV
to the $u$ quark SMG mass.

Some of the important energy scales for \Table{tab:current-mass} (b):\\
\noindent
1. SMG scale $\Lambda_{\SMG}$, much higher, like GUT scale or other scales.

\noindent
2. Electroweak Higgs scale 246 GeV.

\noindent
3. UV renormalization scale $E_{\overline{\text{MS}}} =  2$ GeV, the $\overline{\text{MS}}$ scale for the lattice QCD.

\noindent
4. $\Lambda_{\QCD}$ scale 200 MeV.

\begin{table}[!t] 
\begin{center}
(a) \; \begin{tabular}{c|c|| c | c c}
\hline
quark
&  \begin{tabular}{c}
current\\
quark\\
mass
$m_q$
\end{tabular}
 & 
 \begin{tabular}{c}
 Higgs \\
 contribution
 \end{tabular}
 & 
  \begin{tabular}{c}
 SMG  \\
 contribution
 \end{tabular}
 &\\
\hline
$u$   & 2.16 MeV  & 
\multirow{6}{*}{$
\begin{array}{c}
\text{All}
\end{array}
$} 
&  
\multirow{6}{*}{$
\begin{array}{c}
\text{No}\\
\end{array}
$
}
\\
\cline{1-2}
$d$   & 4.67 MeV  &  &  &\\
\cline{1-2}
$c$   & 1.27 GeV &
 & 
&\\
\cline{1-2}
$s$   & 93.4 MeV  & & &\\
\cline{1-2}
$t$   & 172.69 GeV  & & &\\
\cline{1-2}
$b$   & 4.18 GeV  & & &\\
\hline
\hline
lepton & 
\begin{tabular}{c}
lepton\\
mass
$m_q$
\end{tabular}
 &  \begin{tabular}{c}
 Higgs \\
 contribution
 \end{tabular}
 & 
  \begin{tabular}{c}
 SMG  \\
 contribution
 \end{tabular}
 &\\
\hline
$e$ & 0.511 MeV & 
\multirow{3}{*}{$
\begin{array}{c}
\text{All}
\end{array}
$} 
&  
\multirow{3}{*}{$
\begin{array}{c}
\text{No}
\end{array}
$} 
\\
\cline{1-2}
$\mu$ & 105.66 MeV  
& 
& 
&
\\
\cline{1-2}
$\tau$ & 1776.86 MeV  & & &\\
\hline
\end{tabular}
(b) \; \begin{tabular}{c|c|| c | c c}
\hline
quark
&  \begin{tabular}{c}
current\\
quark\\
mass
$m_q$
\end{tabular}
 & 
 \begin{tabular}{c}
 Higgs \\
 contribution
 \end{tabular}
 & 
  \begin{tabular}{c}
 SMG  \\
 contribution
 \end{tabular}
 &\\
\hline
$u$   & 2.16 MeV  & 0 & 2.16 MeV & \\
\hline
$d$   & 4.67 MeV  & 4.67 MeV $- m_d^{\SMG}$ & $m_d^{\SMG}$ &\\
\hline
$c$   & 1.27 GeV &
\multirow{4}{*}{$
\begin{array}{c}
\text{Higgs condensation}\\
\text{make up the}\\
\text{non-SMG part} 
\end{array}
$
}
 & 
\multirow{4}{*}{$
\begin{array}{c}
\text{No SMG}\\
\text{required, but}\\
\text{still possible.} 
\end{array}
$
}
&\\
\cline{1-2}
$s$   & 93.4 MeV  & & &\\
\cline{1-2}
$t$   & 172.69 GeV  & & &\\
\cline{1-2}
$b$   & 4.18 GeV  & & &\\
\hline
\hline
lepton & 
\begin{tabular}{c}
lepton\\
mass
$m_q$
\end{tabular}
 &  \begin{tabular}{c}
 Higgs \\
 contribution
 \end{tabular}
 & 
  \begin{tabular}{c}
 SMG  \\
 contribution
 \end{tabular}
 &\\
\hline
$e$ & 0.511 MeV & 0.511 MeV $- m_e^{\SMG}$ & $m_e^{\SMG}$ &\\
\hline
$\mu$ & 105.66 MeV  & 
\multirow{2}{*}{$
\begin{array}{c}
\text{Higgs condensation}\\
\text{makes up the remains}
\end{array}
$}
& 
\multirow{2}{*}{$
\begin{array}{c}
\text{No SMG}\\
\text{required}
\end{array}
$}
&
\\
\cline{1-2}
$\tau$ & 1776.86 MeV  & & &\\
\hline
\end{tabular}
\caption{(a) The conventional mass-generating mechanism is attributed to the Higgs mechanism, see \Fig{fig:Higgs-induced}.
(b) On the left-hand side of the table, 
the current
quark
mass $m_q$ from \cite{PDG-quark} is based on fitting the quark mass
as an input to the QCD lagrangian under $\overline{\text{MS}}$ at the energy scale 2 GeV
for the lattice simulation, in order to produce the correct hadron mass (e.g. meson such as the pion mass 135 MeV or baryon such as the proton mass 938 MeV)
observed at the low energy (say the $\Lambda_{\QCD}$ scale 200 MeV)
in Nature confirmed by experiments.
On the right-hand side of the table, we show the portion of the Higgs contribution (\Fig{fig:Higgs-SMG-induced}'s red color)
and the portion of the hypothetical SMG contribution (\Fig{fig:Higgs-SMG-induced}'s blue color).
}
\label{tab:current-mass}
\end{center}
\end{table}
\begin{figure}[!h] 
\includegraphics[width=.6\textwidth]{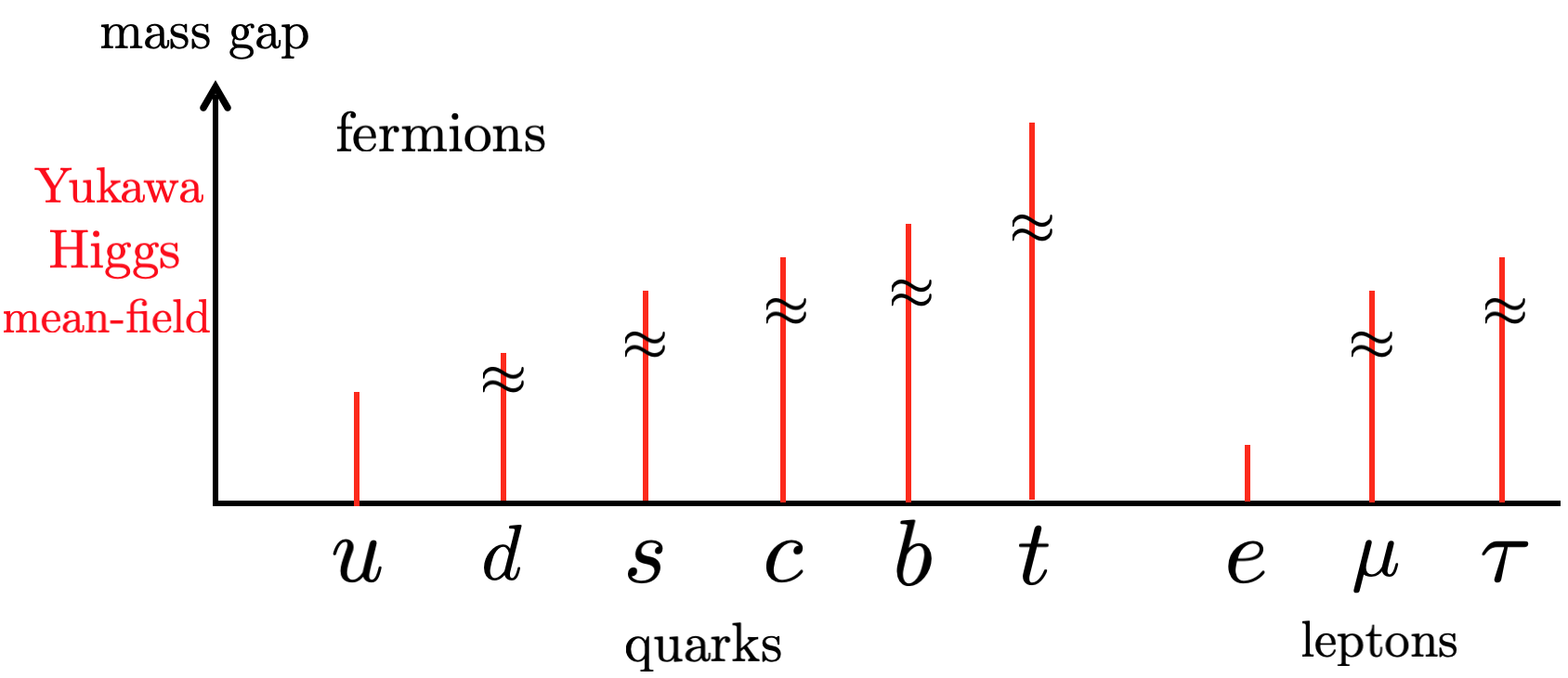}
\caption{A schematic plot on the Higgs-induced contribution (red)
to quark and lepton masses in Table \ref{tab:current-mass}(a).
This data is meant to compare with the current quark mass of the lattice QCD lagrangian input at the RG energy scale $E_{\overline{\text{MS}}} =  2$ GeV.}
\label{fig:Higgs-induced}
\end{figure}
\begin{figure}[!h] 
\includegraphics[width=.6\textwidth]{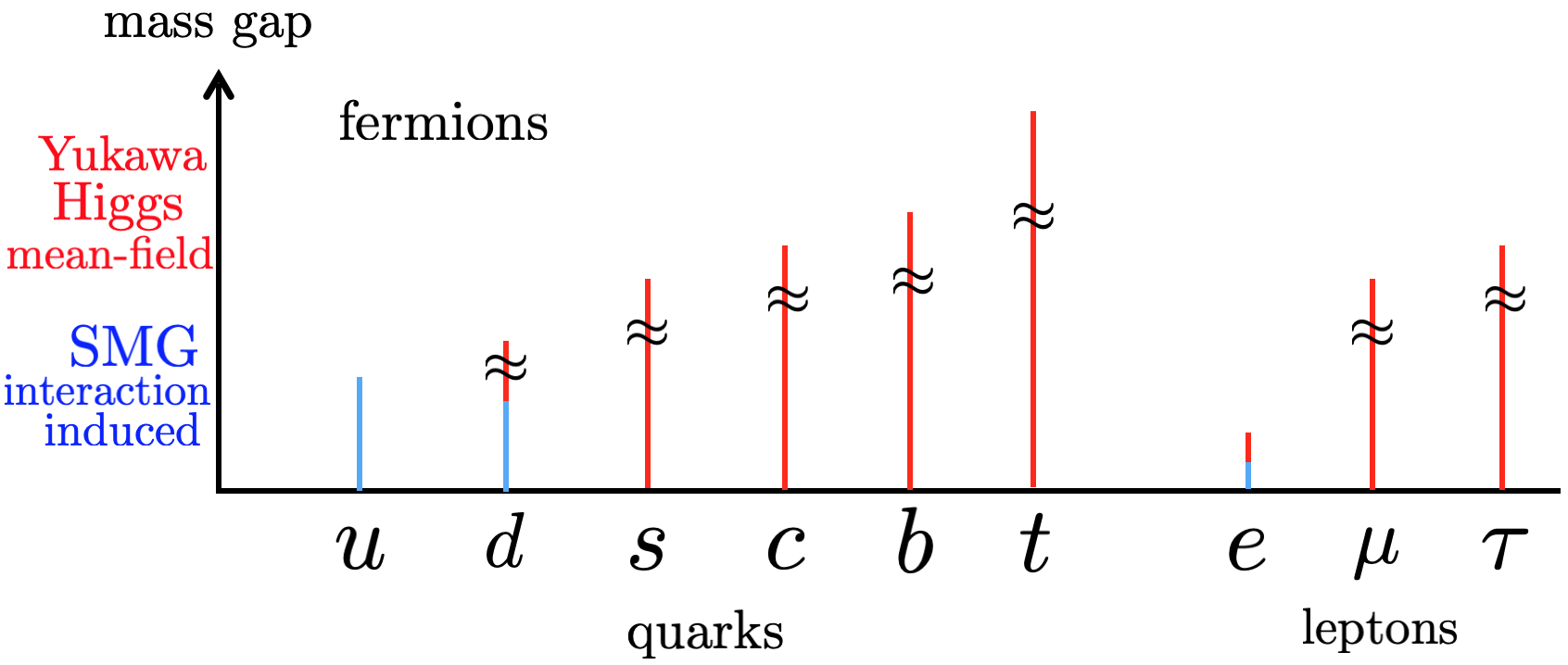}
\caption{A schematic plot on the SMG-induced (blue) and Higgs-induced (red) contribution to quark and lepton masses in Table \ref{tab:current-mass}(b).
This data is meant to compare with the current quark mass of the lattice QCD lagrangian input at the RG energy scale $E_{\overline{\text{MS}}} =  2$ GeV.
The SMG-induced contribution to a part of some generation of quark and lepton
masses implies that we need to modify the conventional QCD lagrangian to include the SMG interaction term 
$\xi_{\rm I} \CO_{\rm SMG, IJ} \psi_{\rm J}$ into the UV lagrangian. In the scenario presented here,
although the fermions $u$, $d$, and $e$ receive SMG-induced mass from the same SMG interaction term,
their SMG-induced masses do \emph{not} need to be in the same value. 
Their SMG-induced masses depend on the multi-fermion-pairing structure of the
SMG interaction term  $\xi_{\rm I} \CO_{\rm SMG, IJ} \psi_{\rm J}$ (see the related discussion in \cite{WangCTorPProblem2207.14813}).
}
\label{fig:Higgs-SMG-induced}
\end{figure}

Below the SMG scale, the multi-fermion interaction gives the set of fermion SMG-induced mass.
Below the electroweak Higgs scale, the fermions also obtain the Higgs-induced mass.
We note that just like the Higgs scale is not directly equal to the Higgs-induced mass.
the SMG scale (say $g_{c,{\rm SMG}}$) is not directly equal to the SMG-induced mass.
Only below or around the $\Lambda_{\QCD}$ scale, the confinement happens and the chiral condensate $\< \bar{\psi}_{q} \psi_q \> \neq 0$ kicks in.
Then the hadrons form and get about 98\% of their mass from the confinement, while only a few 2\%
is from the Higgs mechanism and possibly also some from the SMG mechanism (\Table{tab:current-mass} (b)).
However, the lattice QCD lagrangian input is at $E_{\overline{\text{MS}}} =  2$ GeV, which is way above the
$\Lambda_{\QCD}$ scale, so the chiral condensate $\< \bar{\psi}_{q} \psi_q \> =0$ at $E_{\overline{\text{MS}}}$ and does not contribute 
to the initial input of the lattice QCD lagrangian. In fact, the chiral condensate $\< \bar{\psi}_{q} \psi_q \> \neq0$
is generated spontaneously under the RG flow from running $E_{\overline{\text{MS}}} =  2$ GeV to 
$\Lambda_{\QCD}$ scale in the lattice QCD simulation.

 {\bf Falsifiable prediction}: We can make a falsifiable test of our proposal.
Suppose we start with a modified QCD and SM plus SMG lagrangian,
and start with the mean-field current quark mass $m_u=0$ at $E_{\overline{\text{MS}}} =  2$ GeV,
we can still have the SMG-induced mass for the $u$ quark and other fermions.
Then we predict that running the simulation in terms of this \Table{tab:current-mass} (b) scenario,
we could still reproduce the correct hadron spectrum at low energy to match the experimental
data (say pion mass 135 MeV or proton mass 938 MeV).
Although this simulation is a very challenging task, if this is indeed verified numerically,
it will give a support to the \Sec{sec:toy-SM}'s solution.

{\bf Other challenges}: If $u$ quark does get involved in a multi-fermion interaction to gain its SMG mass,
it means that at high enough energy, the nonperturbative multi-fermion interaction will gradually
dominate over the asymptotic free behavior of quarks. 
The good news for this nonperturbative multi-fermion interaction is that it gives another falsifiable prediction
at higher energy that modifies the asymptotic freedom at much deeper UV.
The bad news is that experiments already rule out any physics deviated
from the QCD asymptotic freedom at least for energy as high as 10 TeV.
QCD is a very good description of quarks at least to 10 TeV or even higher.
This means that the $\Lambda_{\SMG}$ has to be higher if this scenario works.

{\bf Variants of scenarios}: Even if the \Sec{sec:toy-SM}'s solution (on the first family of fermions get SMG-induced mass) scenario fails,
theoretically we could still propose some modification of similar kinds.
Another theoretical solution is that a hypothetical fourth family of quarks and leptons get the SMG-induced mass.
Some comments are this scenario:\\
$\bullet$ It could be that the fourth family of fermions get the full SMG-induced mass. 
(It is less likely any fourth family of fermions receives Higgs-induced mass, because that would cause observable effects 
on the Higgs channel that should be already tested and observed.)\\
$\bullet$ The fourth family of fermions should however get a large SMG-induced mass larger than the mass scale that has been tested.
So the mass scale of the fourth family of fermions should be in principle larger than $t$ quark mass above 173 GeV or TeV scales.\\
$\bullet$ The fourth family of quarks must couple to the same $\su(3)$ strong force so that those quarks 
can absorb $\bar{\theta}$ away into the complex phase of its zero mean-field mass. 
(Namely, as emphasized previously, 
we must have at least one of the quarks receive its mass only from SMG but not from Higgs. 
We may as well just have the full fourth family of fermions get the full SMG-induced mass.)
But this also implies some
possible channels on the gluon-gluon interaction to observe of the evidence fourth family of quarks.
This shall be a falsifiable statement with more phenomenological constraints.

In fact, 
because we directly apply the SMG scenario to at least one family of the chiral fermion sector of the SM,
the phenomenological constraints (regardless known or unknown to the contemporary experiments) may more 
easily falsify or rule out all the models presented in \Sec{sec:FirstSolution}.
If all these theoretical proposals are not favored by phenomenology,
we can still propose another type of new scenario, in the next \Sec{sec:SecondSolution}, 
to hide the SMG mechanism in the SMG-induced gapped mirror fermion sector.

\newpage

\section{Second Solution: Symmetric Gapped Mirror Fermion}
\label{sec:SecondSolution}

\begin{figure}[!h] 
\centering
\includegraphics[height=6.75cm]{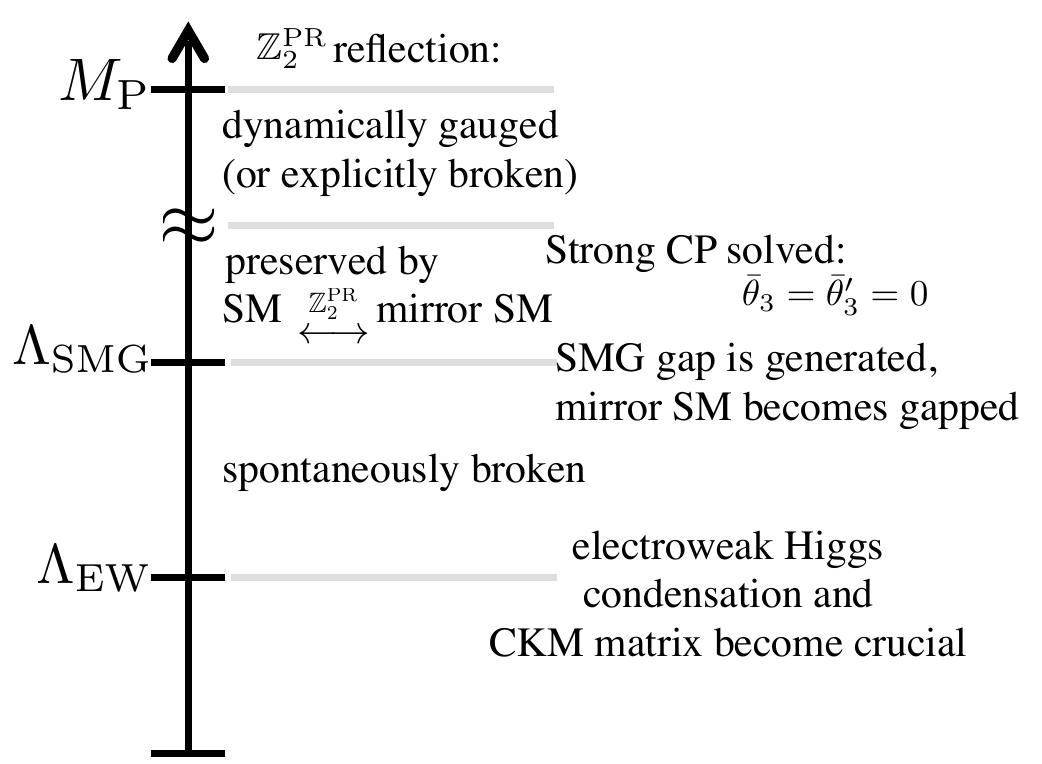}
\caption{The relations between the energy scale $E$ and the \Sec{sec:SecondSolution}'s Strong CP solution ${\bar{\theta}_{3}}= {\bar{\theta}_{3}}' =0$ 
imposed by $\Z_2^{\rm PR}$ above the SMG scale, are explained in the main text. 
Since the energy scale ${\rm E} \sim  t^{-1}_{\rm period}$ is inverse of the time scale $ t_{\rm period}$,
this shows a schematic time evolution of some of the processes 
of this quantum universe from the early universe to the later universe.}
\label{fig:2}
\end{figure}

Our second SMG solution to the Strong CP problem 
is meant to be more flexible to fit experimental constraints. 
We will however still provide only the general strategy, but will not seek for the SM phenomenological fitting here.
Here are step-by-step constructions and highlights of this solution, schematically shown in \Fig{fig:1} and \Fig{fig:2}: 
\begin{figure}[!h] 
\centering
{{(a)}}\includegraphics[height=6.75cm]{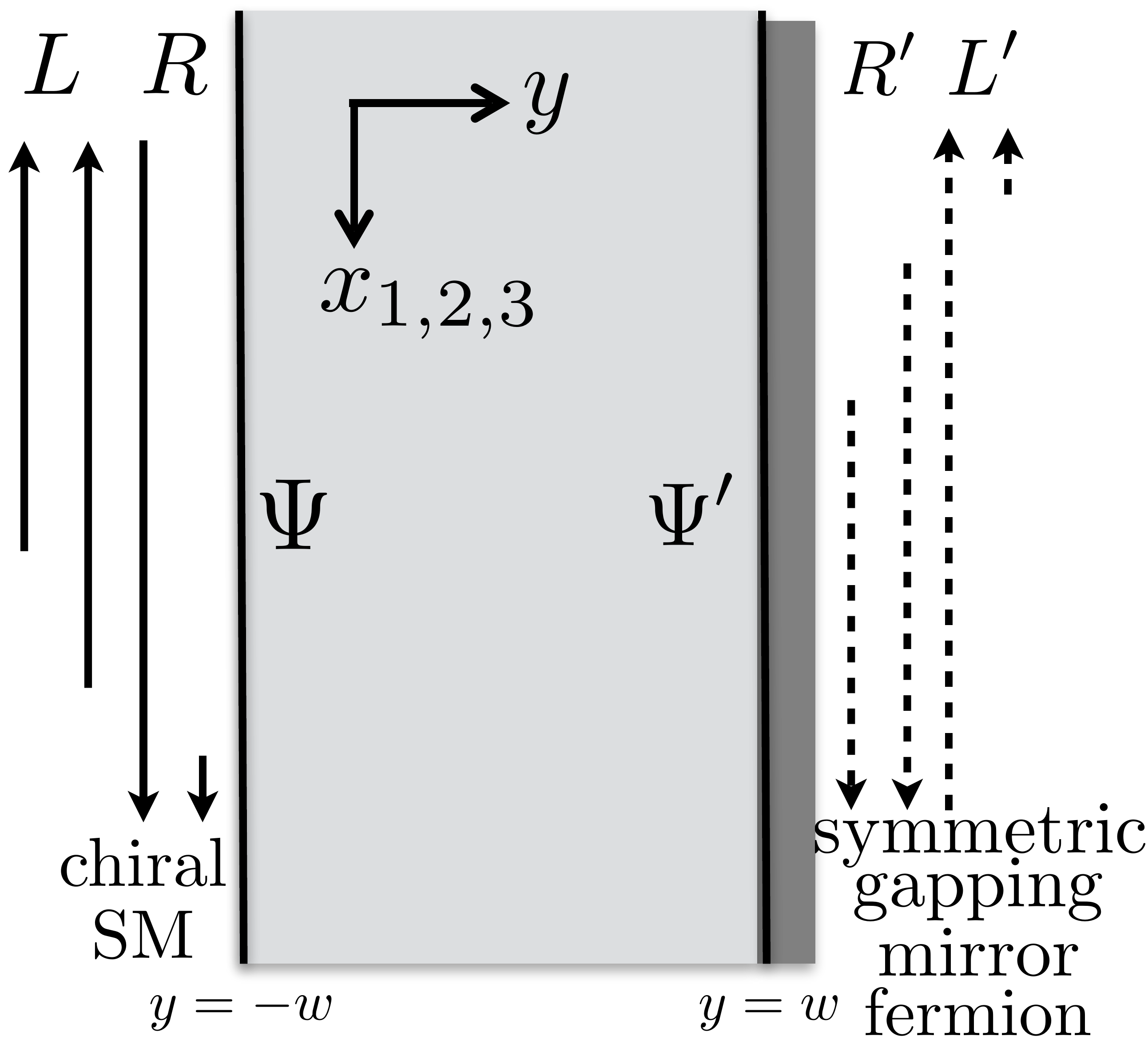} 
{{(b)}}\includegraphics[height=6.75cm]{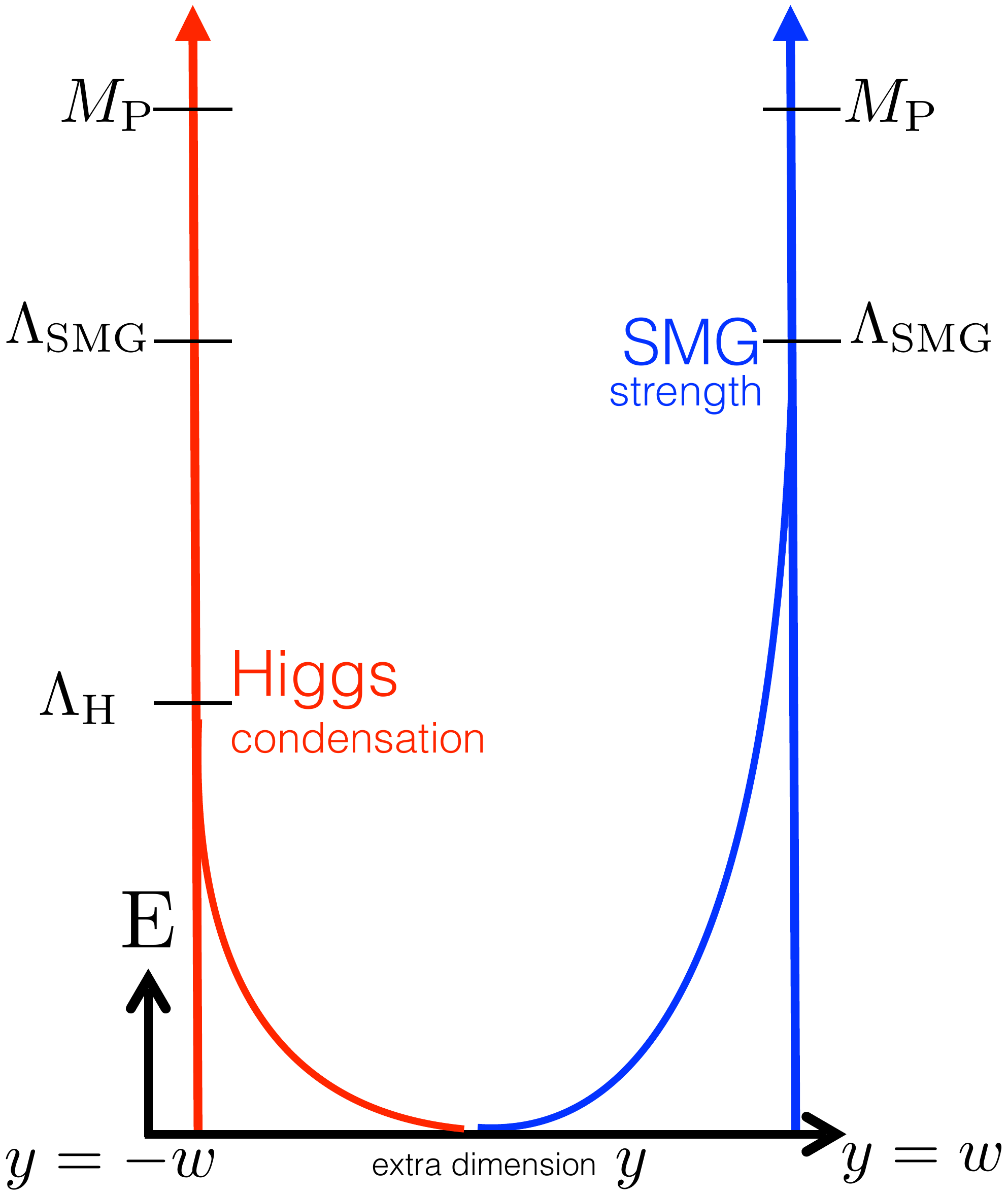}
\caption{(a) Our model has a chiral SM and a mirror SM sharing the same gauge group $G_{\SM_p}$,
placed on two domain walls, left-handed $L$ on $y=-w$ or right-handed $R$ on $y=w$ respectively shown on a spatial manifold
(e.g., two ends of a 5d regularizable manifold $\CM^4 \times I^1_{y}$). 
The $\CM^4$ contains the 4d spacetime $t, x_1, x_2, x_3$ coordinates, while the $I^1_{y}$ contains the extra finite-width fifth-dimensional $y$ coordinate.
The cause of a chiral SM and a mirror SM is due to fermion doubling. 
But the mirror SM is eventually gapped by SMG. 
Thus the fermion doubler is removed by SMG and not observed at low energy.
In the limit when the SM $\su(3) \times  \su(2) \times \u(1)_Y$ gauge field is treated as background gauge field, 
the 5d theory is gapped, either a trivial gapped vacuum or a topological field theory at low energy.
(b) The Higgs condensation profile (red curve) becomes dominant at the energy scale $\Lambda_{\rm H}$
but becomes exponentially small when going into the bulk ($+\hat y$ direction).
On the other hand, the SMG strength becomes dominant at a higher energy scale $\Lambda_{\rm SMG}$,
also it becomes exponentially small when going into the bulk ($-\hat y$ direction). 
The horizontal axis labels the bulk direction,
while the vertical axis labels both the energy scale and also schematically the strength of (Higgs condensate or SMG) interaction terms. 
In principle, it is preferred that the Higgs and SMG dynamics do not interfere with each other in any spacetime region 
(i.e., the red and blue curves do not both have nonzero values at the same region). 
However, even if the Higgs condensation and SMG interfere at the same spacetime region, 
as long as the SMG does not generate any mean-field mass as Higgs condensation does,
we can still maintain our solution of the Strong CP problem
to the low energy (IR below the $\Lambda_{\rm H}$ scale).}
\label{fig:1}
\end{figure}

\begin{enumerate}[leftmargin=-0mm, label=\textcolor{blue}{\arabic*}., ref={\arabic*}]
 
\item
Dictated by the {\bf Nielsen-Ninomiya (NN) fermion-doubling} \cite{NielsenNinomiya1981hkPLB},
typically a chiral fermion theory with a UV completion on a Planck scale or lattice regularization
can suffer from the mirror fermion doubling, which renders the full theory vector-like instead of being chiral. 
{\bf Domain wall fermions} ---
The chiral and mirror fermions localized on the two domain walls are related to the lattice domain wall fermion construction \cite{Kaplan1992A-method},
thus there is a UV completion of these models on a lattice \cite{Wen2013ppa1305.1045, Wang2013ytaJW1307.7480, You2014oaaYouBenTovXu1402.4151, YX14124784,
BenTov2015graZee1505.04312, Wang2018ugfJW1807.05998, WangWen2018cai1809.11171, RazamatTong2009.05037, Tong2104.03997}. 
%
It could be helpful to regard the 4d chiral SM and 4d mirror SM as two 4d boundaries of a 5d topological field theory
with a finite-width extra bulk dimension, see \Fig{fig:1}.\footnote{For each 4d complex Weyl fermion
with a U(1) charge $k$ global symmetry $\psi_L \mapsto \re^{\ii k \upalpha} \psi_L$, the 4d Weyl fermion theory is a 4d boundary of a 5d bulk partition function
\bea \label{eq:Z5U1}
\bZ_{5}^{\U(1)_{}} 
\equiv \exp(\ii 
((k^3 \int_{M^5} {A} c_1^2) + \frac{k}{48}  \rCS_3^{T(\PD(c_1))})
).
\eea
This is a 5d iTFT evaluated on a 5d manifold ${M^5}$ 
that captures the cubic pure gauge $\U(1)^3$ with a $k^3$ coefficient and 
mixed gauge-gravity $\U(1)$-(gravity)$^2$ anomalies with a $k$ coefficient.
In the SM, the anomaly is related to $\U(1)_{{ \mathbf{B}-  \mathbf{L}}}^3$ and 
mixed gauge-gravity $\U(1)_{{ \mathbf{B}-  \mathbf{L}}}$-(gravity)$^2$.
The $A$ is the $\Spin \times_{\Z_2^F} \U(1)_{} \equiv \Spin^c$ gauge field.
The gravitational Chern-Simons 3-form is 
$\rCS_3^{TM^3} \equiv \frac{1}{4 \pi} \int_{M^3= \prt M^4} \Tr( \omega \dd \omega + \frac{2}{3} \omega^3)
=\frac{1}{4 \pi} \int_{M^4} \Tr( R(\omega) \wedge R(\omega))$ where $M^3$ is evaluated as a boundary $\prt M^4$,
{while $\omega$ is the 1-connection of tangent bundle $TM$ and $R(\omega)$ is the Riemann curvature 2-form of $\omega$}.
In \eq{eq:Z5U1}, we take ${M^3= \PD(c_1)}$ to be a 3-manifold Poincar\'e dual (PD) to the degree-2 first Chern class $c_1$ on the 5d $M^5$.
On a closed oriented $M^4$, we further have $\frac{1}{4 \pi} \int_{M^4} \Tr( R(\omega) \wedge R(\omega)) = {2 \pi} \int_{M^4} p_1 (TM) = {2 \pi} \cdot 3  \sigma$,
where $p_1$ is the first Pontryagin class of $TM$ and the $\sigma$ is the $M^4$'s signature. 
Overall, 
the 5d term \eqq{eq:Z5U1} descends from a 6d anomaly polynomial on $M^6$ at the U(1)-valued $\theta = 2 \pi$: 
$
\exp(\ii \theta 
( (\int_{M^6} \frac{k^3}{2}  c_1^3) + \frac{k}{16}  \sigma(\PD(c_1)))
).$
If we have an appropriate number of Weyl fermions (e.g., 8 left-moving and 8 right-moving in the 16-fermion SM), the anomaly coefficient
in \eq{eq:Z5U1} cancels. This means that we need no 5d bulk. If we do not require any ${ \mathbf{B}- \mathbf{L}}$ symmetry, 
then the 15-fermion SM can also have no anomaly (because then $A=0$ and $c_1=0$ in \eqq{eq:Z5U1}), 
that also requires no 5d bulk.
One can also write the $\su(3) \times  \su(2) \times \u(1)_Y$ version of 5d Chern-Simons type theory of \eq{eq:Z5U1}, but the 4d SM has no anomaly
from the spacetime Spin and internal $\su(3) \times  \su(2) \times \u(1)_Y$ symmetry. So that 5d bulk is trivial and not presented. 
\\[2mm]
}
But because we are dealing with a particular \emph{anomaly-free} 4d SM,
it is not entirely important nor necessary to have a 5d bulk, 
because the 5d bulk topological field theory becomes a trivial gapped vacuum.

Let us denote the left-handed and right-handed Weyl fermions as, 
$\psi_L$ and $\psi_R$ on the domain wall (say, at an extra-dimensional coordinate $y=-w$ of \Fig{fig:1} (a)),
while $\psi'_L$ and $\psi'_R$ on the mirror domain wall (at $y=+w$ of \Fig{fig:1} (a)).
Let us denote the $F$ and $F'$ as the $\su(3)$ field strength on the domain wall and the mirror domain wall respectively. 
Then we clarify the $\Z_2^{\rm P}$ parity symmetry,
the $\Z_2^{\rm R}$ reflection symmetry,
and the combined generalized $\Z_2^{\rm PR}$ parity-reflection symmetry as \emph{active} transformations acting on \emph{fields}:
\bea
\Z_2^{\rm P} &:& \psi_L (t,x_j) \mapsto  \psi_R (t,-x_j), \quad \psi_R (t,x_j)   \mapsto \psi_L (t,-x_j), \quad F \wedge F(t,x_j) \mapsto -F \wedge F (t,-x_j).\\ 
\Z_2^{\rm R} &:& \psi_L (t,x_j) \mapsto  \psi'_L (t,x_j), \quad\quad \psi_R (t,x_j)   \mapsto \psi'_R (t,x_j), \quad \quad F \wedge F(t,x_j) \mapsto F' \wedge F' (t,x_j).\\
\label{eq:PR}
\Z_2^{\rm PR} &:& \psi_L (t,x_j) \mapsto  \psi'_R (t,-x_j), \; \quad \psi_R (t,x_j)   \mapsto \psi'_L (t,-x_j), \; \quad F \wedge F(t,x_j) \mapsto - F' \wedge F' (t,-x_j).
\eea
The domain wall fermion construction (like \Fig{fig:1}) is similar but somehow different from the Parity Solution \cite{BabuMohapatra1989, BabuMohapatra1989rbPRD, BarrChangSenjanovic1991qxPRL}.
In the sense that the domain wall fermion has the chiral and mirror fermions 
respecting the combined generalized $\Z_2^{\rm PR}$ parity-reflection symmetry.
The $\Z_2^{\rm PR}$ is a discrete spacetime symmetry, which, in terms of a \emph{passive} transformation that acts on the \emph{coordinates}, mapping the spatial coordinates:
\bea
\Z_2^{\rm PR} &:& x \equiv (t, {x_1}, {x_2}, {x_3}, y) \mapsto x'_{\rR} \equiv (t, -{x_1}, -{x_2}, -{x_3}, -y).
\eea
For example, in \Fig{fig:1} (a), one domain wall with chiral fermions at $y=-w$ and another domain wall with mirror fermions 
at $y=+w$ are mapped to each other under 
the reflection $\Z_2^{\rm R}$ (without flipping ${x_1}, {x_2}, {x_3}$) and the parity-reflection $\Z_2^{\rm PR}$ (flipping also ${x_1}, {x_2}, {x_3}$).
In contrast, the generalized {parity} symmetry in the Parity Solution \cite{BabuMohapatra1989, BabuMohapatra1989rbPRD, BarrChangSenjanovic1991qxPRL}
is typically a combined \emph{spatial} parity symmetry with an \emph{internal} discrete $\Z_2$ symmetry that exchanges the two copies of the SM gauge group.

In fact, the original chiral fermion $\psi_L(x)$ and the mirror doubling $\psi_R'(x)$ are parity-reflection 
PR paired 
with each other under the \emph{active} $\Z_2^{\rm PR}$ transformation:
\bea
\psi_L(x) \overset{\mathbb{Z}_2^{\rm PR}}{\longleftrightarrow} 
{\rm P R} \psi_L(x){\rm R}^{-1} {\rm P }^{-1}=\psi_R'(x'_{\rR}).
\eea

In \Fig{fig:1}(a), 
this PR symmetry sends the chiral fermion SM theory (left-handed $L$) on one domain wall A to the mirror fermion theory (mirror SM$'$, right-handed $R$)
on another domain wall B,
along the $y$-axis via ${y} \leftrightarrow - {y}$. For illustration, we can imagine such a 
full PR-invariant vector-like system
with two 4d domain walls placed on a finite-width strip 
(a two-brane model in a finite-width 5th dimension, on a 5d smooth triangulable manifold $\CM^4 \times I^1_{x_4}$) 
as \Fig{fig:1}(a) shows.


\item
{\bf Energy scales and parity-reflection PR symmetry}:

In \Fig{fig:2}, we suggest that different physics are dominated at different energy scales $E$:

\begin{enumerate}[leftmargin=5.0mm, 
label=\textcolor{blue}{(\arabic*)}., ref={(\arabic*)}]

\item
Near the Planck scale $M_{\rm P} \sim 10^{19}$GeV, the PR must be dynamically gauged or explicitly broken due to the quantum gravity no-global-symmetry argument.\\

\item
For $M_{\rm P} > E >\Lambda_{\rm SMG}$,
the preserved $\Z_2^{\rm PR}$ symmetry dictates
the chiral SM and mirror SM are mapped into each other under the $\Z_2^{\rm PR}$-symmetry.
Then the ${\bar{\theta}_{3}}$ of SM and the
${\bar{\theta}_{3}}'$ of the mirror SM$'$ has the constraint:
${\bar{\theta}_{3}}' = - {\bar{\theta}_{3}}$, so that
${\bar{\theta}_{3}}+ {\bar{\theta}_{3}}' =0$ for the full theory.
Since the full theory couples to the same $\su(3)$,
we may simply rotate to ${\bar{\theta}_{3}} ={\bar{\theta}_{3}}'  =0$.
So the Strong CP problem is solved at this stage.\\

\item
For $\Lambda_{\rm SMG} > E > \Lambda_{\rm EW}$, 
when the mirror SM is gapped at $\Lambda_{\rm SMG}$, the $\Z_2^{\rm PR}$ symmetry is spontaneously broken.\footnote{If the
${\Z_2^{\rm PR}}$ is dynamically gauged (such as $E \sim M_P$ due to the quantum gravity reasoning on no global symmetry), 
then the ``gauge symmetry'' is not a physical global symmetry, but a gauge redundancy 
(that many seemly distinct states are indeed the same state in the Hilbert space), 
thus the gauged ${\Z_2^{\rm PR}}$ rigorously cannot be ``\emph{spontaneously} broken.''
The modern quantum interpretation of ``\emph{spontaneously} broken discrete gauge symmetry'' can be found in \cite{Verresen2211.01376} and references therein.
The quantum gravity interpretation of ``\emph{spontaneously} broken discrete parity-like symmetry'' can be found in \cite{McNamaraReece2212.00039}.
\label{ft:PR-borken}}
Crucially the whole SMG process has the SM internal $\su(3) \times \su(2) \times \u(1)_{Y}$ symmetry preserved,
thus we only need the same single copy of SM gauge group (without doubling the SM gauge group in the mirror sector).\\

\item
For $E \leq \Lambda_{\rm EW}$, 
the electroweak Higgs condensation and CKM matrix become crucial.
 
\end{enumerate}

Below we will clarify more on the reasonings in depth.

\item {\bf Total Strong CP angle ${\bar{\theta}_3 +\bar{\theta}'_3}=0$ at UV}:
Let us first explain why the $\Z_2^{\rm PR}$ symmetry 
imposes 
\bea
\text{ 
$\frac{\theta_3}{8 \pi^2} \Tr[F \wedge F] + \frac{\theta_3'}{8 \pi^2} \Tr[ F' \wedge F']=
0$ \quad  by  \quad $\theta_3'=- \theta_3$.
}
\eea
The $F$ and $F'$ are the same kind of gauge fields (SM's $\su(3)$) on the two domain walls on $y=-w$ and $y=w$.
Next we will also explain why ${\bar{\theta}_3 +\bar{\theta}'_3}=0$.

{\bf Theta term in the parent theory}:
The parent theory can include a generic $\theta$ term (in particular, hereafter we mean $\theta=\theta_3$) with the field strength $F$ and $F'$ on two domain walls as
\bea \label{eq:thetaF-theta'F}
\frac{\theta}{8 \pi^2} \int_{\cM^4} \Tr[F \wedge F] \big\vert_{y=-w} + 
\frac{\theta'}{8 \pi^2} \int_{\cM'^4} \Tr[F' \wedge F']  \big\vert_{y=w}.
\eea
One domain wall has its 4d spacetime ${\cM^4}$, the other has its 4d spacetime ${\cM'^4}$. 
This 4d spacetime can be generally curved manifolds, 
where the ${\cM^4} \sqcup {{\overline{\cM'}^4}}= \partial \cM^4$ as the two boundaries of a 5d bulk manifold $\cM^5$,
with the overline ${\overline{\cM'}^4}$ implying the orientation reversal ${\cM'}^4$. 
For the specific configuration in \Fig{fig:1} (a),
we can choose ${\cM^4}={\cM^3_{\text{space}}} \times {\cM^1_{\text{time}}}$ and ${\cM'^4}={\cM'^3_{\text{space}}} \times {\cM^1_{\text{time}}}$,
where the 1d time ${\cM^1_{\text{time}}}$ is shared 
and the ${\cM^3_{\text{space}}} \sqcup {\overline{\cM}^3_{\text{space}}}$ is the boundary of a 4-dimensional strip or a cylinder.
The generic $\theta$ and $\theta'$ vacua violate
the R, C, P, and RP symmetries, if we treat the gauge fields on both domain walls distinctly.

However, the gauge fields from $F$ and $F'$ on both sides are the same gauge field,
so we could even combine the effects into
a combined theta term in the limit when the width $w$ is small:
\bea \label{eq:theta-theta'}
\frac{\theta +\theta'}{8 \pi^2} \int_{\cM^4} \Tr[F \wedge F] .
\eea
In this case, we have to consider a finite-width strip as \Fig{fig:1} (a), 
while ${\cM^2}$ is identified as ${\cM'^2}$ in the spacetime integration range.

{\bf Impose the $\Z_2^{\rm P R} \equiv \Z_2^{\cP}$ parity-reflection symmetry on the theta term}: 
\label{Sol-Remark:5} 
If we impose the PR symmetry
on the parent theory with chiral and mirror fermions 
and with the ${\theta+\theta'}$ term in \eq{eq:theta-theta'}
(above the energy $\rE > \Lambda_{\SMG}$),
then the PR symmetry in \eq{eq:PR}
demands that the $\theta$ term on the domain wall on ${\cM^4}$
maps to the $\theta'$ term on the domain wall on ${\cM'^4}$
\bea
\frac{\theta}{8 \pi^2} \int_{\cM^4} \Tr[F \wedge F (t,x_j)] \big\vert_{y=-w}  
\mapsto 
\frac{\theta}{8 \pi^2} \int_{\cM'^4} -\Tr[F' \wedge F' (t,-x_j)] \big\vert_{y=w} 
=\frac{-\theta}{8 \pi^2} \int_{\cM'^4} \Tr[F \wedge F (t,x_j)] \big\vert_{y=w} .
\eea
Note that the right-hand side also defines the 
$\frac{\theta'}{8 \pi^2} \int_{\cM'^4} \Tr[F' \wedge F']  \big\vert_{y=w}$ 
term constrained by 
the $\Z_2^{\rm P R}$ symmetry on ${\cM'^4}$. 
Thus, the PR symmetry for the full parent theory 
implies that ${\theta'} =- \theta$ in \eq{eq:theta-theta'}, 
\bea \label{eq:theta=0}
\theta + \theta' =0 
\quad 
\Rightarrow 
\quad 
\frac{\theta +\theta'}{8 \pi^2} \int_{\cM^4} \Tr[F \wedge F] .
 =0.
\eea
Namely \eq{eq:thetaF-theta'F} and \eq{eq:theta-theta'} vanishes.

{\bf UV theory with ${\bar{\theta}_3 +\bar{\theta}'_3}=0$}:
Moreover, to respect a $\Z_2^{\rm PR}$ symmetry for an energy $E >\Lambda_{\rm SMG}$ (or likely within $M_{\rm P} > E >\Lambda_{\rm SMG}$),
we must demand a full parent field theory with an action that maps back to itself under ${\Z_2^{\rm PR}}$:
\bea \label{eq:UV-PR-action}
{\Z_2^{\rm PR}}: S_{\rm UV}=\int \dd t \dd^3 x \big( 
\cL_{\rm UV}(t, x_j, -w ) + \cL_{\rm UV}(t, x_j ,w)
\big) \;  {\mapsto} \;  S_{\rm UV}
\eea 
with its Lagrangian density $\cL_{\rm UV}(x,y,t)$ on $y = -w$ and $y = w$
at UV.
This implies that the $\Z_2^{\rm PR}$ symmetry is preserved at some UV scale, for the fermion doubling, 
all field contents, the kinetic, theta, and interaction terms in the Lagrangian at $y=-w$ and $y=w$, 
at least kinematically.

{\bf \emph{Without even assuming the
 mean-field interpretation} of the theta angle} in \eq{eq:def-mass},
the imposed ${\Z_2^{\rm PR}}$ symmetry demands the UV theory
\eq{eq:UV-PR-action} satisfying
${\theta_{}}
=-
{\theta_{}'}$
and 
$\arg (\det 
 M + g_{\rm SMG} \CO_{\rm SMG}
 )
=-
\arg (\det 
 M'^\dagger + g'^*_{\rm SMG} \CO'^\dagger_{\rm SMG} )$ in general.
So all together 
${\bar{\theta}_{}}
+{\bar{\theta}'_{}} =0$.

Note that (1) the chiral fermions on ${\cM^4}$ and the mirror fermions on ${\cM'^4}$ have the opposite chirality,
and (2) their chiral symmetries (with the opposite chiralities on the two domain walls) 
coupled to the same $\su(3) \times \su(2) \times \u(1)_{Y}$ gauge field, 
the chiral U(1) symmetry transformation will rotate $\bar\theta \mapsto \bar\theta + \upalpha$
and $\bar\theta' \mapsto \bar\theta' - \upalpha$ oppositely, but keeps the
$\bar\theta + \bar\theta' \mapsto \bar\theta + \bar\theta'$ invariant.

Because the above two reasons,
this means that the PR $\equiv \cP$ symmetry
at the parent theory solves the zero theta angle problem at a high energy,
since $\bar\theta + \bar\theta' =0$ and the chiral transformation with an appropriate $\upalpha$ like \eq{eq:bartheta0}
allows us to choose both 
${\bar{\theta}_{}}= {\theta_{}} - \upalpha =0$
and ${\bar{\theta}'_{}}= {\theta'_{}} + \upalpha =0$.


\item
We avoid Nielsen-Ninomiya 
fermion-doubling
to solve the 
\emph{nonperturbative chiral fermion 
problem} \cite{NielsenNinomiya1981hkPLB}
via SMG, by lifting the mirror SM fermion doubling spectrum 
to a finite energy gap $\Lambda_{\rm SMG}$ larger than the electroweak scale $\Lambda_{\rm EW}\sim246$GeV, 
leaving only the SM chiral fermion theory at the low energy (so to agree with phenomenology).
The criteria to fully gap the mirror doubling via SMG
through a \emph{symmetric deformation of QFT} \cite{WangWen2018cai1809.11171, NSeiberg-Strings-2019-talk, WangWanYou2112.14765}
is that the QFT of $G_{\SM_p}$ is fully anomaly-free (or in the zeroth $G_{\SM_p}$-cobordism class) 
--- luckily Ref.~\cite{GarciaEtxebarriaMontero2018ajm1808.00009, DavighiGripaiosLohitsiri2019rcd1910.11277, WW2019fxh1910.14668, JW2006.16996} checked all local or global anomalies of SM vanish in $G_{\SM_p}$.
We had done the same analysis in \Sec{sec:toy-SM}. 
Just like \Sec{sec:toy-SM}, we can use the modification of \cite{RazamatTong2009.05037, Tong2104.03997}'s model to
deform a gapless mirror SM to a featureless fully gapped mirror sector.
\begin{enumerate}[leftmargin=5.0mm, 
label=\textcolor{blue}{(\arabic*)}., ref={(\arabic*)}]
\item
\emph{15n or 16n Weyl-fermion SM and mirror SM}: 
We could consider the three-family $\su(3) \times \su(2) \times \u(1)_{Y}$ SM
with the 15 or 16 left-handed Weyl fermions (in \eqq{eq:SMrep})
$$
(\bar{d}_R \oplus {l}_L  \oplus q_L  \oplus \bar{u}_R \oplus   \bar{e}_R) \oplus n_{{\nu}_R} \bar{\nu}_R
$$ 
per family
on one domain wall, including the right-handed neutrino or not, $n_{{\nu_j}_R} \in \{0,1\}$.
We like to gap the mirror SM with the right-handed Weyl fermion
$$
(\bar{d}'_L \oplus {l}'_R  \oplus q'_R  \oplus \bar{u}'_L \oplus   \bar{e}'_L ) \oplus n_{{\nu}_R} \bar{\nu}'_L 
$$
but in the same $\su(3) \times \su(2) \times \u(1)_{Y}$ representation on the other domain wall.
{Note that chiral SM and mirror SM are only mapped onto each other under ${\Z_2^{\rm PR}}$ symmetry transformation; 
each particle has its own anti-particle under the charge conjugation C.}

We follow the Razamat-Tong model \cite{RazamatTong2009.05037}
to embed the 15 (or 16) Weyl-fermion SM into the 27 Weyl-fermion
with the following fermion matter content \eq{eq:SM-nonSUSY}
by introducing additional vector-like theory (3 right-handed and 3 left-handed Weyl fermions, 
in the second and third rows of \eq{eq:SM-nonSUSY}).

Following the Tong model \cite{Tong2104.03997},
we can find an enlarged symmetry \eq{eq:GSMLR}
$
G_{\text{SM-LR}_{p,p'}} \equiv
\frac{G_{\SM_q}\times\SU(2)_\mathrm{R}\times\U(1)_\mathrm{R}}{\Z_{p'}}
\equiv
\frac{\SU(3)\times\SU(2)_\mathrm{L}\times\SU(2)_\mathrm{R}\times\U(1)_\mathrm{L}\times\U(1)_\mathrm{R}}{\Z_p\times\Z_{p'}}
$
with Lie algebra 
$\su(3) \times  \su(2) \times \u(1)_Y \times  \su(2)_{\rm R} \times \u(1)_{\rm R}$
$\equiv$
$\su(3) \times$ $\su(2)_{\rm L} \times  \su(2)_{\rm R}$ 
$\times$ $\u(1)_{\rm L}  \times \u(1)_{\rm R}$ 
such that the 27-fermion
representation in \eq{eq:SM-nonSUSY}
becomes
\bea \label{eq:SMrepLR}
&&\hspace{-2mm}\text{chiral SM $\psi$ }
\big(\bar{d}_R \oplus {l}_L  \oplus q_L  \oplus \bar{u}_R \oplus   \bar{e}_R \oplus \bar{\nu}_R \big) 
\oplus {\bar{d}'}_R \oplus l'_L \text{ or}\cr
&&\hspace{-2mm}\text{mirror SM $\psi'_M$ }
\big(\bar{d}'_L \oplus {l}'_R  \oplus q'_R  \oplus \bar{u}'_L \oplus   \bar{e}'_L \oplus \bar{\nu}'_L \big) 
\oplus {\bar{d}''}_L \oplus l''_R: \cr
&&\sim\big((\overline{\bf 3},{\bf 1},{\bf 2})_{2,-1} \oplus ({\bf 1},{\bf 2},{\bf 2})_{-3,3}  
\oplus
({\bf 3},{\bf 2},{\bf 1})_{1,-2} \oplus (\overline{\bf 3},{\bf 1},{\bf 1})_{-4,2}  
\oplus ({\bf 1},{\bf 1},{\bf 1})_{6,-6} 
\oplus
{({\bf 1},{\bf 1},{\bf 2})_{0,-3}}\big)
\oplus
({\bf 3},{\bf 1},{\bf 1})_{-2,4}  
\oplus ({\bf 1},{\bf 2},{\bf 1})_{3,0}
.\cr
&&
\eea 
The chiral fermion multiplet is $\psi$, while the mirror fermion multiplet is $\psi'_{\rm M}$.
(In the first line of \eq{eq:SMrepLR}, the prime $'$ field is meant to indicate the extra addition to the original SM content \eq{eq:SMrep}. 
While in the second line of \eq{eq:SMrepLR}, 
the last prime is meant to indicate the field's mirror doubling partner.)
{See footnote \ref{ft:LR}, for the left and right notations,
we use the italic font $L$ and $R$ to denote that of spacetime symmetry,
while use the text font L and R for that of internal symmetry.}

\item
\emph{Gapping 15n vs 16n Weyl-fermion mirror SM}:\\
$\bullet$ If we only preserve $G_{\SM_p}$,
then without or with the sterile neutrino $\bar{\nu}_R={({\bf 1},{\bf 1})_{0}}$ (namely, 15 or 16 Weyl fermions per family)
can lead to a short-range entangled SMG phase (without any low energy TQFT).\\
$\bullet$
However, if we preserve not only $G_{\SM_p}$ but also some ${{\bf B} - {\bf L}}$ symmetry,
either we need a 16n Weyl-fermion model to achieve SMG,
or we need a 15n Weyl-fermion model plus SGTO with a low energy TQFT \cite{JW2006.16996, JW2012.15860}
 to achieve a symmetric gapped phase. 

\end{enumerate}

\item {\bf Mean-field vs non-mean-field SMG gapping}:

\emph{Mean-field gapping}:
Notice both Razamat-Tong model \cite{RazamatTong2009.05037}
and Tong model \cite{Tong2104.03997}
propose certain SMG gapping terms, however
 there are either mean-field condensates  \cite{Tong2104.03997}
or generic complex coupling coefficients \cite{RazamatTong2009.05037}
--- those make solving the Strong CP problem by 
Razamat-Tong or Tong's SMG deformation  \cite{RazamatTong2009.05037,Tong2104.03997} either difficult or impossible.\footnote{We had explained in \Sec{sec:toy-SM}
why the mean-field condensates $\<\phi\>\neq 0$ fails to solve the Strong CP problem.\\
Here we explain why the smooth confinement deformation of $\CN=1$ supersymmetrized SM's model in \cite{RazamatTong2009.05037}
is also difficult to solve the Strong CP problem.
First, one introduces three copies of \cite{RazamatTong2009.05037,Tong2104.03997}'s model
with family indices ${\rm I},{\rm J}\in \{1,2,3 \}$, 
and a new $\SU(2)' \times \SU(2)'' \times \SU(2)'''$ gauge theory.
Precisely, 
for the supersymmetrized $\CN=1$ chiral multiplets of \eq{eq:SMrepLR}
denoted in their superfield forms ${\rm D}, {\rm L}, {\rm Q}, {\rm U}, {\rm E}, {\rm D}', {\rm L}', {\rm N}$,
their supersymmetric $\CW$ potential has three family mixing:
\bea
&&\CW_{\rm UV}=\lambda_{{\rm I}{\rm J}}^{\rm U}{\rm D}^{\rm I}{\rm D}^{\rm I}{\rm U}^{\rm J} +
\lambda_{{\rm I}{\rm J}}^{\rm Q}{\rm L}^{\rm I}{\rm D}^{\rm I}{\rm Q}^{\rm J} +
\lambda_{{\rm I}{\rm J}}^{\rm D}{\rm D}^{\rm I}{\rm N}^{\rm I}{\rm D'}^{\rm J}
+\dots, \cr
\quad
&&\CW_{\rm IR}=\lambda_{{\rm I}{\rm J}}^{\rm U} \tilde{\rm U}^{\rm I}{\rm U}^{\rm J}+
\lambda_{{\rm I}{\rm J}}^{\rm Q} \tilde{\rm Q}^{\rm I}{\rm Q}^{\rm J}+
\lambda_{{\rm I}{\rm J}}^{\rm D} \tilde{\rm D}^{\rm I}{\rm D'}^{\rm J}+\dots.
\quad
\eea
Each $\SU(2)$ couples to one of three families of 
an extended internal $\su(2)_{{\rm R}_1} \times \su(2)_{{\rm R}_2} \times \su(2)_{{\rm R}_3}$ symmetry
of the $\su(3) \times  \su(2)_{\rm L} \times \u(1)_Y$ SM,
which guarantees the mirror SM to reach the fully gapped SMG phase
by the s-confinement once $\su(2)_{{\rm R}_1} \times \su(2)_{{\rm R}_2} \times \su(2)_{{\rm R}_3}$ are dynamically gauged. 
Any ${\theta}$ of $\su(2)_{{\rm R}_1} \times \su(2)_{{\rm R}_2} \times \su(2)_{{\rm R}_3}$ can be also removed
by the chiral and mirror sectors imposed by the $\Z_2^{\rm R}$ symmetry, so ${\theta}=-{\theta}'=0$.
But there is another worry that
any complex coefficient $\lambda_{{\rm I}{\rm J}}$ etc.~in the superpotential
 can affect the ${\bar{\theta}'_{3}}$.
Even if any $\lambda_{{\rm I}{\rm J}} \in \C$
is rotated away via
$\lambda_{{\rm I}{\rm J}} \equiv 
U_{{{\rm I}{\rm I}'}}^{\dagger}\lambda_{{\rm I'}{\rm J}'}^{\diag} U_{{\rm J}'{\rm J}}^{\dagger}$,
we need to further ask whether this basis redefinition by $U$ rotation results in any analogous CKM matrix to the
quarks coupling to $\su(2)_{\rm R}$ or $\su(2)_{\rm L}$. In general, there are too many complex coefficients in the \cite{RazamatTong2009.05037}'s model.
Generally, the complex coefficients in \cite{RazamatTong2009.05037}'s model violate the CP and T symmetry, thus the Strong CP problem cannot be solved directly within  \cite{RazamatTong2009.05037}.
}

\begin{figure}[!h] 
\includegraphics[width=.8\textwidth]{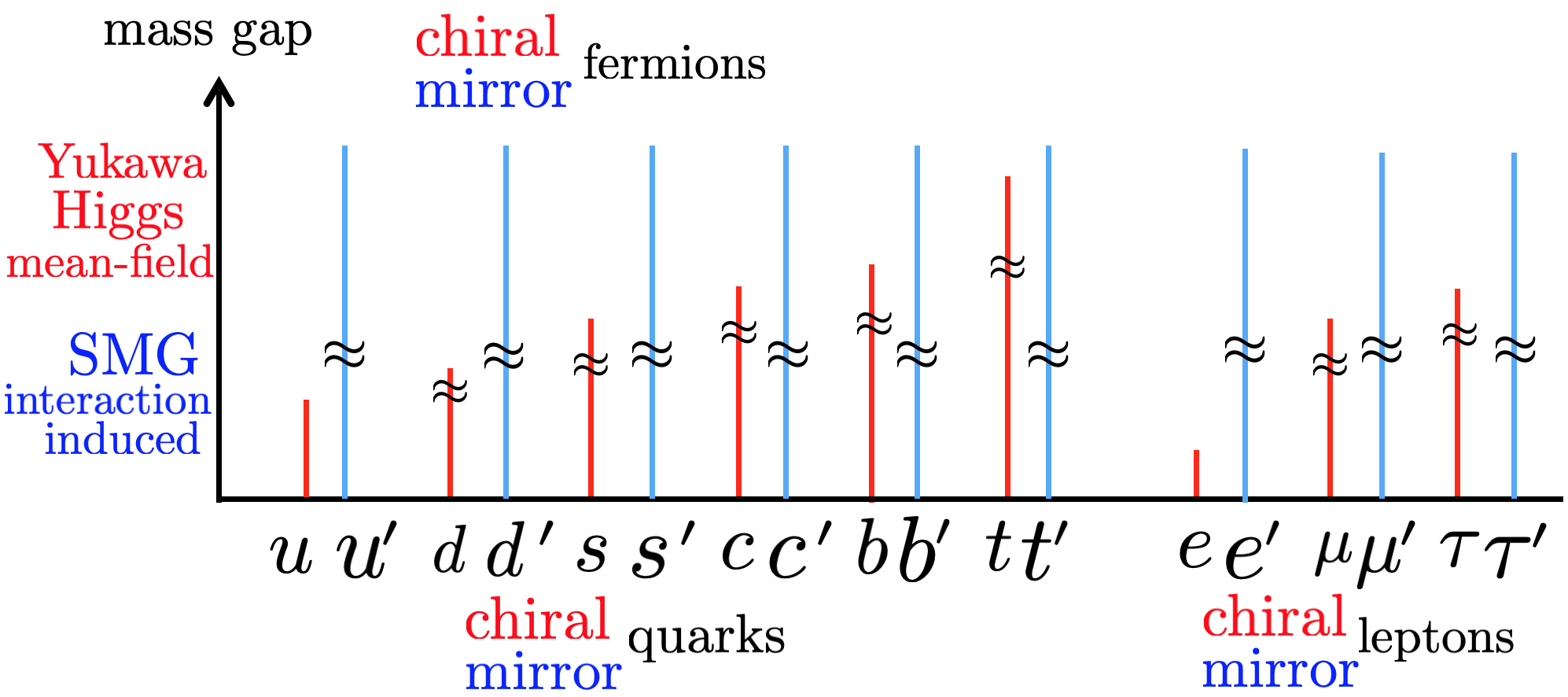}
\caption{A schematic plot on the Higgs-induced contribution to SM's quark and lepton masses in Table \ref{tab:current-mass}(a)
(shown in the red color)
while the SMG-induced mirror SM's quark and lepton masses entirely (shown in the blue color).
The SMG-induced mirror SM's mass gap the fermion doubling completely hidden at UV, 
so only the original chiral SM is observed at low energy.}
\label{fig:Higgs-SMG-induced-mirror}
\end{figure}

\emph{Non-mean-field gapping}:
In contrast, in \Sec{sec:toy-SM},
we already propose the non-mean-field multi-fermion interactions in \eq{eq:Yukawa-full-gap-multi-fermion} and \eq{eq:Yukawa-full-gap-multi-fermion-2}
that can gap one generation of quarks and leptons. Then here we just need to apply the same non-mean-field multi-fermion interactions
to gap the three generations of the mirror fermions. This mirror fermion gapping implies 
the mass spectrum like \Fig{fig:Higgs-SMG-induced-mirror}.

 \item {\bf Solve the Strong CP problem ${\bar{\theta}_3 =\bar{\theta}'_3}=0$ from UV to IR}:
 
Earlier we had explained that the imposed ${\Z_2^{\rm PR}}$ symmetry makes the UV theory
\eq{eq:UV-PR-action} with no Strong CP angle because ${\bar{\theta}_3 =\bar{\theta}'_3}=0$ at that $M_{\rm P} > E >\Lambda_{\rm SMG}$ scale.

However, \\
(1) at the lower energy $E <\Lambda_{\rm SMG}$, the dominant SMG interaction only occurs at the mirror fermion sector;\\
(2) while even at a further lower energy $E <\Lambda_{\rm EW}$, the dominant Higgs condensation only occurs at the original chiral fermion sector.\\
--- both cases mean to ``\emph{spontaneously}'' breaks the ${\Z_2^{\rm PR}}$ symmetry at IR (see footnote \ref{ft:PR-borken}).
Once ${\Z_2^{\rm PR}}$ is broken at low energy, one may be worried that 
both ${\bar{\theta}_3}$ and ${\bar{\theta}'_3}$ can be deviated from zeros.
We need to argue why ${\bar{\theta}_3}$ and/or ${\bar{\theta}'_3}$ stays close to zero at IR.

Similar to the argument of the UV parent theory,
now at low energy IR, we include
the generalized mass matrix $ M_{\rm total}$  that
\bea
\< M_{\rm total}\> =
\begin{pmatrix} 
\< M + g_{\rm SMG} \CO_{\rm SMG}
\> & 0 \\
0 & \< M'^\dagger + g'^*_{\rm SMG} {\CO'}_{\rm SMG}^\dagger \>
\end{pmatrix} 
=
\begin{pmatrix} 
\< M\> & 0 \\
0 & \< M'^\dagger\>
\end{pmatrix} 
\eea
that includes the mean-field mass matrix $M=\< M \>$
as well as the multi-fermion interaction type of SMG that offers no mean-field mass $\<\CO_{\rm SMG}\>=0$.
The computation of the 
${\bar{\theta}_{}}+{\bar{\theta}'_{}}$ for the full theory with the total fermion mass $M_{\rm total}$ contribution of the whole system
is:
\bea \label{eq:thetathetaMF}
{\bar{\theta}_{}}
+{\bar{\theta}'_{}} 
&\equiv&
{\theta_{}}
+
{\theta_{}'}
+
\arg (\det 
\< M_{\rm total}\>)
\cr
&\equiv&
{\theta_{}}
+
{\theta_{}'}
+
\arg (\det 
\< M + g_{\rm SMG} \CO_{\rm SMG}
\>  )
+
\arg (\det 
\< M'^\dagger + g'^*_{\rm SMG} \CO'^\dagger_{\rm SMG} \>)
\cr
&=&{\theta_{}}
+
{\theta_{}'}
+
\arg (\det 
\< M + g_{\rm SMG} \CO_{\rm SMG} \>
\cdot \det 
\< M'^\dagger + g'^*_{\rm SMG} \CO'^\dagger_{\rm SMG} \>) 
\cr
&=&
{\theta_{}}
+
{\theta_{}'}
+
\arg (\det 
(\< M + g_{\rm SMG} \CO_{\rm SMG} \>
\cdot
\< M'^\dagger + g'^*_{\rm SMG} \CO'^\dagger_{\rm SMG} \>))\cr
&=&
{\theta_{}}
+
{\theta_{}'}
+
\arg (\det 
(\< M\>
\< M'^\dagger \>
+
g_{\rm SMG} \< \CO_{\rm SMG}\>\< M'^\dagger \>
+
g'^*_{\rm SMG} \< \CO'^\dagger_{\rm SMG}\>\< M \>
+g_{\rm SMG}  g'^*_{\rm SMG} \< \CO_{\rm SMG}\>\< \CO'^\dagger_{\rm SMG}\>
)).
\eea
Here are some comments concerning the
${\bar{\theta}_{}}
+{\bar{\theta}'_{}}$ of the whole system at different energy scales $E$:
\begin{enumerate}[leftmargin=5.0mm, 
label=\textcolor{blue}{(\arabic*)}., ref={(\arabic*)}]
\item $M_{\rm P} > E >\Lambda_{\rm SMG}$:

$\bullet$ \emph{Without even assuming the
 mean-field interpretation} of the theta angle in \eq{eq:def-mass},
the imposed ${\Z_2^{\rm PR}}$ symmetry demands the UV theory
\eq{eq:UV-PR-action} satisfying
${\theta_{}}
=-
{\theta_{}'}$
and 
$\arg (\det 
 M + g_{\rm SMG} \CO_{\rm SMG}
 )
=-
\arg (\det 
 M'^\dagger + g'^*_{\rm SMG} \CO'^\dagger_{\rm SMG} )$ in general.

$\bullet$
Moreover, at this high energy scale $M_{\rm P} > E >\Lambda_{\rm SMG}$ at UV,
there is no mean-field condensation yet (neither SM Higgs nor the SMG's $\xi_{\rm I} \CO_{\rm SMG, IJ} \psi_{\rm J}$ condenses yet.
So $\< M \>=\< M'\>=0$ within $M_{\rm P} > E >\Lambda_{\rm SMG}$.  
In fact 
$\< \xi_{\rm I} \psi_{\rm J}\>=
\< \CO_{\rm SMG, IJ} \>=0$ is always strictly zero).

$\bullet$
\emph{Assuming the
 mean-field interpretation} 
 of the theta angle in \eq{eq:def-mass},
 then
$\arg (\det 
\< M + g_{\rm SMG} \CO_{\rm SMG}
\>  )
=-
\arg (\det 
\< M'^\dagger + g'^*_{\rm SMG} \CO'^\dagger_{\rm SMG} \>)=0$ is strictly zero within $M_{\rm P} > E >\Lambda_{\rm SMG}$.
An appropriate chiral transformation $\upalpha$ like \eq{eq:bartheta0}
allows us to choose both $\bar\theta=0$ and $\bar\theta'=0$

\item $\Lambda_{\rm SMG} > E > \Lambda_{\rm EW}$: Below and around the SMG scale, the ${\Z_2^{\rm PR}}$ symmetry is  ``\emph{spontaneously}'' broken
due to the SMG interaction $\< \xi'_{\rm I} \CO'_{\rm SMG, IJ} \psi'_{\rm J}\>\neq 0$ on the mirror fermion sector, but
$\< \xi_{\rm I} \CO_{\rm SMG, IJ} \psi_{\rm J}\>=0$ on the original SM sector
(also strictly 
$\< \xi_{\rm I} \psi_{\rm J}\>=
\< \CO_{\rm SMG, IJ} \>=
\< \xi'_{\rm I} \psi'_{\rm J}\>=
\< \CO'_{\rm SMG, IJ} \>=0$ always for non-mean-field SMG).

$\bullet$ \emph{Without even assuming the
 mean-field interpretation} of the theta angle in \eq{eq:def-mass},
we can still read that
${\bar{\theta}_{}}
+{\bar{\theta}'_{}} 
=
 {\theta_{}}
+
{\theta_{}'}
+
\arg (\det 
( M
 M'^\dagger 
+
g_{\rm SMG}  \CO_{\rm SMG}  M'^\dagger 
+
g'^*_{\rm SMG} \CO'^\dagger_{\rm SMG}  M 
+g_{\rm SMG}  g'^*_{\rm SMG}  \CO_{\rm SMG}  \CO'^\dagger_{\rm SMG}
))
$.
\Refe{WangCTorPProblem2207.14813}
argues that in terms of the configuration in
\Fig{fig:1} (b), the most dominant mass term in 
\eq{eq:thetathetaMF} 
is from $g'^*_{\rm SMG} \CO'^\dagger_{\rm SMG}  M$.

$\bullet$
\emph{Assuming the
 mean-field interpretation} 
 of the theta angle in \eq{eq:def-mass},
$
{\bar{\theta}_{}}
+{\bar{\theta}'_{}} 
\equiv
{\theta_{}}
+
{\theta_{}'}
+
\arg (\det 
(\< M\>
\< M'^\dagger \>
+
g_{\rm SMG} \< \CO_{\rm SMG}\>\< M'^\dagger \>
+
g'^*_{\rm SMG} \< \CO'^\dagger_{\rm SMG}\>\< M \>
+g_{\rm SMG}  g'^*_{\rm SMG} \< \CO_{\rm SMG}\>\< \CO'^\dagger_{\rm SMG}\>
)).$
Because $E > \Lambda_{\rm EW}$, Higgs does not yet condense in the SM sector,
so 
$\< M\> =0$, also $\< M'^\dagger \> = \< \CO_{\rm SMG}\>=\< \CO'^\dagger_{\rm SMG}\>=0$.

\item $E < \Lambda_{\rm EW}$: Below and around $\Lambda_{\rm EW}$, 
the ${\Z_2^{\rm PR}}$ symmetry is further ``\emph{spontaneously}'' broken, 
because the Higgs induces $\< M \>\neq 0$ on the SM sector.

$\bullet$ \emph{Without even assuming the
 mean-field interpretation} of the theta angle in \eq{eq:def-mass},
\Refe{WangCTorPProblem2207.14813}
argues that in terms of the configuration in
\Fig{fig:1} (b), the most dominant mass term in \eq{eq:thetathetaMF} 
is from $g'^*_{\rm SMG} \CO'^\dagger_{\rm SMG}  M$.
The Higgs-induced mass $M$ is in the chiral fermion sector,
while the SMG-induced mass with operator $g'^*_{\rm SMG} \CO'^\dagger_{\rm SMG}$ is in the mirror fermion sector.
Different mass-generating mechanisms on the chiral and mirror sector may give a generic complex nonzero phase.\\[-2mm]

{However, because the SMG \emph{multi-fermion interaction} and disorder scalar interaction are \emph{highly irrelevant operators} at IR. 
The lower the energy, the weaker effects are these interactions on the IR correction of $\bar{\theta}$.}
There gives another reason that the IR correction to our UV solution $\bar{\theta} = 0$ is extremely small. \\

$\bullet$
\emph{Assuming the
 mean-field interpretation} 
 of the theta angle in \eq{eq:def-mass} which implies \eq{eq:thetathetaMF}.
Then, Higgs condense $\< M\> \neq 0$
in the SM sector, but others have no mean-field values $\< M'^\dagger \> = \< \CO_{\rm SMG}\>=\< \CO'^\dagger_{\rm SMG}\>=0$.
Thus \eq{eq:thetathetaMF} 
only contains 
${\bar{\theta}_{}}
+{\bar{\theta}'_{}} 
\equiv
{\theta_{}}
+
{\theta_{}'}$
while the
$(\< M\>
\< M'^\dagger \>
+
g_{\rm SMG} \< \CO_{\rm SMG}\>\< M'^\dagger \>
+
g'^*_{\rm SMG} \< \CO'^\dagger_{\rm SMG}\>\< M \>
+g_{\rm SMG}  g'^*_{\rm SMG} \< \CO_{\rm SMG}\>\< \CO'^\dagger_{\rm SMG}\>)=0$.\\[-2mm]

As long as
${\bar{\theta}_{}}
+{\bar{\theta}'_{}} 
\equiv
{\theta_{}}
+
{\theta_{}'}=0$ at UV 
as \eq{eq:theta=0} set by ${\Z_2^{\rm PR}}$,
our solution has a smaller or zero
${\bar{\theta}_{}} +{\bar{\theta}'_{}} \approx 0$ at IR, 
even better than the higher-loop calculation arguments given in Parity Solution \cite{BabuMohapatra1989, BabuMohapatra1989rbPRD, BarrChangSenjanovic1991qxPRL}.
Because \Refe{BabuMohapatra1989, BabuMohapatra1989rbPRD, BarrChangSenjanovic1991qxPRL} requires only the Higgs mechanism, 
thus Higgs mixing on both the chiral and mirror sectors can generate complex phases at the higher-order quantum corrections 
beyond the tree-level semiclassical analysis.
Here instead we have two mechanisms in our solution: Higgs mechanism dominates on the chiral sector and the SMG dominates on the mirror sector, 
and there is \emph{no mean-field mass matrix mixing} to generate quantum corrections to $\bar{\theta} +{\bar{\theta}'_{}}  \approx 0$.\\[-2mm]

We can take any fermion (here any mirror fermion) that receives its entire mass from SMG
and do a chiral transformation with an appropriate $\upalpha$ like \eq{eq:bartheta0}
allows us to choose both 
${\bar{\theta}_{}}= {\theta_{}} - \upalpha \approx 0$
and ${\bar{\theta}'_{}}= {\theta'_{}} + \upalpha \approx0$ even including the IR correction.
Theoretically this Scenario in \Sec{sec:SecondSolution} provides a candidate Strong CP solution.
\end{enumerate}

\end{enumerate}

\newpage
\section{Conclusion and Comparison to Other Strong CP Solutions}
\label{sec:Conclusion}

To conclude, let us summarize and compare the two SMG-based Strong CP solutions, of \Sec{sec:FirstSolution} and \Sec{sec:SecondSolution},
with other well-known Strong CP solutions.\\

\noindent
{\bf First Solution: Symmetric Mass
Gap within the Chiral Fermion}:

\begin{enumerate}[leftmargin=-0mm, label=\textcolor{blue}{\arabic*}., ref={\arabic*}]
 
\item The first solution in \Sec{sec:FirstSolution} is related to the SMG modification of {\bf 't Hooft massless $u$ quark solution} \cite{tHooft1976ripPRL}.
In 't Hooft solution \cite{tHooft1976ripPRL}, if any of the quark $\psi$ (say $u$ quark) has no mean-field nor Higgs mass,
then we can do the chiral transformation on this quark $\psi$ alone to rotate the $\theta$ away but without gaining a complex phase 
in the mean-field mass matrix (since there is no mean-field mass for $\psi$).
This sets $\bar{\theta}=0$. Our SMG version in \Sec{sec:FirstSolution}  modifies the  't Hooft solution
by allowing the SMG-induced non-mean-field mass on the $u$ quark and a set of $G$-symmetry anomaly-free fermions.
For the SM example, we can take $G=G_{\SM_p} \equiv({\SU(3)} \times {\SU(2)} \times {\U(1)_Y})/{\Z_p}$, $p=1,2,3,6$ with the Lie algebra $\su(3) \times  \su(2) \times \u(1)_Y$.

\item The first solution in \Sec{sec:FirstSolution} is also related to the SMG modification of {\bf Peccei-Quinn axion solution} \cite{PecceiQuinn1977hhPRL,PecceiQuinn1977urPRD, Weinberg1977ma1978PRL, Wilczek1977pjPRL}.
In the weakly gauge limit or global symmetry limit of $G$, 
the Peccei-Quinn solution can be regarded as an ``approximate symmetry breaking'' solution (see footnote \ref{ft:Peccei-Quinn-transition})
that uses the mean-field Yukawa-Higgs symmetry breaking mass term.
The arbitrariness of $\bar\theta \in [0, 2 \pi)$ can be relaxed by the symmetry-breaking dynamics to $\bar\theta \simeq 0$.
The fluctuation around a fixed $\bar\theta = 0$ gives a low-energy Goldstone mode or an axion.
The transition from the mean-field mass to the interacting SMG mass in \Sec{sec:FirstSolution}
is analogous to the phase transition between the \emph{ordered} phase and the \emph{disordered} phase
--- for example, the famous superfluid-to-insulator type of phase transition in the condensed matter \cite{Fisher1989zza}.

\end{enumerate}

\noindent
{\bf Second Solution: Symmetric Gapped Mirror Fermion}:

We have combined four physics together to build our second model in \Sec{sec:SecondSolution}:\\
(i) \emph{Nielsen-Ninomiya}  
\emph{fermion-doubling} \cite{NielsenNinomiya1981hkPLB} 
 imposes a discrete spatial parity-reflection $\Z_2^{\rm PR}$ symmetry between the chiral SM and mirror SM.\\
(ii) \emph{Strong CP problem} is solved by respecting the $\Z_2^{\rm PR}$ symmetry so ${\bar{\theta}_{3}}= {\bar{\theta}_{3}}' =0$ for the chiral and mirror SM.\\
(iii) SMG gaps the mirror SM fermion doubler to leave only a chiral SM at low energy, which also solves the 
{Nielsen-Ninomiya} fermion-doubling and chiral fermion problems altogether.\\
(iv) \emph{Parity-reflection  $\Z_2^{\rm PR}$ symmetry is maximally broken} in the weak force
due to gapping out the mirror fermion, which also gives a reason for parity violation in the weak force \cite{LeeYang1956}.

{We compare our second model with other previously proposed Strong CP solutions.}
One class of solutions is based on imposing discrete CP \cite{Nelson1983zb1984PLB,Barr1984qx1984PRL} 
or P symmetries \cite{BabuMohapatra1989, 
BabuMohapatra1989rbPRD,  BarrChangSenjanovic1991qxPRL, Hook2014cda1411.3325}:

\begin{enumerate}[leftmargin=-0mm, label=\textcolor{blue}{\arabic*}., ref={\arabic*}]
\item
Nelson-Barr \cite{Nelson1983zb1984PLB,Barr1984qx1984PRL}
starts with a CP invariant theory then spontaneously breaks the CP.
\item
Barr-Chang-Senjanovic  \cite{BarrChangSenjanovic1991qxPRL}
implements the left-right model \cite{SenjanovicMohapatra1975} or two copies of $\su(2)_{\rm L} \times \su(2)_{\rm R}$ for the weak force,
and imposes a diagonal symmetry mixing 
between the discrete CP or P and the discrete \emph{internal} symmetry exchanges the left and right $\su(2)$.
Typically, the gauge group needs to be doubled at least for the weak force (so $\su(2)_{\rm L} \times \su(2)_{\rm R}$).
\item
In comparison, 
our second solution in \Sec{sec:SecondSolution} is still different from all of the above ---  
we impose the discrete parity-reflection spatial symmetry $\Z_2^{\rm PR}$ only due to {Nielsen-Ninomiya} fermion doubling \cite{NielsenNinomiya1981hkPLB}  
between the energy $M_{\rm P} > E >\Lambda_{\rm SMG}$, 
but without imposing any extra discrete \emph{internal} symmetry.
The gauge group in principle needs not to be doubled, see for example the analogous 1+1d toy model does not double the gauge group \cite{WangCTorPProblem2207.14813}.
However, it turns out that the convenient SMG deformation of the SM \cite{RazamatTong2009.05037, Tong2104.03997} naturally 
happens to introduce $\su(2)_{\rm L} \times \su(2)_{\rm R}$.

In comparison, our second model is still different from all of the above --- 
we respect a discrete reflection R symmetry only due to {Nielsen-Ninomiya} fermion doubling between the energy $M_{\rm P} > E >\Lambda_{\rm SMG}$, 
also we keep the original chiral SM gauge group with \emph{only} a single $\su(2)_{\rm L}$, \emph{not} the vector-like left-right model,
\emph{nor} need two copies of the gauge group.
In addition to the SM, our second Strong CP solution
also works for other chiral fermion model 
(e.g., Georgi-Glashow \cite{Georgi1974syUnityofAllElementaryParticleForces} or flipped $\su(5)$ \cite{Barr1982flippedSU5}) 
and for the vector-like (e.g., Pati-Salam $\su(4) \times \su(2)_{\rm L} \times \su(2)_{\rm R}$ \cite{Pati1974yyPatiSalamLeptonNumberastheFourthColor} or left-right \cite{SenjanovicMohapatra1975}) 
models, 
by imposing the $\Z_2^{\rm PR}$ then gapping the mirror fermion sector via SMG.\\

\end{enumerate}

One predictive signature 
of both SMG-based models, in  \Sec{sec:FirstSolution} and \Sec{sec:SecondSolution}, 
is that even if we turn off the conventional SM Higgs or SM gauge interactions,
some SM fermions or mirror fermions can still be highly interacting,
mediated through hypothetical direct multi-fermion or disordered mass-field interactions.

If our \emph{theoretical solutions} in  \Sec{sec:FirstSolution} and \Sec{sec:SecondSolution}
turn out to be also favored as valid \emph{phenomenological solutions} to the Strong CP problem,
then this implies that the fermions in our Nature can indeed be \emph{more interacting} among fermions themselves
than what we used to think of in the SM.


\section{Acknowledgments}

JW especially thanks Yuta Hamada for the precious collaboration on \cite{HamadaWang2209.15244},
for several crucial discussions at the initial stage, and for various conversations during the development of this project.
JW also thanks William Detmold, Daniel Jafferis, Dmitri Kharzeev, Zohar Komargodski, 
Matthew Reece, Matthew Strassler, Edward Witten, and Yi-Zhuang You for very helpful conversations.
{JW appreciates the generous feedback from many participants of Harvard CMSA Phase Transitions and Topological Defects in the Early Universe workshop (August 2-5, 2022)
and Generalized Global Symmetries, Quantum Field Theory, and Geometry (September 19-23, 2022).}
JW is supported by Harvard University CMSA.



\bibliography{BSM-CP.bib,BSM-CP-2d.bib}

\end{document}